\documentclass{aa}  

\usepackage{graphicx}
\usepackage{txfonts}
\usepackage{natbib}
\usepackage{subcaption}
\usepackage{lscape} 
\usepackage{adjustbox}
\usepackage{xcolor}
\bibpunct{(}{)}{;}{a}{}{,} % to follow the A&A style

\newcommand{\farc}{$^{\prime\prime}$}
\newcommand{\farcmin}{$^{\prime}$}

\newcommand{\kalfa}{Fe $\rm K\alpha$}
\DeclareMathOperator{\sinc}{sinc} % sinc function

\usepackage{dcolumn}
\newcolumntype{L}{D{.}{.}{2,3}}
\newcolumntype{P}{D{\pm}{\pm}{6,6}}
\newcommand{\mcdash}[1]{\multicolumn{1}{#1}{---}}

\newcommand{\noop}[1]{}

\begin{document}

\title{Localizing narrow Fe $\rm K\alpha$ emission within bright AGN \thanks{The full version of Table\,\ref{t:ObsInfo} is available in electronic form at the CDS via http://cdsweb.u-strasbg.fr/cgi-bin/qcat?J/A+A/} }

   \author{
    Carolina Andonie\inst{1,2,3}
    \and
    Franz E. Bauer\inst{1,2,4} %\ORCID{0000-0002-8686-8737}
    \and
    Rosamaria Carraro\inst{5}
    \and
    Patricia Arevalo\inst{5}
    \and
% Collaborators (alphabetical)
    David M. Alexander\inst{3}
    \and
    William N. Brandt\inst{6,7,8}
    \and
    Johannes Buchner\inst{9}
    \and
    Adam He\inst{10,11}
    \and
    Michael J. Koss\inst{12,4} %\ORCID{0000-0002-7998-9581}
    \and
    Claudio Ricci\inst{13,14,15}
    \and
    Vicente Salinas\inst{5,1}
    \and
    Manuel Solimano\inst{13,1}
    \and
    Alessia Tortosa\inst{13}
    \and
    Ezequiel Treister\inst{1}
}
   \institute{  Instituto de Astrof\'\i{}sica and Centro de Astroingenier{\'{\i}}a, Facultad de F\'\i{}sica, Pontificia Universidad Cat\'{o}lica de Chile, Casilla 306, Santiago 22, Chile
              %\email{cpandonie@uc.cl}
         \and
   Millennium Institute of Astrophysics (MAS), Nuncio Monse{\~{n}}or S{\'{o}}tero Sanz 100, Providencia, Santiago, Chile
         \and
   Centre for Extragalactic Astronomy, Department of Physics, Durham University, South Road, Durham DH1 3LE, UK  
         \and 
   Space Science Institute, 4750 Walnut Street, Suite 205, Boulder, Colorado 80301
         \and
   Instituto de F\'\i{}sica y Astronom\'\i{}a, Universidad de Valpara\'\i{}so, Gran Bretaña 1111, Playa Ancha, Valpara\'\i{}so, Chile   
         \and
    Department of Astronomy and Astrophysics, 525 Davey Lab, The Pennsylvania State University, University Park, PA 16802, USA
         \and
    Institute for Gravitation and the Cosmos, The Pennsylvania State University, University Park, PA 16802, USA
         \and
    Department of Physics, 104 Davey Laboratory, The Pennsylvania State University, University Park, PA 16802, USA
         \and
    Max-Planck-Institut für Extraterrestrische Physik, Giessenbachstrasse, 85748 Garching, Germany
         \and
    Department of Physics and Astronomy, University of Southern California, Los Angeles, CA 90089, USA
         \and
    Department of Astronomy, Columbia University, New York, NY 10027, USA
         \and
    Eureka Scientific, 2452 Delmer Street Suite 100, Oakland, CA 94602-3017, USA
         \and
    N{\'u}cleo de Astronom{\'i}a de la Facultad de Ingenier{\'i}a, Universidad Diego Portales, Av. Ej{\'e}ercito Libertador 441, Santiago 22, Chile
         \and
    Kavli Institute for Astronomy and Astrophysics, Peking University, Beijing 100871, People’s Republic of China
         \and
    George Mason University, Department of Physics \& Astronomy, MS 3F3, 4400 University Drive, Fairfax, VA 22030, USA}

   \date{Received October 20, 2021; accepted April 12, 2022}

% \abstract{}{}{}{}{} 
% 5 {} token are mandatory
 
  \abstract
 {The 6.4\,keV \kalfa{} emission line is a ubiquitous feature in X-ray spectra of active galactic nuclei (AGN), and its properties track the interaction between the variable primary X-ray continuum and the surrounding structure from which it arises.}
  % aims heading (mandatory)
{  We clarify the nature and origin of the narrow \kalfa{}  emission using X-ray spectral, timing, and imaging constraints, plus possible correlations to AGN and host galaxy properties, for 38 bright nearby AGN ($z < 0.5$) from the Burst Alert Telescope AGN Spectroscopic
Survey.}
 % methods heading (mandatory)
 {Modeling Chandra and XMM-Newton spectra, we computed line full-width half-maxima (FWHMs) and constructed \kalfa{} line and 2–10 keV continuum light curves. The FWHM provides one estimate of the \kalfa{} emitting region size, $R_{\rm Fe K\alpha}$, assuming virial motion. A second estimate comes from comparing the degree of correlation between the variability of the continuum and line-only light curves, compared to simulated light curves. Finally, we extracted Chandra radial profiles to place upper limits on $R_{\rm Fe K\alpha}$.}
  % results heading 
 {For 90$\%$ (21/24) of AGN with FWHM measurements, $R_{\rm Fe K\alpha}$ is smaller than the fiducial dust sublimation radius, $R_{\rm sub}$. From timing analysis, 37 and 18 AGN show significant continuum and \kalfa{} variability, respectively. Despite a wide range of variability properties, the constraints on the \kalfa{} photon reprocessor size independently confirm that $R_{\rm Fe K\alpha}$ is smaller than $R_{\rm sub}$ in 83$\%$ of AGN. Finally, the imaging analysis yields loose upper limits for all but two sources; notably, the Circinus Galaxy and NGC 1068 show significant but subdominant extended \kalfa{} emission out to $\sim$100 and $\sim$800 pc, respectively.}
 {Based on independent constraints, we conclude that the majority of the narrow \kalfa{} emission in typical AGN predominantly arises from regions smaller than and presumably inside $R_{\rm sub}$, and thus it is associated either with the outer broad line region or outer accretion disk. However, the large diversity of continuum and narrow \kalfa{} variability properties are not easily accommodated by a universal scenario.}
 

\maketitle
%
%-------------------------------------------------------------------

\section{Introduction}

X-ray emission is a universal characteristic of active galactic nuclei (AGN), thought to arise from inverse Compton scattering of optical-UV photons from the accretion disk by hot electrons in the corona  \citep[e.g.,][]{1991ApJ...380L..51H}. The intrinsic X-ray emission takes a power-law spectral form ($f(E){\propto}E^{-\Gamma}$, with  typical photon indices of ${\langle}\Gamma{\rangle}{\sim}1.8$--$2.1$; e.g., \citealt{1994MNRAS.268..405N, 2009WinterL, 2011A&A...530A..42C}), but it can be modified due to an interaction with matter in the vicinity of the central supermassive black hole (SMBH). In particular, Compton scattering and photoelectric absorption of the primary X-ray continuum lead to two important features in the X-ray spectrum: the \kalfa{} emission line and the so-called Compton-hump. By studying these reprocessed features, together known as the so-called AGN reflection component, we can infer the physical properties of the matter from which they originate, and hence probe the circumnuclear environments of central SMBHs.  

The \kalfa{} feature at 6.4 keV is produced by fluorescence processes related to the absorption of higher energy X-ray photons by neutral Fe atoms. Its spectral profile is generally comprised of broad and narrow components. The narrow component of the \kalfa{} line \citep[
$\rm FWHM \lesssim 10,000\:km\:s^{-1}$; e.g.,][]{2001MNRAS.323L..37L,2004ApJ...604...63Y,2010ApJS..187..581S} is a ubiquitous spectral feature of AGN, and in a majority of cases the only component immediately visible, while the broad component is harder to pin down since it requires exceptional statistics and broad energy coverage to decouple the line from the underlying continuum and absorption components \citep[e.g.,][]{2006AN....327.1032G, 2014ApJ...787...83M}.
Nonetheless, when present, reverberation studies suggest that the broad component originates from a compact zone, only a few Schwarzschild radius ($\rm r_g$) in extent, around the SMBH \citep[e.g.,][]{2014MNRAS.438.2980C}, and hence it is strongly affected by Doppler and gravitational broadening \citep[e.g.,][]{1995MNRAS.272L...9M,1995Natur.375..659T,1995ApJ...453L..81Y}. On the other hand, the narrow component has been thought to be produced somewhere amongst the outer accretion disk, the broad line region (BLR), and the torus clouds \citep[e.g.,][]{1994MNRAS.267..743G,1994ApJ...420L..57K,1995ApJ...453L..81Y}, corresponding to months-to-year variability timescales; although, contributions from the more distant narrow line region (NLR) and ionization cone have been observed \citep[e.g.,][]{2011ApJ...736...62W, 2017ApJ...842L...4F}. In practice, the line likely has contributions from all of the above structures.

One of the more straightforward narrow-line constraints that provided sufficient spectral resolution is the full width at half maximum (FWHM), which traces the spatially unresolved kinematics of the circumnuclear matter and hence can be used to estimate the average reprocessor location. Different studies have arrived at different conclusions with respect to the location of the \kalfa{} emitting regions. Based on a sample of 14 bright AGN observed with the high energy grating (HEG) of the \textit{Chandra} X-ray observatory \citep{2000SPIE.4012....2W}, 
\citet{2006MNRAS.368L..62N} found a lack of correlation between the \kalfa{} FWHM and either the optically derived BLR line width or the SMBH mass, and they concluded that the \kalfa{} core likely has a mix of contributions from the outer accretion disk, BLR, and torus, in differing proportions depending on the source, but it predominantly originates in regions outside the BLR, possibly near the inner edge of the torus. 
\citet{2010ApJS..187..581S} expanded upon this, using a sample of 36 nearby AGN with HEG spectra (27 with FWHM constraints), and arrived at similar conclusions. 
Later, \citet{2015Gandhi} analyzed 13 local type 1 AGN, also using HEG spectra, and found that the \kalfa{} sizes, as estimated from the \kalfa{} FWHM, appeared to be bounded by the dust sublimation radii (i.e., the inner wall of the torus), and they may predominantly originate in clouds associated with either the inner edge of the torus, the BLR, or even further inside. Among type 2 AGN, \citet{2011ApJ...738..147S} found no obvious differences compared to type 1 AGN, while \citet{2012MNRAS.423L...6M} presented a time, spectral, and imaging analysis of NGC\,4945, showing that the reflecting structure is at a distance $\geqslant$30--50\,pc, which is much larger than the typical torus scales.

Rapid X-ray continuum variability is commonly observed in unobscured, obscured \citep[e.g.,][]{Uttley2005}, and even some heavily obscured AGN \citep{Puccetti2014} and suggests that the primary X-ray continuum emitting source (i.e., the corona) is produced in a compact zone very near to the SMBH \citep[e.g.,][]{1993ARA&A..31..717M, 2013MNRAS.431.2441D}. The X-ray light curve can be analyzed via the power-spectral density (PSD) function, which is typically characterized as a power-law of the form $P_{\nu} \propto \nu ^{\alpha}$, where $\nu$ is the temporal frequency and $\alpha$ is the power-law slope \citep[e.g.,][]{1993MNRAS.265..664G, 1999ApJ...514..682E, 2003Vaughan}. Typical values for the power-law slope in AGN are $\alpha{\sim}{-}1$ at low frequencies, indicative of pink noise, and $\alpha{\la}{-}2$ at high frequencies, indicative of red-noise \citep[e.g.,][]{2003Vaughan,  Markowitz2003,Mchardy2004,Uttley2005,Mchardy2005,Summons2007, Arevalo2008}. The transition in the PSD between these two regimes is denoted as $\nu_B = 1/t_B$, and is related to the characteristic X-ray variability timescales of the system \citep[e.g.,][]{Markowitz2003,2006Natur.444..730M,2012A&A...544A..80G}. 

Many studies have investigated the correlated variability from the X-ray continuum and the broad \kalfa{} via reverberation mapping to constrain SMBH spin \citep[see recent review by][]{2014A&ARv..22...72U}, but far fewer have investigated the relation between the variability of the X-ray continuum and the narrow \kalfa{} line. The latter have focused on campaigns of individual nearby sources like MCG-6-30-15 \citep{1996MNRAS.282.1038I}, 
MRK\,509 \citep{2013A&A...549A..72P},
NGC\,2992 \citep{2018MNRAS.478.5638M}, 
NGC\,4051 \citep{2003MNRAS.338..323L}, 
NGC\,4151 \citep{2019ApJ...884...26Z}, and
NGC\,7314 \citep{1996ApJ...470L..27Y}. In at least a few of these, the authors were able to place constraints on the location and size of the \kalfa{} emitting region by studying the reaction of the \kalfa{} line to continuum variations \citep[e.g.,][]{2013A&A...549A..72P, 2019ApJ...884...26Z}. Surprisingly, some studies found a tight correlation between the narrow \kalfa{} line and the continuum on observational timescales of a few days, implying that the narrow \kalfa{} component predominantly arises from regions interior to the BLR.

While timing and multiepoch spectral investigations can place important constraints on reflection close to a SMBH (e.g., light hours-to-years scales), a number of studies based on \textit{Chandra} observations have also found \kalfa{} emission extending out hundreds to thousands of light years in galaxies such as NGC\,4151 \citep{2011ApJ...736...62W}, NGC\,6240 \citep{2014ApJ...781...55W}, NGC\,4945  \citep{2017MNRAS.470.4039M}, and ESO 428-G014 \citep[e.g.,][]{ 2017ApJ...842L...4F}. These scales are much larger than the putative size range of the dusty torus, which ALMA and IR reverberation studies have shown to be $\lesssim 10$ pc \citep[e.g.,][]{2016ApJ...829L...7G,2016ApJ...822L..10I,2016ApJ...823L..12G,2021LyuJ}. The fractional contribution from such highly extended reflection, however, is generally not dominant.

The goal of the present work is to enhance our understanding of the reflecting cloud structure in AGN. In particular, we aim to constrain the location(s) and size(s) where the narrow \kalfa{} line is produced, and understand how such regions may vary among a large AGN sample, particularly as a function of various AGN and host galaxy properties. To this end, we carry out spectral, timing, and imaging analyses on a large ensemble of \textit{Chandra} and \textit{XMM-Newton} observations, investigating the FWHM of the \kalfa{} line, its temporal properties both alone and as they relate to those of the X-ray continuum, and its potential spatial extent. The paper is organized as follows. In $\S$\ref{sec:obs} we describe the observations and data reduction, in $\S$\ref{sec:spec_fitting} we explain the X-ray the spectral fitting, while in $\S$\ref{sec:lc_analysis} we present an analysis of the light curves. In $\S$\ref{subsec:res_fwhm} we analyze the \kalfa{} line FWHM of the sample, in $\S$\ref{sec:NXS} and~\ref{sec:rep_sims} we investigate the \kalfa{} and continuum variability, and in $\S$\ref{sec:image} we outline our assessment of the \textit{Chandra} images. We conclude with some discussion in $\S$\ref{sec:discussion} and a summary in $\S$\ref{sec:conclusion}.

\section{Observations and data reduction} \label{sec:obs}

In the following subsections, we describe how we select the sample of bright, mostly local AGN, and the observations and data reduction for \textit{Chandra} and \textit{XMM-Newton}.

\subsection{Sample selection}

Our broad goal is to quantify the spectral and temporal properties of the \kalfa{} line in AGN, and relate these to the variable X-ray continuum; we focus on the 2--10\,keV continuum band in order to minimize spectral complexity associated with the soft excess and host contamination \citep[e.g.,][]{Fabbiano2006, Done2012}.
To start, we consider that the typical \kalfa{} equivalent width of AGN ranges from $\approx$0.1--1\,keV \citep[e.g.,][]{2010ApJS..187..581S, 2011ApJ...738..147S}, which implies that $\sim$1-10\% of the total photons in the 2--10\,keV band will be \kalfa{} photons assuming typical AGN spectra. Thus a clear requirement emerges such that the AGN be observed by a facility with a high 2--10\,keV sensitivity. We therefore focus on observations from \textit{Chandra} \citep{2000SPIE.4012....2W} and \textit{XMM-Newton} \citep{2001A&A...365L...1J}, whereby even a relatively short 10\,ks exposure of a typical AGN ($\Gamma{=}1.9$ power-law) with a flux of $f_{\rm \: 2-10\,keV}{=}4\times10^{-12}{\rm \:erg\:s^{-1}\:cm^{-2}}$ yields $\sim$1700 (ACIS-I) and $\sim$5000 ($pn$) 2--10\,keV photons, respectively. Such limits produce what we consider to be the bare minimum in terms of \kalfa{} photon statistics (i.e., $>$20--50 counts in the line) to enable spectroscopic constraints for typical observations (${\geqslant}$10--20\,ks).\footnote{Calculated using PIMMS v4.11b; \url{https://heasarc.gsfc.nasa.gov/cgi-bin/Tools/w3pimms/w3pimms.pl}} 
Clearly \textit{XMM-Newton} observations provide better spectral and timing statistics, but suffer substantial background flaring and potential contamination from off-nuclear emission.
On the other hand, \textit{Chandra} observations can be more versatile since they offer higher spatial resolution to search for extended \kalfa{} emission on $\sim$100-pc to kpc scales, and, when the HEG is deployed, sufficient spectral resolution to resolve the \kalfa{} line.

To obtain the broadest possible sample, we adopt as our parent input sample the most recent 105-month \textit{Swift}-Burst Alert Telescope (BAT) Survey (BASS, \citealp[]{2018ApJS..235....4O}), an all-sky survey in the ultra-hard X-ray band (14--195 keV), which provides a relatively unbiased AGN sample at least up to $N_{\rm H}{\lesssim}10^{24}\: \rm cm^{-2}$ \citep{2009WinterL,2015ApJ...815L..13R,2016Koss}. The 105-month Swift-BAT catalog is a uniform hard X-ray all-sky survey with a sensitivity of $8.4 {\times} 10^{-12}\: {\rm erg\:s^{-1}\:cm^{-2}}$ over $90\%$ of the sky in the 14–195 keV band. The survey catalogs 1632 hard X-ray sources, 947 of which are securely classified as AGN. We include one additional target in our sample, the well-known narrow-line Sy1 1H0707$-$495, which is relatively bright in the 2--10\,keV band and is undetected in the BAT 105-month catalog, probably due to its very steep spectral index \citep[e.g.,][]{2007DoneC,2021Boller}.  
To limit our analysis to only those observations for which we have a high likelihood of constraining the \kalfa{} line, we apply a flux cut of $f_{\rm 14-195\: keV} \geqslant  10^{-11}\,{\rm erg\:s^{-1}\:cm^{-2}}$ or $f_{\rm 2-10\: keV} \geqslant  4\times 10^{-12}\, {\rm erg\:s^{-1}\:cm^{-2}}$ (these are roughly equivalent for a $\Gamma{=}1.9$ power-law); this resulted in the selection of 252 sources in the local universe ($z<0.1$), and 28 more distant galaxies with redshifts between 0.1 and 0.56.
In order to assess off-nuclear contamination and carry out both imaging and timing analysis (to assess long-term variability), we require a minimum of at least five \textit{Chandra} observations; this reduces our final sample to 38 objects. However, for historical and technical reasons which we outline below, we reduce and extract {\it Chandra} spectra for all 280 sources that satisfy the flux cut previously described, in order to calibrate the flux of the annular spectra (see Appendix \ref{ap:pileup} for details). We complement our final sample of 38 sources with \textit{XMM-Newton} pn observations when available. Table\,\ref{t:ObsInfo} in the Appendix summarizes the observations analyzed in this work.

\begin{figure}[h]
\resizebox{\hsize}{!}{\includegraphics[trim=10 5 55 40, clip,width=\textwidth]{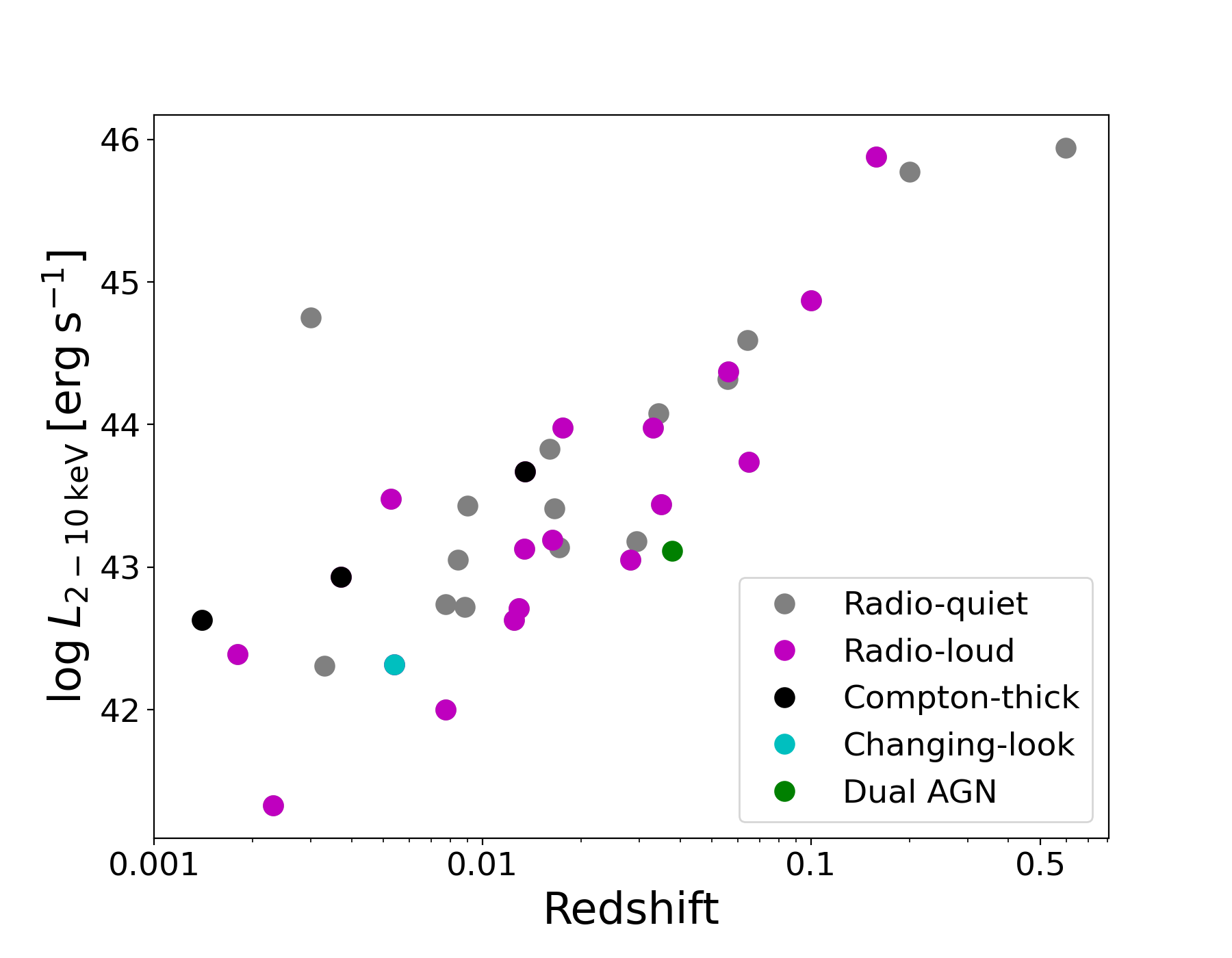}} %
\resizebox{\hsize}{!}{\includegraphics[trim=10 5 55 40, clip,width=\textwidth]{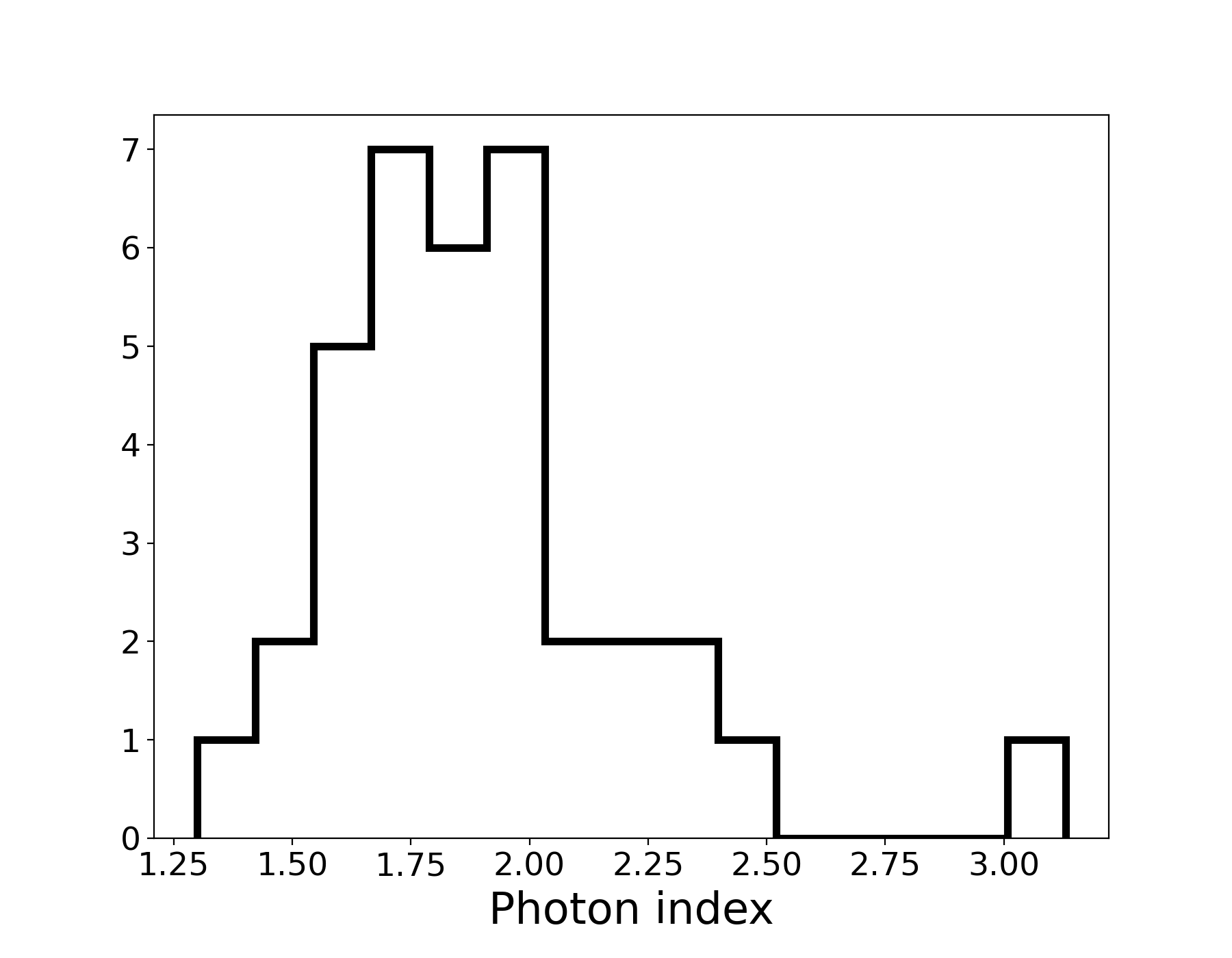}} 
\caption{Some X-ray properties of our sample. {\it Top figure}: Intrinsic 2--10\,keV luminosity versus redshift of our sample. Magenta circles are radio-loud sources (see Table\,\ref{t:properties} for details), black circles are Compton-thick AGNs, green circles are dual AGNs, and gray circles are normal, radio-quiet AGNs. {\it Bottom figure}: Distribution of X-ray photon indices of our sample. } \label{fig:Lx_z}    

\end{figure}

We stress that while this final sample is by no means complete, it spans a reasonable range of the parameter space to be considered representative of local hard X-ray selected AGN. The top panel of Figure\,\ref{fig:Lx_z} shows the intrinsic X-ray luminosities and redshifts of our sample, while the bottom panel shows the X-ray photon index distribution. The X-ray luminosity and photon index values are taken from \citet{2017ApJS..233...17R}, while redshifts and AGN properties used in this work are from \citet{2017ApJ...850...74K} and BASS DR2 (\citeauthor{2022Koss} (submitted)), which extends the DR1 results of \citet{2017ApJ...850...74K} and \citet{2017ApJS..233...17R}. A large majority of the sample reside in the local universe ($z<0.1$) and have modest X-ray luminosities ($L_{\rm 2-10 \: keV}<10^{44} \rm \: erg \: s^{-1}$). Additionally, the photon indices of our sample range from 1.3 to 3.13, with a median value of $\Gamma = 1.82$, which is broadly consistent with the range and median photon index of the nonblazar AGN in the BASS survey \citep[$\Gamma=1.78 \pm 0.01$][]{2017ApJS..233...17R}. The distributions of other AGN properties such as the 
line-of-sight column density [$\log(N_{\rm H}/{\rm cm^{-2}})$=20--25.9, with median of 21.5], AGN type (predominantly Seyfert types Sy1, Sy1.2, Sy1.5, Sy1.8 and Sy2), black hole mass [$\rm \log(M_{BH}/M_{\odot}) = 6.13$--9.83, with median 7.6] and Eddington ratio ($\rm ER=3.4\times10^{-7}$--1.13, with median 0.09) also span comparable ranges to the $z<0.85$ nonblazar AGN in the BASS survey, and hence should be as representative (compare Figs.\,\ref{fig:ratio_rel} and \ref{fig:slope_rel} of the Appendix with Fig.\,13 of \citealp{2017ApJ...850...74K}).

\subsection{\textit{Chandra} X-ray Observatory}\label{subsec:chandra}

We downloaded 1001 observations with the Advanced CCD Imaging Spectrometer (ACIS), both the ACIS-S and \hbox{ACIS-I} CCD configurations, available from the \textit{Chandra} archive associated with the 280 sources in our sample; we excluded the small number of High Resolution Camera observations, due to that instrument's much lower spectral resolution and poorer high-energy sensitivity. Among the 1001 ACIS observations, 279 were acquired in the High-Energy Transmission Grating mode (\citealp{2000ApJ...539L..41C}, HETG). The HETG consists of two grating assemblies: the High Energy Grating (HEG, 0.8--10 keV) and the Medium Energy Grating (MEG, 0.4--5 keV); we only use the HEG spectra, since our focus is on the study of the 2--10 keV spectra, and in particular the region around the 6.4 keV \kalfa{} line to make our analysis. The observations were acquired between 2000 and 2018.

The data reduction follows standard procedures with the CIAO software (v4.11) and calibration files (CALDB v4.8.3), using the \texttt{chandra\_repro} script. The X-ray peaks in images of the same object are occasionally shifted up to $\approx$1\farc{} from each other, or from the established optical centroid of the galaxy. To correct this, we manually choose the center of each observation and estimate the alignment offset with respect to the observations with higher exposure time, create a new aspect solution using the \texttt{wcs\_update} task, and update the astrometry. We resort to manual alignment for two reasons. First, in some observations, there are not enough point sources in the field in common to match and align them automatically, and second, nuclear saturation (i.e., pileup) is very high in some observations, producing a hole in the center of the source which can confuse simple detection and alignment codes.  Therefore, we apply new aspect solutions based on the different X-ray observations. Then, we remove background flares from the event files with the script \texttt{deflare} and subtract the readout streak, if present above the background, with the \texttt{acisreadcorr} task, to allow for more precise spectral and imaging analysis. 

Finally, we reproject the events to a common tangent point using the \texttt{reproject\_events} task and merge them for image analysis.

\subsubsection{Spectral extraction}

For both normal ACIS and zeroth-order HETG observations, we extract spectra with the \texttt{specextract} task, which creates spectra and responses files (ARF and RMF). We generate source spectra for each observation using both a 1\farcs5 radius circular aperture and an annulus of 3\farc{}--5\farc{}, adopting an annular region of 20\farc{}--35\farc{} aperture to estimate the background spectra. We mask all obvious off-axis point sources using the observation with the highest exposure time as our reference image. We use the \texttt{wavdetect} task to detect all the point sources in the field, setting the wavelet scale parameters to one and two pixels. By visual inspection, we confirmed that all point sources were detected and masked correctly. Then, we mask the off-axis point sources in the field using the default 3$\sigma$ elliptical regions from the \texttt{wavdetect} output file before the spectral extraction.\footnote{For Circinus Galaxy and Cen A, we manually adjusted some mask regions based on visual inspection.} We use the circular extracted spectra for the observations unaffected by pileup and the annular spectra otherwise. We apply aperture corrections to both the 1.5\farc{} circular spectrum and the annular spectra using the \texttt{arfcorr} task, which are applied through modified responses files.

For HETG observations, we extract 1st-order dispersed spectra using a 6-pixel width ($\approx$3\farc{}) aperture. We first create a mask to delineate events in the HEG and MEG arms using the \texttt{tg\_create\_mask} task, and then we resolve the spectral orders using the \texttt{tg\_resolve\_events} task. Lastly, we generate spectra and response files using \texttt{tgextract} and \texttt{mktgresp}, respectively.

A key concern is that the spectra are not highly affected by off-axis sources or extended emission. As mentioned above, before extracting the spectra, we remove all contaminating sources in the field. For the nuclear 1\farcs{}5 and grating spectra, we confirm that there is generally little contamination, due to the small aperture. However, for sources affected by extended emission (see Appendix~\ref{ap:pileup} for further details), contamination can be larger and we do not use the 3\farc{}--5\farc{} spectra in such cases. 

\subsubsection{Imaging analysis}

To complement the spectral analysis, we also investigate the high angular resolution \textit{Chandra} images to see whether some sources are spatially extended and compare this with the spectral results. To assess this, we extract radial profiles. Specifically, for each source, we merge all the event files with an off-axis angle $< 2^{\prime}$ (to avoid observations affected by strong distortions to the point spread function, or PSF), using the task \texttt{reproject\_obs}, which reprojects the observations to a common tangent point and then merges them. We mask off-nuclear point sources in the field, as well as dispersed photons related to the 1st-order spectrum for grating-mode observations. Then, we extract radial profiles for each source with \texttt{dmextract}, using sequential annuli, accounting for the masked area from each annulus. 

\subsection{\textit{XMM-Newton} Observatory}\label{subsec:xmm}

When available, we complement our data with observations from the pn camera of \textit{XMM-Newton} for 32 sources among our sample of AGN with more than five \textit{Chandra} observations.  We do not include observations from the EPIC-MOS cameras, as many of the observations are likely affected by pileup due to the high typical readout time (2.6\,s). The pn camera has a full-frame time resolution of $73.3\, \rm ms$ per CCD \citep{2001A&A...365L...1J}, such that the observations generally do not suffer much from pileup, making them well suited for variability analysis. When available, we use 118 spectra of 13 AGN, extracted from 30\farc{} aperture provided by Tortosa et al. (in prep). Additionally, we downloaded 172 60\arcsec{}-aperture pn spectra for 29 sources from the 4XMM--DR9 catalog \citep{2020Webb}. The 30\farc{} aperture spectra are preferred, when available, to minimize host contamination. The procedure for the extraction and data reduction in Tortosa et al. (in prep) is described below. For details of the extraction and data reduction of sources in the 4XMM--DR9 catalog we refer to \citet{2020Webb}.

The extraction of calibrated \textit{XMM-Newton} spectra is performed by means of the Science Analysis System (SAS) software package (v.18.0.0). 
The \textit{XMM-Newton} pn data are processed using the task \texttt{epchain} in order to obtain calibrated and concatenated event lists.
\textit{XMM-Newton} can also focus charged particles on the detection plane. Since these background flares have a large effect on the detected X-ray spectrum of all three EPIC cameras aboard \textit{XMM-Newton}, it is imperative to remove these events during the data reduction process. To this end, the \texttt{evselect} command examines the count-rate
of such events, selecting all single-pixel events (i.e., PATTERN==0) 
in the energy range sensitive to soft proton flares. Source and background spectra are extracted again using \texttt{evselect}, adopting a 30\farc{} radius circular source region and a nearby source-free 50\farc{} radius circular background region.
The Redistribution Matrix Files (RMF) are generated using the command \texttt{rmfgen}, the Auxiliary Response Files (ARF) are generated with the command \texttt{arfgen}. 
When the input spectrum is multiplied by the ARF, the result will be the distribution of counts as would be seen by a detector with ideal resolution in energy. Then, the RMF is needed, in order to produce the final spectrum. All these files are compressed into a single file, easily readable by \textsc{XSPEC} \citep{1996ASPC..101...17A}, using the SAS tool \texttt{specgroup}. EPIC spectra were binned in order to over-sample the instrumental resolution by at least a factor of 3 and to have no less than 30 counts in each background-subtracted spectral channel.

The 30\farc{} and 60\farc{} aperture \textit{XMM-Newton} spectra are more likely to be affected by potential galaxy contamination than the nuclear \textit{Chandra} spectra. We estimated the amount of contamination in the \textit{XMM-Newton} spectra by computing the ratio between the encircled counts in a 1\farcs{}5 aperture and the encircled counts in a 30\farc{} and 60\farc{} aperture of the merged \textit{Chandra} images for each galaxy, after proper PSF correction.  We found that the galaxy contamination for \object{Centaurus A} (Cen\,A) and \object{Cygnus A} dominates the spectra, contributing more than $60\%$ of the total flux, and therefore the \textit{XMM-Newton} spectral epochs for those AGN were not used. Four additional sources (IC 4329A, M51a, H1821+643 and 1H0707-495) have host galaxy contributions between 20--40\% of the total flux, while for the rest of the sample the host galaxy contributions are less than 10--20\%. We expect that the contribution of the host contamination will be roughly constant between the observations, and thus apply a correction to the measured flux by a constant factor. Thus, spatially resolved host contamination should be not have a significant impact on subsequent variability results.

\section{X-ray spectral analysis} \label{sec:spec_fitting}

We fit the X-ray spectra of 652 {\it Chandra} and {\it XMM-Newton} observations in our sample. We describe our spectral fitting approach and the results for the Fe K$\alpha$ line profile below.

\subsection{Spectral fitting}

\begin{figure}[h]

\resizebox{\hsize}{!}{\includegraphics[trim=10 20 30 50, clip, width=\textwidth]{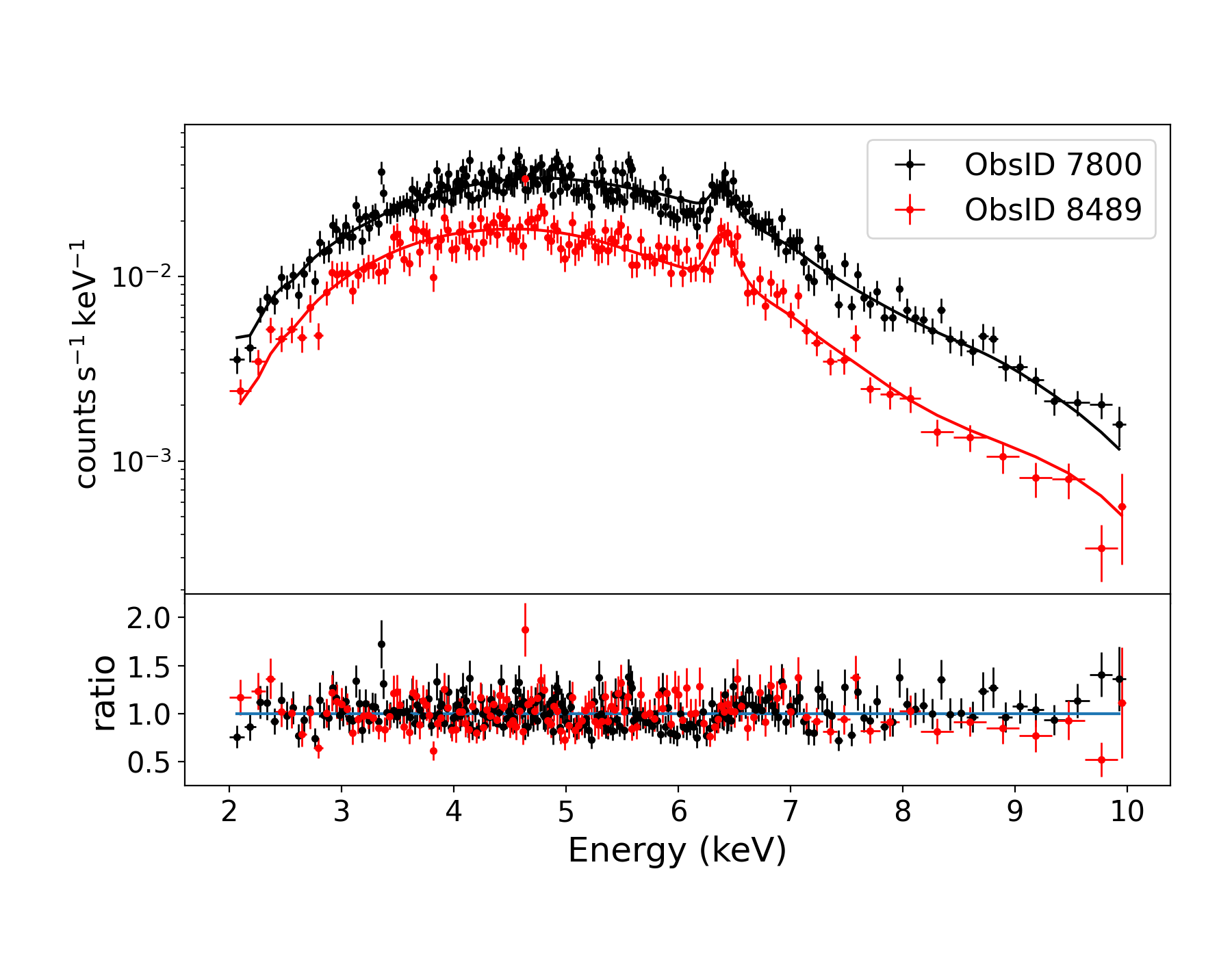}}

\resizebox{\hsize}{!}{\includegraphics[trim= 10 20 30 10, clip, width=\textwidth]{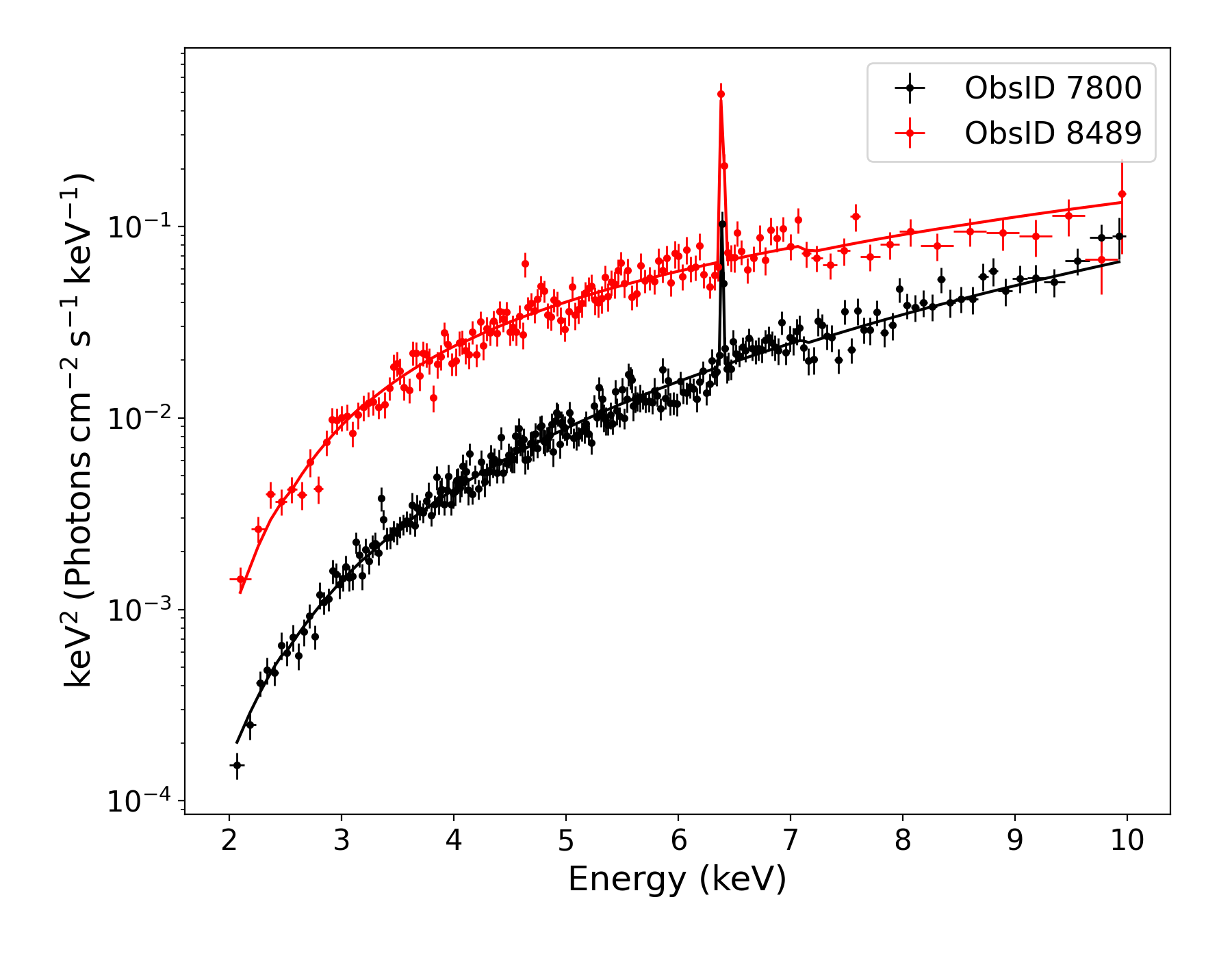}} 

\caption{2--10\,keV spectra of the Type 2 AGN Centaurus A, extracted and aperture-corrected from 3\arcsec--5\arcsec\
annuli in ACIS-I ObsIDs 7800 (black) and 8489 (red), fitted by the model described in Eq.~\ref{eq:spec}. The top panel shows the spectra and model-to-data ratio in instrument units while the bottom panel shows the unfolded spectra. The spectra are attenuated by strong absorption. The sole emission line is the 6.4\,keV Fe $\rm K\alpha$ line.}\label{fig:CenA_spec}    

\end{figure}

We fit the spectra using the package PyXspec,\footnote{\url{https://heasarc.gsfc.nasa.gov/docs/xanadu/xspec/python/html/index.html}} which is a python version of the popular standalone X-ray spectral fitting package XSPEC (version 12.10.1, \citealp[]{1996ASPC..101...17A}). We apply a simple model to compute the fluxes of the Fe K$\alpha$ line and the continuum between 2--10 keV, given by:
\begin{equation}
\textsc{phabs}_{1} \times \textsc{phabs}_{2}  \times (\textsc{cpflux}_{1} \times \textsc{powerlaw}+\textsc{cpflux}_{2} \times \textsc{zgauss}),
\label{eq:spec}
\end{equation}
\noindent where the $\textsc{phabs}_{1}$ and $\textsc{phabs}_{2}$ components account for the Galactic and intrinsic AGN line-of-sight photoelectric absorption, respectively, the \textsc{powerlaw} component corresponds to the continuum emission, and the \textsc{zgauss} term reproduces the Fe $\rm K\alpha$ line emission at 6.4 keV. The Galactic absorption is fixed at the value obtained from the HI 4$\pi$ survey \citep{2016A&A...594A.116H}. We leave free to vary the normalization and spectral index of \textsc{powerlaw}, and line center (within 10$\%$ of its rest-frame energy, adopting redshifts fixed at their published values), the normalization and line width of \textsc{zgauss}. The $\textsc{cpflux}_{1}$ and $\textsc{cpflux}_{2}$ components provide the unabsorbed photon fluxes of the continuum emission and the \kalfa{}, respectively. To compute unabsorbed fluxes, we first freeze the free parameters after minimizing the fit and set the absorbing column densities to zero. Figure~\ref{fig:CenA_spec} shows an example of the fit for two observations of Cen A in instrumental (top panel) and unfolded (bottom panel) units. The model reproduces fairly well the 2--10 keV spectra, which are strongly attenuated by absorption, and yields a secure measure of the well-known emission line at 6.4 keV. We can see strong variability between the observations of about one order of magnitude for both the continuum and the line. Table\,\ref{t:ObsInfo} in the Appendix lists the best-fitting parameters of the observations analyzed in this work. The fits are carried out with Cash statistics \citep{1979Cash}.

This model is simple but sufficient to measure the fluxes of unobscured sources. We confirmed this by replacing \textsc{zgauss} by the \textsc{pexmon} model, which reproduces the Compton neutral reflection with self-consistent Fe K$\alpha$, Fe K$\beta$ and Ni K$\alpha$ lines. We applied this model to three sources with different levels of obscuration: \object{Circinus Galaxy} ($\rm N_H= 4\cdot 10^{24}\:cm^{-2}$, \citealp[]{2017ApJS..233...17R}), Cen A ($\rm N_H= 1\cdot 10^{23}\:cm^{-2}$, \citealp[]{2017ApJS..233...17R}), and \object{NGC 4051} ($\rm N_H \lesssim 10^{20}\:cm^{-2}$, \citealp[]{2017ApJS..233...17R}). For the last two sources, the continuum and Fe K$\alpha$ line fluxes are consistent at a 90$\%$ confidence level, which is expected since at $\rm  N_H < 10^{24}\: cm^{-2}$ the spectrum is dominated by the transmitted component. For the Circinus Galaxy, however, the model with reflection predicts an unabsorbed continuum flux twice that of the simple model, although the Fe K$\alpha$ fluxes are consistent. The \textsc{pexmon} model predicts a higher column density and softer photon index, such that an increase in the continuum flux is expected. At $\rm  N_H > 10^{24}\: cm^{-2}$ and energies lower than 10 keV, the spectrum is dominated by the reflected component, therefore it is not surprising that the \textsc{pexmon} model predicts a different continuum flux. We do not consider this to be a problem, since the majority of our sources have $\rm  N_{\rm H} < 10^{24}\:cm^{-2}$ (see Fig.~\ref{subfig:ratio-nh}) and we take special care in the interpretation of the results for more heavily obscured sources.

To properly compare the fluxes measured from different spectra, we need to know whether they are affected by pileup, which occurs when two or more photons fall in the same pixel within a single readout "frame" and, in consequence, the information in that pixel is altered. If the source is very bright or the readout time is high, there is a higher probability to be affected. Normal CCD observations with the ACIS camera of \textit{Chandra} are commonly affected by pileup given the sharp on-axis PSF and the nominal 3.2\,s frame time. \textit{Chandra} first-order grating spectra are much less affected by pileup since the photons are more dispersed. Therefore, if an observation is performed in HETG grating mode, we rely on the HEG 1st-order spectrum. If an observation is performed in normal CCD mode, we first estimate the amount of pileup to decide whether to use the 1.5\farc{} circular aperture spectrum or the annular spectrum. The \textit{XMM-Newton} pn observations are less affected by pileup due to the much shorter readout time of that instrument and the fact that the PSF is spread over many more pixels. 

One issue we discovered during this process is a large inconsistency between the fluxes measured by the HEG first-order spectrum and the annular spectrum from the zeroth-order spectrum of the same data, indicating that the aperture correction performed by the task \texttt{specextract} is not accurate. Appendix~\ref{ap:pileup} explains in detail how we calibrate an aperture correction factor between the HEG and 3\farc{}-5\farc{} spectra, in order to incorporate the fluxes measured by the annular spectra in our analysis.

\subsection{Fe K$\alpha$ line FWHM} \label{subsec:res_fwhm}

\begin{table}
\caption{Key properties of the sample.}
%\begin{adjustbox}{width=\textwidth}
\scriptsize
% \centering
\scalebox{0.83}{
\begin{tabular}{l L L L L L L l} 
 \hline
 \hline
\noalign{\smallskip}
\multicolumn{1}{l}{Source} & \multicolumn{1}{c}{${\rm log}(M_{BH})$} & \multicolumn{1}{c}{${\rm log}(L_{bol})$} & \multicolumn{1}{c}{$\nu_{b}$} & \multicolumn{1}{c}{Ref. $\nu_{b}$} & \multicolumn{1}{l}{RL} & \multicolumn{1}{l}{$R_{\rm H\beta}$} & \multicolumn{1}{c}{Ref. $R_{\rm H\beta}$} \\
% \noalign{\smallskip}
& \multicolumn{1}{c}{$(M_{\odot})$} & \multicolumn{1}{c}{$(\rm erg s^{-1})$} &  \multicolumn{1}{c}{(10$^{-6}$ Hz)} & & &  \multicolumn{1}{l}{(pc)} & \\
\multicolumn{1}{l}{(1)} & \multicolumn{1}{c}{(2)} & \multicolumn{1}{c}{(3)} & \multicolumn{1}{c}{(4)} & \multicolumn{1}{l}{(5)} & \multicolumn{1}{l}{(6)} & \multicolumn{1}{l}{(7)} & \multicolumn{1}{l}{(8)} \\
\noalign{\smallskip}
\hline
\object{1H0707-495}$^{*}$      & 6.6 & \mcdash{c} & 398 & (1) & \mcdash{l}& \mcdash{l}& \mcdash{l} \\ % Zoghbi2010 yes, deLaCalle2010 marginal
2MASXJ23444$^{\dagger}$  & 10.1   & 47.76 & 0.0225 & (2) & -3.99& \mcdash{l}& \mcdash{l}\\
\object{3C120}$^{*}$           & 7.74 & 45.17 & 7.8 & (2) & -3.11 & 38.1^{+15.3}_{-21.3}& (1) \\ % Lohfink2013 marginal
\object{3C273}           & 8.84 & 47.02 & 0.7 & (2) & -2.78 & 146.8^{+12.1}_{-8.3} & (2) \\ % 2009Brenneman marginal
\object{3C445}          & 8.39 & 45.50 & 1.36 & (2) & -2.75 &\mcdash{l}& \mcdash{l}\\ 
\object{4C+29.30}        & 8.28 & 44.91 & 1.3 & (2) & -3.59 &\mcdash{l}& \mcdash{l}\\ 
\object{4C+74.26}       & 9.83 & 46.01 & 0.02 & (2) & -3.30&\mcdash{l}& \mcdash{l}\\ % Tzanavaris2019 no
Cen A           & 7.77 & 43.10 & 2.3 & (2) & -3.35&\mcdash{l}& \mcdash{l}\\ 
Circinus Galaxy & 6.23 & 43.55 & 107.15 & (1) & -4.50&\mcdash{l}& \mcdash{l}\\ 
\object{Cygnus A}        & 9.43 & 45.56 & 0.05 & (2)& -0.73&\mcdash{l}& \mcdash{l}\\ 
\object{H1821+643}      & 9.26 & 47.52 & 0.291 &  (2) & -4.13&\mcdash{l}& \mcdash{l}\\
\object{IC 4329A}$^{*}$          & 7.81 & 45.04 & 5.9 & (2) & -5.43&1.5^{+1.8}_{-2.7}& (1) \\ % deLaCalle2010 marginal
\object{M 81}             & 7.90 & 39.55 & 0.2 & (2) & -3.99&\mcdash{l}& \mcdash{l}\\ 
\object{MCG-6-30-15}$^{*}$      & 6.14 & 43.86 & 38 & (1) & -6.59&\mcdash{l}& \mcdash{l}\\  % deLaCalle2010 clear
\object{MR 2251-178}      & 8.20 & 45.78 & 0.25 & (1) & -5.59&\mcdash{l}& \mcdash{l}\\ 
\object{MRK 1040}         & 7.41 & 44.57 & 15.5 & (2) & -5.47 &\mcdash{l}& \mcdash{l}\\ 
\object{MRK 1210}         & 6.76 & 44.30 & 99.4 &  (2) & -4.53&\mcdash{l}& \mcdash{l}\\ 
\object{MRK 273}          & 8.78 & 44.13 & 0.2 &  (2) & -3.70&\mcdash{l}& \mcdash{l}\\ 
\object{MRK 290}          & 7.28 & 44.36 & 20.6 &  (2) & -5.44& 8.7^{+1}_{-1.2}&(3)\\ 
\object{MRK 3}            & 6.72 & 44.84 & 151.3 &  (2) & -4.08&\mcdash{l}& \mcdash{l}\\ 
\object{MRK 509}$^{*}$          & 8.05 & 45.26 & 0.08 & (1) & -5.50 & 79.6^{+5.4}_{-6.1} &(1) \\ % deLaCalle2010 marginal
\object{MRK 766}$^{*}$           & 6.962 & 43.91 & 290 &  (1) & -4.52&\mcdash{l}& \mcdash{l}\\ % 2009Brenneman % deLaCalle2010 marginal
\object{NGC 1068}         & 6.93 & 43.94 & 48.1 &  (2)& -3.07 &\mcdash{l}& \mcdash{l}\\ 
\object{NGC 1275}         & 7.55 & 45.14 & 13.8 &  (2)& -2.36 &\mcdash{l}& \mcdash{l}\\
\object{NGC 1365}         & 7.84 & 43.44 & 2.2 &  (2) & -3.99&\mcdash{l}& \mcdash{l}\\ 
\object{NGC 2992}         & 8.33 & 43.13 & 0.4 &  (2) & -3.85&\mcdash{l}& \mcdash{l}\\ % 2009Brenneman no
\object{NGC 3393}         & 7.52 & 43.80 & 7.2 &  (2)&-5.06&\mcdash{l}& \mcdash{l}\\ 
\object{NGC 3516}$^{*}$          & 7.39 & 43.85 & 6.6 &  (1)& -5.36& 6.7^{+3.8}_{6.8} &(1) \\ % deLaCalle2010 clear
\object{NGC 3783}$^{*}$          & 7.37 & 44.57 & 13 &  (1) &-5.41 & 10.2^{+2.3}_{-3.3} & (1) \\ % 2009Brenneman marginal, deLaCalle2010 super marginal
\object{NGC 4051}$^{*}$          & 6.13 & 42.42 & 510 &  (1)&-4.40 &5.6^{+1.8}_{-2.6} & (1)\\  % 2009Brenneman clear, deLaCalle2010 clear
\object{NGC 4151}$^{*}$         & 7.56 & 43.44 & 0.58 &  (1) & -4.96 &6.6^{+0.8}_{-1.1}& (1) \\ % Zoghbi2012 clear
\object{NGC 4388}         & 6.94 & 44.22 & 54.4 &  (2) & -5.23&\mcdash{l}& \mcdash{l}\\ 
\object{NGC 5548}         & 7.72 & 44.34 & 1.3 &  (1) & -5.23& 7.2^{+0.35}_{-1.33} &(4) \\  % deLaCalle2010 no
\object{NGC 6300}         & 6.57 & 43.02 & 88 & (3) & \mcdash{l} &\mcdash{l}& \mcdash{l}\\
\object{NGC 7469}         & 6.96 & 44.38 & 56.0 &  (2) & -4.35&4.5^{+0.8}_{-0.7}& (1) \\ % deLaCalle2010 no
\object{NGC 7582}         & 7.74 & 44.66 & 5.9 & (2) & -4.63 &\mcdash{l}& \mcdash{l}\\ 
\object{Pictor A}         & 6.80 & 44.63 & 105.6 &  (2) & -2.77&\mcdash{l}& \mcdash{l}\\  
\object{PKS2153-69}       & 7.23  &44.23 & 22.6  &  (2) & -1.49&\mcdash{l}& \mcdash{l}\\  
\hline
 \end{tabular} }

\tablefoot{Col.(1): Object name. Sources denoted by * have potential relativistically broadened K$\alpha$ components reported in the literature \citep[e.g.,][]{2009Brenneman, deLaCalle2010, Zoghbi2010, Zoghbi2012, Lohfink2013, Tzanavaris2019}; additional care should be exercised when interpreting the fluxes as assessed from our simple continuum and narrow line model components for these sources (see $\S$\ref{sub:slope} for details); Col.(2) Black hole mass obtained from \citet{2017ApJ...850...74K} and BASS DR2 (\citeauthor{2022Koss} (submitted)); Col.(3) AGN bolometric luminosity obtained from \citet{2017ApJ...850...74K} and BASS DR2; Col.(4): Break frequency of the X-ray power spectrum; Col.(5): Reference for the $\nu_{b}$, (1) refers to \citet{summons_thesis} and (2) refers to equation \ref{eq:power_spectrum}; Col.(6): Radio-loudness from \citet{2017ApJS..233...17R}, defined as $\rm RL= log\left( F_{1.4\:GHz}/F_{14-150\:keV} \right)$; Col.(7): optical BLR radius ($R_{\rm H \beta}$) calculated from $H\beta$ reverberation lags; ; Col.(8): reference for $R_{\rm H \beta}$, (1) refers to \citet{2009Bentz}, (2) refers to \citet{2019Zhang}, (3) refers to \citet{2016Du}, and (4) refers to \citet{2016Lu}. $\dagger$ 2MASXJ23444 refers to \object{2MASX J23444387-4243124}.}
\label{t:properties}
\end{table} 

The \kalfa{} line profile provides information about the kinematics of the reflecting clouds from which the line originates. We fit the \kalfa{} FWHM for all sources with available \textit{Chandra} HEG spectra, which provides the best spectral resolution among current observatories ($\rm FWHM \approx 1860\:km\:s^{-1}$, or $\rm 39\: eV$ at $\rm 6.4 \: keV$); we note that the values reported in Table~\ref{t:summary} are deconvolved FWHMs, as the line spread function information from the RMF is used during the fit. Among all the HEG observations, we only consider model fits with a lower limit different from zero and a finite upper
limit at a 90$\%$ confidence. While this could bias our results against sources where the \kalfa{} line is either faint or observed in a low state, we want to avoid fitting poorly detected lines where the profile could attempt to fit the underlying continuum, yielding inaccurate or overestimated line widths. 

We estimate the radius of the narrow \kalfa{} line emission ($R_{\rm Fe\:K\alpha}$) assuming virial motion with the gravitational potential dominated by the SMBH of the emitting material as
\begin{equation}\label{eq:rfe}
    R_{\rm Fe\:K\alpha} =\frac{GM_{\rm SMBH}}{\left( \sqrt{3}/2\: \upsilon_{\rm FWHM} \right)^2},
\end{equation}
where $G$ is the gravitational constant and $M_{\rm BH}$ is the SMBH mass \noindent \citep[e.g.,][]{Netzer1990,2004ApJ...613..682P}. If the \kalfa{} line originates in an outflow instead of the BLR or the dusty torus, the kinematics can no longer be modeled by virial motion. Nevertheless, for an outflow to reach velocities comparable to the typical $\upsilon_{\rm FWHM}$ values of the narrow \kalfa{}, large AGN luminosities are needed ($\rm log(L_{bol}/erg\:s^{-1}) > 45$, \citealp{2017A&A...598A.122B}) and only a few sources of our sample are this powerful.

We compare the $R_{\rm Fe\:K\alpha}$ with the inner radii of the dusty torus, that is the dust sublimation radius, $R_{\rm sub}$, given by \citet{2008IINenkova} as
\begin{equation} \label{eq:rsub}
     R_{\rm sub}=0.4 \left( \frac{L_{bol}}{10^{45} \:erg\:s^{-1}} \right)^{1/2}\:\left( \frac{1500\:K}{T_{sub}} \right)^{2.6}\:pc,
\end{equation}
\noindent  where $\rm L_{bol}$ is the bolometric luminosity and $\rm T_{sub}$ is the dust sublimation temperature, generally assumed to be the sublimation temperature of graphite grains, $\rm T{\approx} 1500 \:K$ \citep[e.g.,][]{2007A&A...476..713K}. We adopt a temperature uncertainty of $\Delta T{=}500$ K to compute $R_{\rm sub}$, which broadly accounts for possible variations due to grain mineralogy,
porosity and size, among others. We get the values of $\rm L_{bol}$ and $M_{\rm BH}$ from the DR2 of BASS (\citeauthor{2022Koss} (submitted)). The values of $L_{bol}$ and $M_{\rm BH}$ are tabulated in Table~\ref{t:properties}. We also compare $R_{\rm Fe\:K\alpha}$ to the optical BLR radius $R_{\rm H\beta}$, inferred from $\rm H\beta $ reverberation studies of eleven sources  \citep{2009Bentz,2016Du,2019Zhang}, as listed in Table~\ref{t:properties}. 

\begin{figure*}
\includegraphics[width=0.55\textwidth]{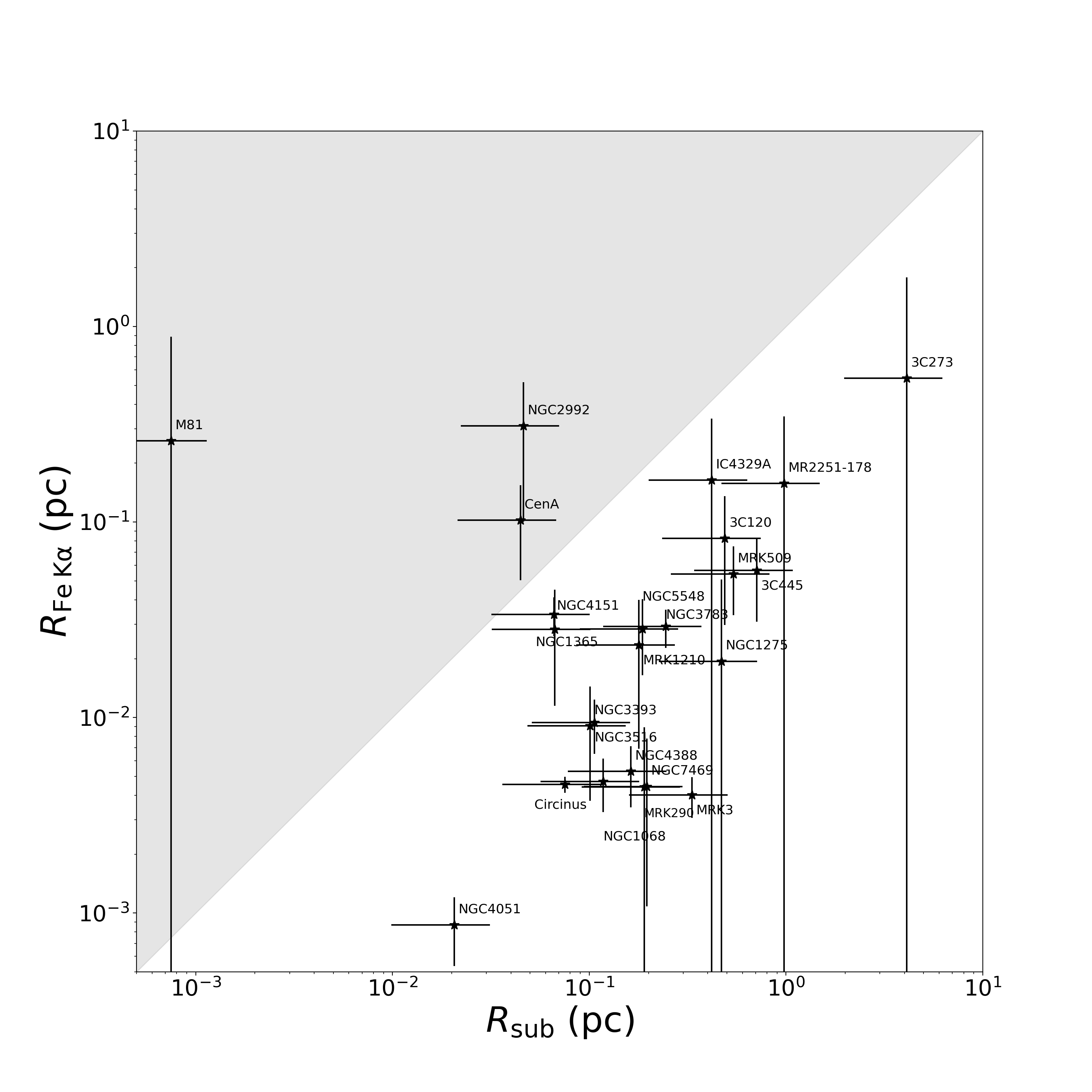}
\hspace*{-1cm}
\includegraphics[width=0.55\textwidth]{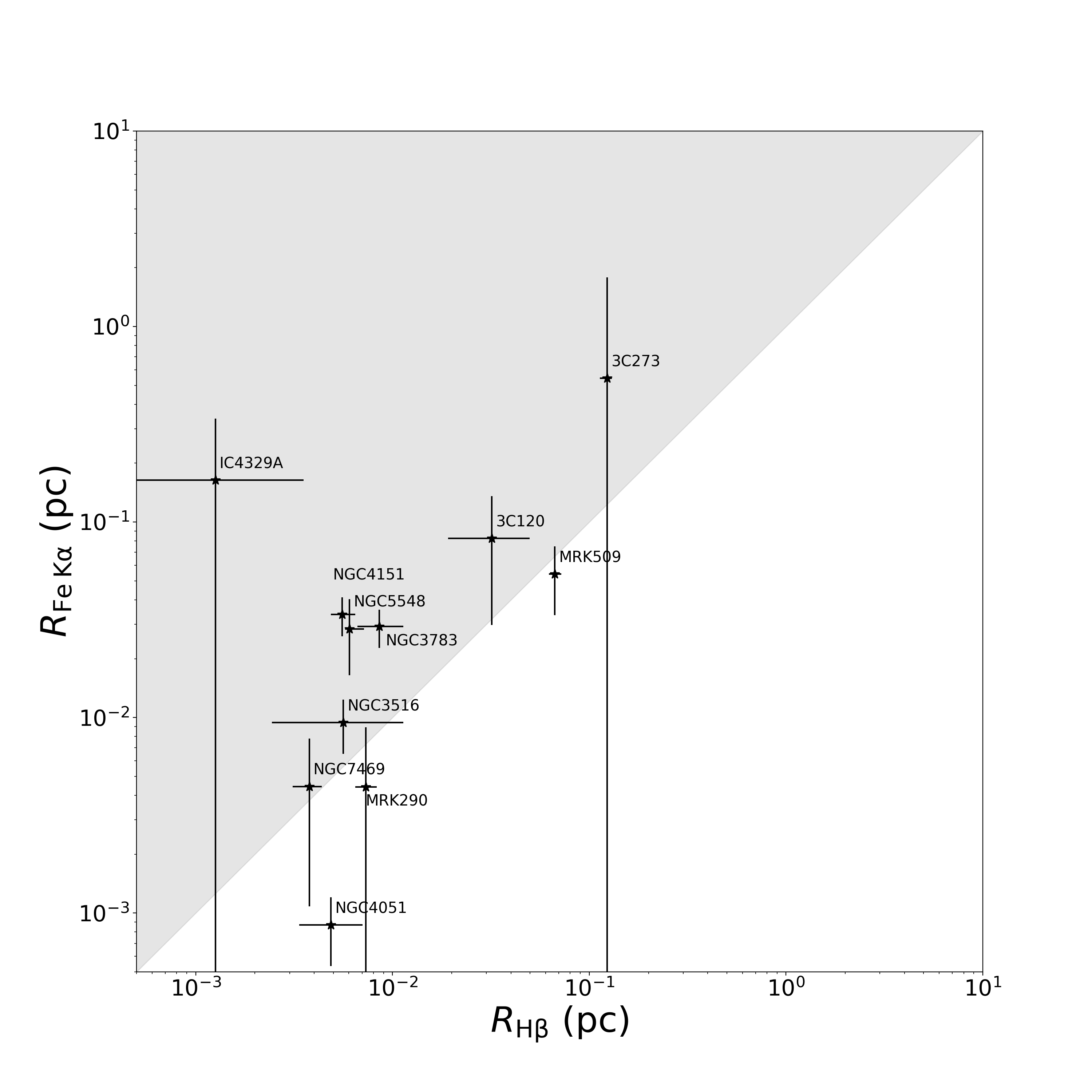}
\caption{{\it Left}: \kalfa{} emitting radius ($R_{\rm Fe\:K\alpha}$) versus dust sublimation radius ($R_{\rm sub}$) in units of parsecs. $R_{\rm sub}$ values are estimated using equation~\ref{eq:rsub} and $R_{\rm Fe\:K\alpha}$ values are estimated using equation~\ref{eq:rfe}. The gray shaded area represents $R_{\rm Fe\:K\alpha} \geqslant R_{sub}$. {\it Right}: \kalfa{} emitting radius ($R_{\rm Fe\:K\alpha}$) versus optical BLR radius ($R_{\rm H\beta}$) in units of parsecs. $R_{\rm H\beta}$ is directly calculated from $\rm H\beta$ reverberation lags. The gray shaded area represents $R_{\rm Fe\:K\alpha} \geqslant R_{\rm H\beta}$. }\label{fig:rfe_rsub_rhb}	

\end{figure*}

We provide an estimate for the \kalfa{} emission location for 24 out of 38 sources in our sample. Table~\ref{t:summary} lists the $R_{\rm Fe\:K\alpha}$, $\upsilon_{\rm FWHM}$ and $R_{\rm sub}$ values. The left and right panels of Figure~\ref{fig:rfe_rsub_rhb} compare the location of the \kalfa{} emitting regions to dust sublimation radius, $R_{\rm sub}$, and $R_{\rm H\beta}$, which we take to be the nominal optical BLR region radius. For 21 out of 24 sources, the bulk of the \kalfa{} emission appears to arise from regions inside the dust sublimation radius, while for eight out of 11 sources, the \kalfa{} emission appears to arise from regions near or a factor of several beyond $R_{\rm H\beta}$. Thus, for most AGN, the narrow \kalfa{} emission appears to originate primarily in the outer BLR. We cannot exclude that small portions of the narrow \kalfa{} flux may arise from the outer accretion disk or the dusty torus, and additionally we note that a small minority ($\sim$20\%) of AGN diverge from this general behavior. 

\section{X-ray light curve analysis} \label{sec:lc_analysis}

We construct light curves for each source using the recalibrated and aperture-corrected Fe $\rm K\alpha$ line ($\rm F_{\rm Fe\:K\alpha}$) and the 2--10 keV continuum ($\rm F_{2-10\:keV}$) fluxes. Appendix Figs.~\ref{fig:lc_cont} and \ref{fig:lc_fe} provide light curves for the continuum and Fe K$\alpha$ line fluxes, respectively, for the entire sample. 

In the following subsections, we characterize the variability of the light curves, estimate the size of the reprocessor from which the Fe K$\alpha$ photons originate, and study possible correlations between the Fe $\rm K\alpha$ line and continuum variability.

\subsection{ Variability features }

\subsubsection{The variability probability $P_{var}$}

To assess whether the light curves are variable or not, we compute $P_{var}=\rm P(\chi^2)$  \citep[e.g.,][]{2004ApJ...611...93P,2014ApJ...781..105L,2017ApJ...849..110S}, which is the probability that a $\chi^2$ lower than that observed could occur by chance, for a nonvariable source. Here chi-squared ($\chi^2$) is calculated as:
\begin{equation}
    \chi^2= \sum^{N_{obs}}_{i=1} \frac{(x_i-\overline{x})^2}{\sigma^2_{err,i}}, 
\end{equation}
\noindent where $N_{obs}$ is the number of observations, $x_i$ is the flux measured at each observation, $\overline{x}$ is the mean flux amongst all observations in the light curve, and $\sigma_{err,i}$ is the flux error. If a source is not variable, we expect that $\chi^2 \sim N_{obs}-1$. The typical threshold to distinguish variable from nonvariable light curves is $P_{var}=0.95$ \citep[e.g.,][]{2004ApJ...611...93P,2008A&A...487..475P}, which indicates a 95\% chance that the source is intrinsically variable, or, alternatively, a 5\% chance that the variability observed is due to Poisson noise. We calculate $P_{var}$ for both the continuum and \kalfa{} light curves, as listed in Table\,\ref{t:summary}.

\subsubsection{The excess variance $\sigma^2_{rms}$}
\label{sec:NXS}
The normalized excess variance $\sigma^2_{rms}$ \citep[e.g.,][]{1990Edelson,1997ApJ...476...70N,Vaughan2003variance,2004ApJ...611...93P,2008A&A...487..475P} is a quantitative measure of the variability amplitude of a light curve, defined as

\begin{equation} \label{eq:NXS}
    \sigma^2_{rms} = \frac{1}{(N_{obs}-1)} \sum^{N_{obs}}_{i=1} \frac{(x_i-\overline{x})^2}{\overline{x}^2}-\frac{1}{N_{obs}} \sum^{N_{obs}}_{i=1}\frac{\sigma^2_{err,i}}{\overline{x}^2}.
\end{equation}

\noindent Effectively, $\sigma^2_{rms}$ is the intrinsic variance of the light curve, normalized by the mean flux, producing a dimensionless quantifier that can be easily compared between objects of different brightness or light curves from different energy bands. The last term in Eq.~\ref{eq:NXS} denotes the contribution of the observational noise to the total variance, which is subtracted in order to find the intrinsic contribution.

Low intrinsic variances compared to the Poisson noise can sometimes lead to negative values of the $\sigma^2_{rms}$ estimate, since the uncertainty in the Poisson noise can be larger than the difference in Eq.~\ref{eq:NXS}. We performed Monte Carlo simulations to estimate more accurately the contribution of the observational noise (second term in Eq. \ref{eq:NXS}) and its asymmetric uncertainties in all our light curves, in order to quantify their variability robustly, even in cases where measured excess variance is negative.

For this purpose, we perform flux randomization of the light curves by adding a Gaussian deviate to each light curve point, with $\sigma$ equal to the error on the flux of each point. This procedure adds variance to the light curves, on average by an amount equal to the observational noise, as shown in Appendix \ref{ap:noise}. In this way, each simulation has the intrinsic variance of the light curve and twice the observational noise. Subtracting the variance of the original light curve from the flux-randomized light curve produces one estimate of the observational noise. Repeating this process 1000 times for each light curve, we obtained the median variance produced by the observational noise and its 16\% and 84\% bounds. We compared the median and bounds of the resulting excess variances (i.e., total variance -noise estimate) to the excess variance and error formula expressed in equations 6, 8, 9, and 11 of \citet{Vaughan2003variance}, obtaining consistent results for most cases.

We consider light curves to be significantly variable if the lower 16\% bound of the excess variance distribution is positive. We caution that this limit may result in a few nonvariable sources being misclassified as variable, but we adopt it nonetheless to improve the completeness of the variable sample at the cost of reducing its purity.

\begin{figure*} 
\centering
$\begin{subfigure}[b]{0.45\textwidth}
 \includegraphics[width=\textwidth]{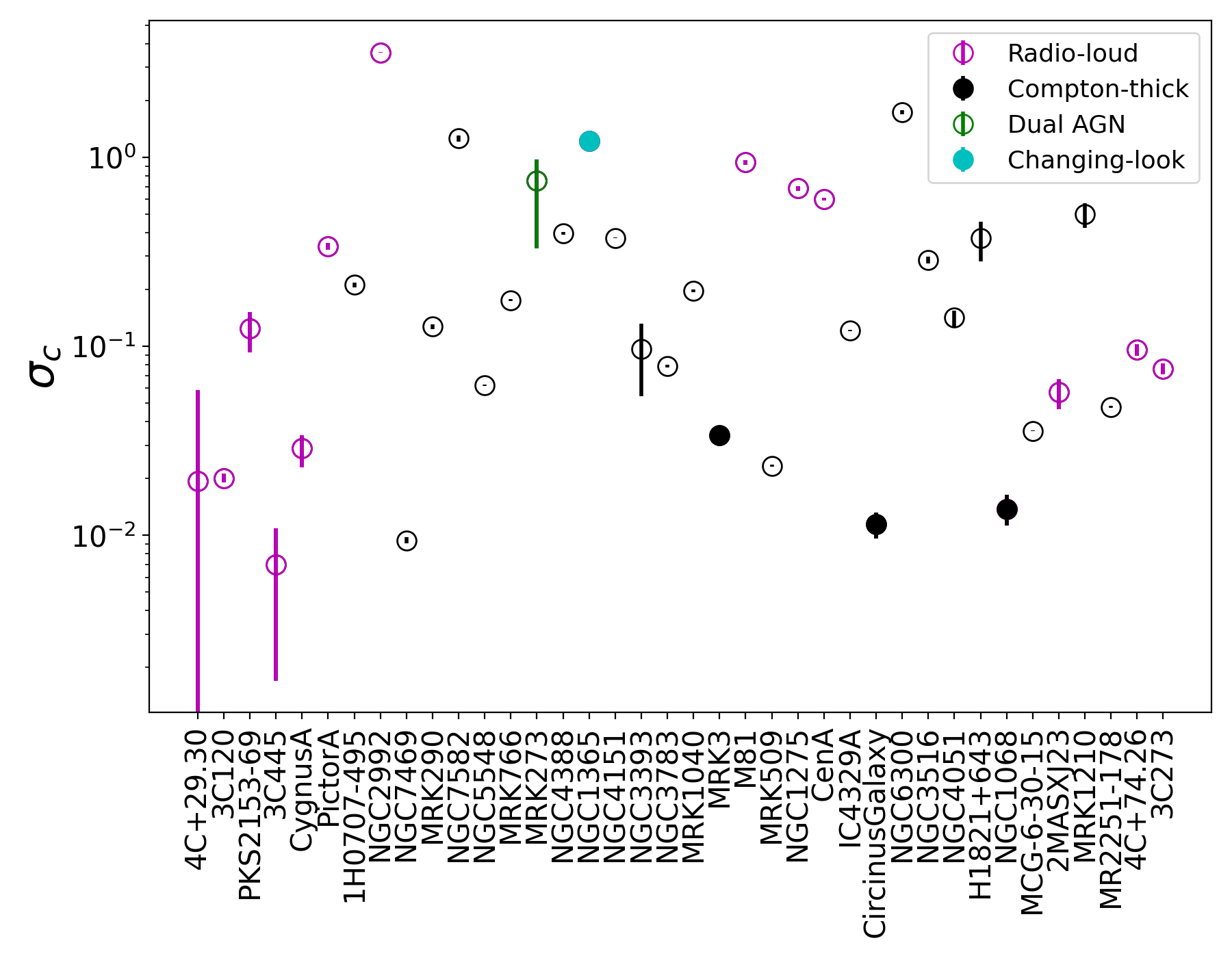}
 \caption{}
 \label{subfig:exvars_cont}
 \end{subfigure}$
$\begin{subfigure}[b]{0.45\textwidth} 
 \includegraphics[width=\textwidth]{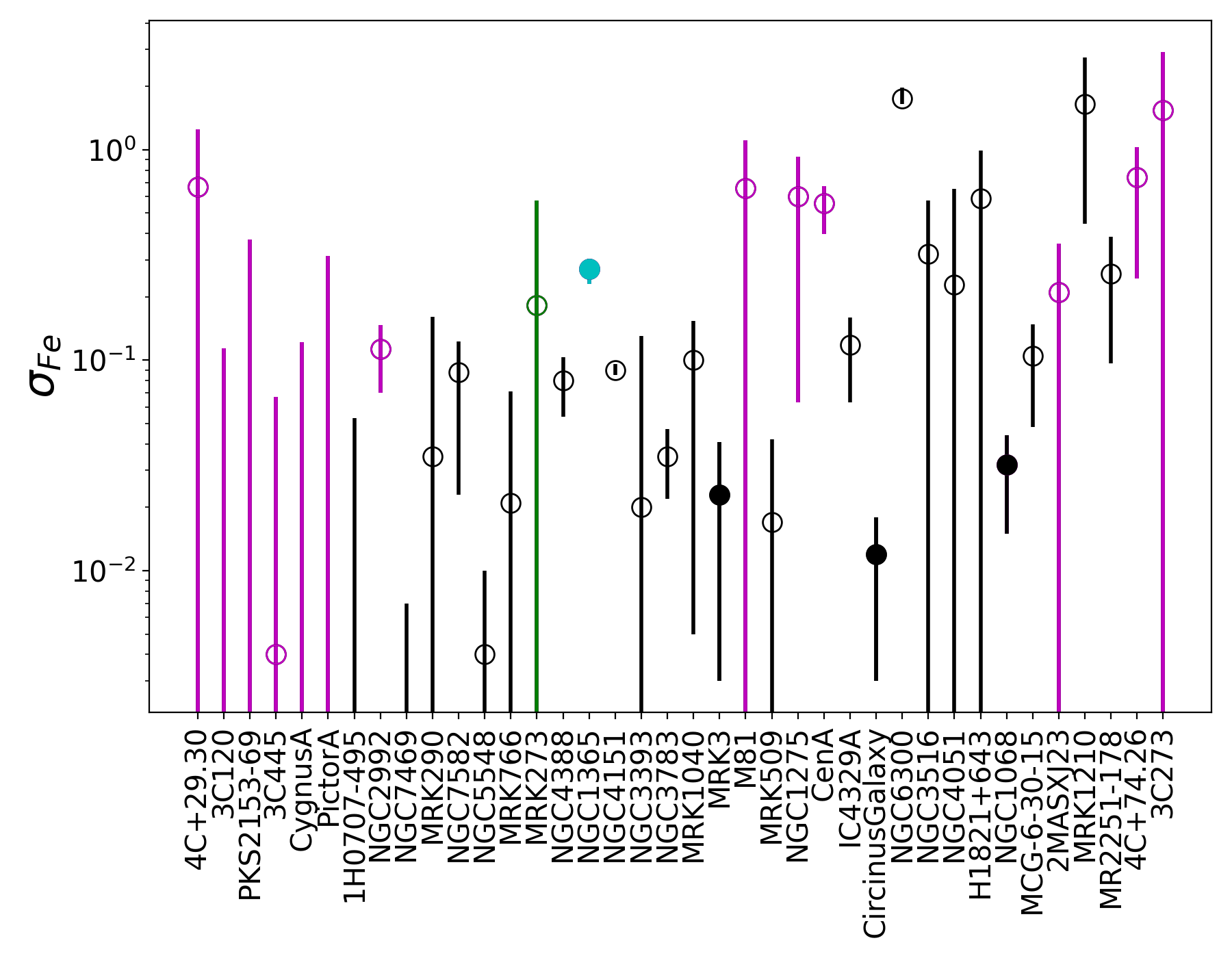}  
 \caption{}
\label{subfig:exvars_fe}
 \end{subfigure}$  

 $\begin{subfigure}[b]{0.45\textwidth}  
 \includegraphics[width=\textwidth]{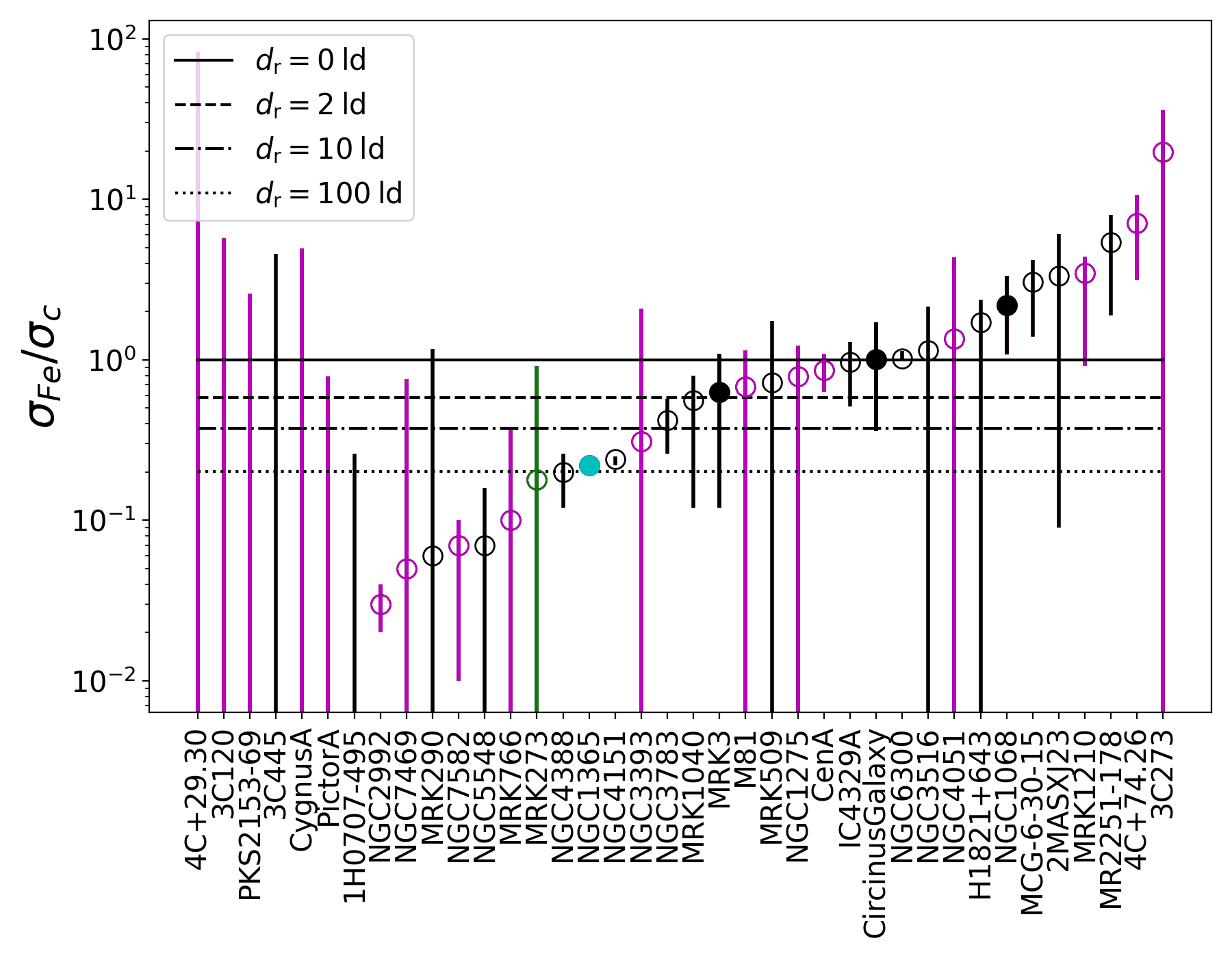}
 \caption{}
 \label{subfig:exvars_fe_cont}
 \end{subfigure}$ 
$\begin{subfigure}[b]{0.45\textwidth}  
 \includegraphics[width=\textwidth]{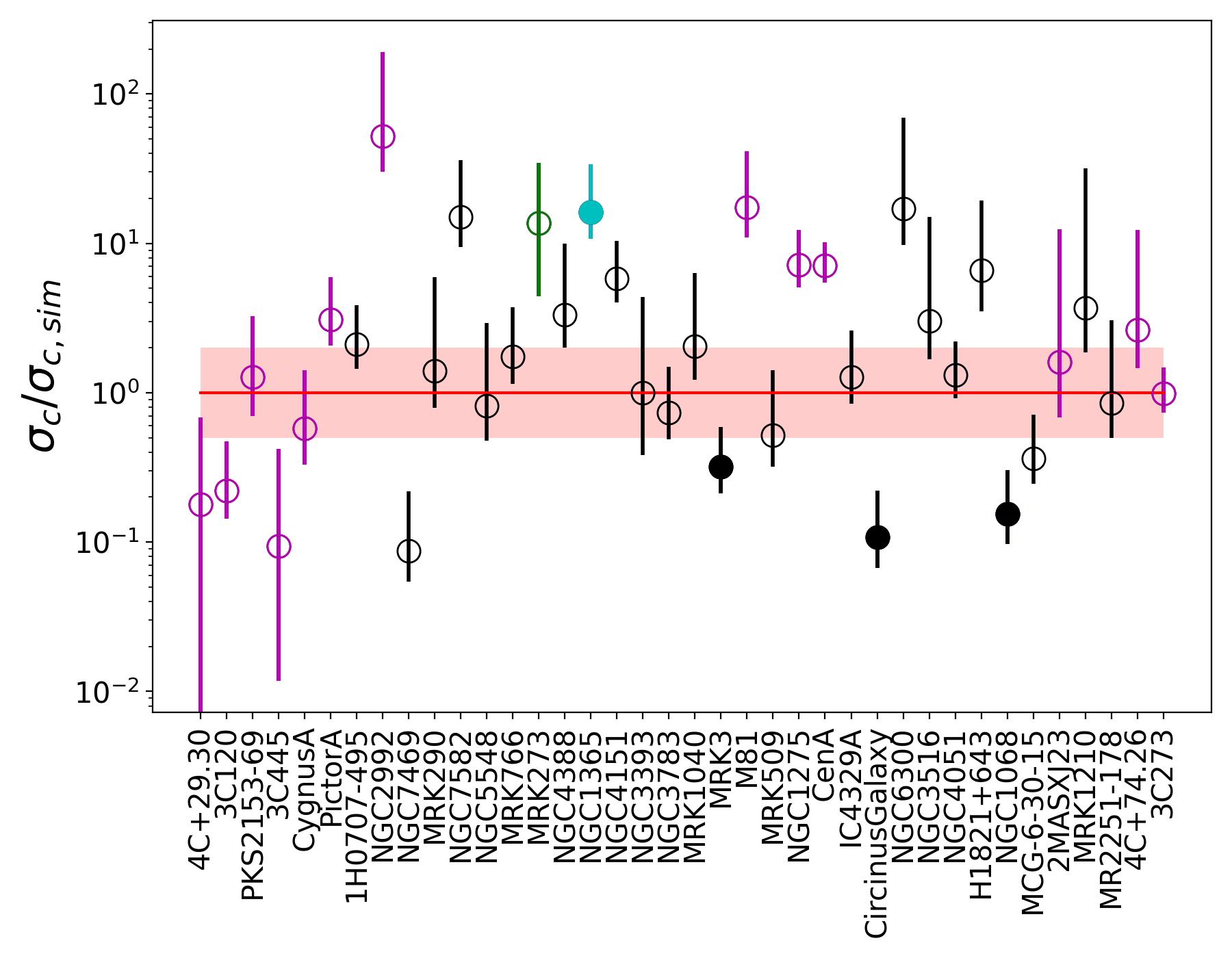}
 \caption{}
 \label{subfig:exvars_cont_sim}
 \end{subfigure}$ 
 
  \caption{Variability properties of the sample. Sources are ordered based on the ratio between the excess variances of the continuum and the \kalfa{} line, $\sigma_{\rm Fe}^2/\sigma_{\rm c}^2$. All four plots adopt the same color code as in Figure\,\ref{subfig:exvars_cont} . Open black circles denote AGN without a special characteristic, magenta empty circles are radio-loud AGN, black filled circles are CT\.AGN, green empty circles are dual AGN, and filled cyan circles are changing-look AGN.
 {\it (a)}: normalized excess variance of the continuum light curves. 
 {\it (b)}: normalized excess variance of the Fe K$\alpha$ line light curves.  
 {\it (c)}: the ratio between the normalized excess variance of the Fe K$\alpha$ line and continuum light curves, where the four horizontal lines denote ratio for simulated reprocessors with diameters $d_{\rm r}=0,2,10,100 \rm \: ld$, assuming identical 10 days power-spectral bend timescales.
 {\it (d)}: ratio between the normalized excess variance of the simulated and real continuum light curves, with unity denoted by the red line. The light red shaded region denotes up to a factor 2 error in our assumed normalization of the simulated power-spectral model. Magenta empty circles are radio-loud sources, black filled circles are Compton-thick sources, green empty circles represents dual AGN, and cyan empty circles are changing-look AGN. Mrk\,3 is both Compton-thick and radio-loud, while NGC\,1365 is both radio-loud and changing-look. 
Plotted values represent the median of each parameter while the error bars correspond to the 16$\%$ and 84$\%$ bounds of the normalized excess variance distributions.\label{fig:exvars}}
 \end{figure*}

Ultimately, we want to compare the variability amplitudes of the continuum and Fe K$\alpha$ line light curves, distinguishing between cases where the Fe K$\alpha$ line variability is not significant because it is small compared to that of the continuum versus cases where the variability amplitudes are similar but the lower count rate in the Fe K$\alpha$ line renders its variability insignificant. For this, we employ a similar set of simulations, with one realization of the noise for the continuum and another for the Fe K$\alpha$ line light curve, to compile the distribution of the ratio between the excess variances of both. Table \ref{t:summary} shows the 50\% percentile and $1\sigma$ uncertainties of the continuum (Col. 10) and \kalfa{} (Col. 11) distributions of the excess variances of each light curve, as well as their ratio (Col. 12). These data are also plotted in Fig.~\ref{fig:exvars} (panels a, b, and c). To illustrate how the ratio changes as a function of reprocessor size, the ratios corresponding to four simulated reprocessors with different diameters ($d_{\rm r}=0,2,10,100 \rm \: ld$) and the same power-spectral bend timescale of 10 days are shown in Fig.~\ref{fig:exvars}c.

Significant variability, denoted by a positive lower bound on the $\sigma^2_{rms}$, is detected in the continuum of 37 out of 38 AGN; the only exception is 4C+29.30.  Significant Fe K$\alpha$ line variability is detected in 18 AGN: 4C+74.26, Cen A, Circinus Galaxy, IC4329A, MCG-6-30-15, MR2251-178, MRK\,1040, MRK\,1210, MRK\,3, NGC\,1068, NGC\,1275, NGC\,1365, NGC\,2992, NGC\,3783, NGC\,4151, NGC\,4388, NGC\,6300 and NGC\,7582. For the rest of the AGN, the lower bound on the excess variance is negative, implying that the Fe K$\alpha$ line variability is consistent with observational noise, within uncertainties. The upper bound on the Fe K$\alpha$ line variability in those AGN is still of interest, depending on how this bound compares to the variance of the continuum.

If the Fe K$\alpha$ line flux tracks the fluctuations of the continuum flux, then we expect their variances to be related. The ratio of the variances should be similar to unity if the reflector is small compared to the timescale of the fluctuations, and smaller than 1 if the reflector is large. Therefore, upper bounds smaller than 1 in the ratio column of Table~\ref{t:summary} can allow us to place a lower limit on the size of the reflector. This is true in the first two objects in Table~\ref{t:summary}. Even though their Fe K$\alpha$ line variance is not significant, its upper limit is so far below the variance of the continuum that a lower limit can be placed on the size of the reflector, because if the reflector was any smaller, the Fe K$\alpha$ line variability would be detectable. Conversely, lower bounds of the variance ratio larger than 0 can put upper limits on the reflector size. One or both of these limits are therefore measurable in many objects in the table. We describe this analysis in detail in $\S$\ref{sec:rep_sims}.

The normalized excess variance of the continuum is generally well constrained in the sources listed in Table~\ref{t:summary}. The median values of the $\sigma^2_{rms}$ distributions range from 0.007 for 3C\,445 to 3.59 for NGC\,2992, as can be seen in Fig.~\ref{fig:exvars}a. This large difference can raise doubts about a common origin of the continuum variability in all sources. We know however, from dedicated monitoring campaigns of radio-quiet AGN, that the X-ray power spectra of different AGN is remarkably uniform in shape, but that the frequency $\nu_{\rm b}$ of the only break in the power spectrum scales inversely with black hole mass and directly with accretion rate \citep{2006Natur.444..730M}. Since the excess variance equals the integral of the power spectrum over the timescales covered by the light curves, we can expect to measure significantly different values for objects of different SMBH mass, even if the light curves have similar lengths.

To test how unusual some of these excess variances are, we compute the expected variance for the respective SMBH masses and accretion rates. Since the variability is a stochastic process, and the underlying power-spectral shape is only realized on average, simulations provide a good way of estimating the possible range of variance measurements, for given SMBH parameters and light curve sampling pattern.
Thus for each source we generate simulated light curves of a red-noise process following the method of \citet{1995A&A...300..707T}. The underlying power-spectral shapes are bending power-laws, with a low-frequency slope $\alpha_{\rm L}{=}-1$ bending to a steeper slope at higher frequencies $\alpha_{\rm H}$ at a characteristic timescale or break frequency $\nu_{\rm b}$:

\begin{equation} \label{eq:power_spectrum}
P(\nu)=
    A \frac{
        \nu^{-\alpha_{\rm L}}
    }{
        1+(\nu/\nu_{\rm b})^{\alpha_{\rm H}-\alpha_{\rm L}}
    },
\end{equation}

where the normalization is $A{=}10^{-2}$\,s with frequencies measured in Hz. This means that at low frequencies, $\nu \ll\nu_{\rm b}$, the combination $\nu P(\nu)$ tends to this constant value of $A$. We adopt this as the 'standard power-spectral density (PSD) model'. The variance scales linearly with the normalization $A$, such that by shifting $A$,  the mean expected variance and its scatter shift by the same factor. Differences in $A$ by a factor of 2 (higher or lower) are consistent with the majority of the monitored sample.

This shape has been shown to represent well the power spectrum of AGN X-ray light curves. When available, we use the high-frequency slopes and break frequencies measured for each object in \citet{summons_thesis} based on the long-term monitoring campaigns performed with the {\it Rossi X-Ray Timing Explorer} (RXTE) observatory. For the rest, we estimate the break frequency using the expression from \citet{2012A&A...544A..80G}:
{\small
\begin{equation} \label{eq:Tbreak}
  \log{T_{\rm b}}{=}(1.09\pm 0.21)\log{M_{\rm SMBH}} + (-0.24\pm0.28)\log{L_{\rm bol}} - 1.88\pm 0.36,
\end{equation}
}

\noindent where $T_{\rm b}$ is the power-spectral bend timescale ($T_{\rm b}{=} 1/\nu_{\rm b}$), $M_{\rm SMBH}$ is the SMBH mass in $10^6\:\rm M_{\odot}$ units, and  $L_{\rm bol}$ is the bolometric luminosity in $10^{44}\:\rm erg\:s^{-1}$ units. The adopted bend frequencies $\nu_{\rm b}$ are tabulated in Table~\ref{t:properties}. The high-frequency slopes $\alpha_{\rm H}$ in \citeauthor{summons_thesis} (\citeyear{summons_thesis}; see also \citealp{Vaughan2003variance,Markowitz2003,Mchardy2004,Uttley2005,Mchardy2005,Summons2007}) range from 1.5 to 3.5. For sources in which $\alpha_{\rm H}$ has not yet been measured, we adopt an intermediate value of $\alpha_{\rm H}=2$. The power-spectral model used herein only considers one break timescale, which is the only one detectable with short-term light curves such as those used by \citet{2012A&A...544A..80G}. It is possible that the power spectra for our AGN sample feature a second break at lower frequencies, flattening further to a value of 0. Dedicated monitoring campaigns on a few sources such as MCG--6-30-15 \citep{Mchardy2004}, NGC4051 \citep{Mchardy2005} and NGC3783 \citep{Summons2007} do cover timescales similar to those covered in this work, and notably have not shown any second, lower-frequency break, so this consideration might not be relevant to the sources studied here. If there were a second break, we would observe lower variances than those predicted by the model, especially for sources with a higher break frequency, for which the second break could be at higher frequencies as well.

We generated 100 realizations of the continuum light curves following this underlying power-spectral model and the expected or measured bend timescales and high-frequency slopes ($\alpha_{\rm H}$) for each object. The simulations are run for at least 10,000 days or 3 times the length of the light curve, whichever is longest, and generated with a time step of 0.01 days or 1/100 times the bend timescale, whichever is smaller. Each simulation is then sampled in an identical manner to the corresponding real light curve, its variance is computed and the mean and root-mean-squared scatter of the resulting variances are recorded. This mean and scatter represent our expectation for the excess variance measured in the continuum light curves. 

The ratio between the real and expected excess variances, $\sigma_{\rm c}/\sigma_{\rm c, sim}$, is plotted in Fig.~\ref{fig:exvars}d, where the error bars represent only the scatter of expected variances produced by the stochasticity of the intrinsic variability; no observational noise is added. Naively, we expect sources to cluster around $\sigma_{\rm c}/\sigma_{\rm c, sim}\sim1$ (solid red line); allowing for up to a factor 2 difference in the normalization of the power-spectral model (denoted by the light red shaded region), we find that a  majority of the measured variances are consistent with this expectation. However, some sources remain significantly above and below this; we distinguish radio-loud sources (magenta empty circles), Compton-thick AGN (black filled circles), dual AGN (green empty circles), and changing look AGN (cyan filled circles), which tend to be outliers.
The radio-loudness (RL) of the sample is computed using the 20\,cm and 14--150\,keV fluxes from \citet{2017ApJS..233...17R} as $\rm RL=\log(f_{\rm 20\,cm}/f_{\rm 14-150\,keV})$ \citep[e.g.,][]{2003ApJ...583..145T, 2013MNRAS.432.1138P, 2015MNRAS.447.1289P}, listed in column 6 of Table\,\ref{t:properties}.  We adopt a separation between radio-loud and quiet sources at $\rm RL=-4$ following \citet{2015MNRAS.447.1289P} and \citet{2017ApJS..233...17R}.
It is possible that the continuum (and Fe K$\alpha$) variability of radio-loud sources could be affected by beamed X-ray emission associated with the powerful jets, leading to stronger or more rapid variations \citep[e.g.,][]{1997Ulrich,2008Chatterjee,2020Weaver}; this could explain why 6/18 radio-loud sources in our sample have $\sigma_{\rm c}/\sigma_{\rm c, sim}{>}$1.
On the other hand, Compton-thick AGN should have reflection-dominated continua with little variation, and thus our simple X-ray spectral fitting approach would measure the flux of this relatively static component. Thus, it is not surprising that the three Compton-thick sources of our sample all have $\sigma_{\rm c}/\sigma_{\rm c, sim} < 1$. Appendix\,\ref{ap:individual_sources} contains notes on individual sources that do not seem to conform to the standard PSD model.

\subsection{Estimating the size of the X-ray reflector} \label{sec:rep_sims}

If we assume that the \kalfa{} line emission is reprocessed from the same X-ray continuum we observe, we can expect the \kalfa{} line flux to track the continuum fluctuations. The light curves of both, primary and reflected components, can still differ by light travel time effects, as the reflected light will travel on different and longer paths to the observer. The \kalfa{} line light curve can therefore be delayed with respect to the continuum and can also be smoothed out, as variations on timescales shorter than the light crossing time of the reflector are damped.

The majority of the light curves obtained in this work are too sparsely sampled, however, to detect directly a delay or lag between the continuum and Fe K$\alpha$ line fluctuations, as exemplified in Appendix Figs.~\ref{fig:lc_cont} and \ref{fig:lc_fe}. The potential reduction of the \kalfa{} line variability amplitude with respect to the continuum, however, can still shed light on the size of the reflector, as larger reflectors will suppress a larger fraction of the intrinsic variance. We quantify the reduction in the observed variability amplitude by calculating the excess variance of the continuum ($\sigma^2_{\rm c}$) and \kalfa{} line ($\sigma^2_{\rm Fe}$) light curves and taking their ratio as $var_{r}=\sigma^2_{\rm Fe}/\sigma^2_{\rm c}$. The longer the light crossing time of the reflector, the smaller the expected value of $var_{r}$ should be.

\begin{figure}
\centering
\resizebox{\hsize}{!}{\includegraphics{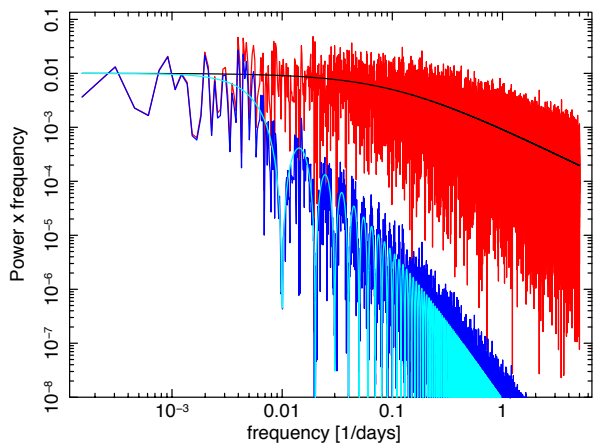}}
\caption{Power spectra of simulated X-ray light curves. The underlying power-spectral shape of the continuum was chosen as a bending powerlaw model with a bend timescale of 10 days (black) and a given realization of this model produces the power spectrum plotted in red. Reflecting the light curves over a spherical reflector of radius $R=50$ light days suppresses the high frequency power and results in the model and realization plotted in cyan and blue, respectively. }\label{fig:sims_pds}	

\end{figure}

\begin{figure}
\centering
\resizebox{\hsize}{!}{\includegraphics{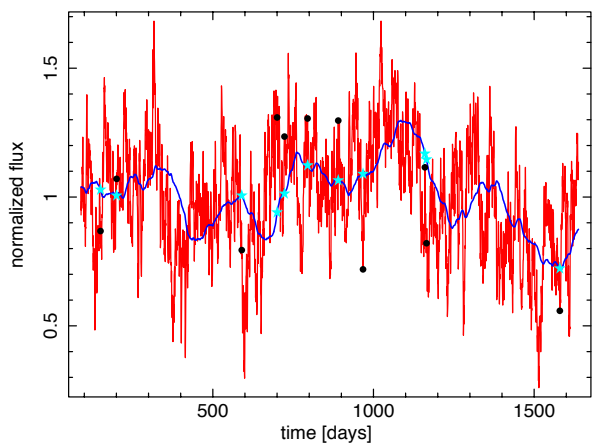}} %[width=0.5\textwidth]
\caption{Example light curve realizations of the continuum (red) and reflected light (blue), which correspond to the power-spectral models plotted in Fig.~\ref{fig:sims_pds}. The reflected light curve reduces variability amplitude and is delayed by 50 days. Increasing the size of the reflector further reduces the amplitude of the reflected light curve and shifts it further forward in time. A random selection of observing epochs are plotted, representing measurements of the continuum (black dots) and Fe K$\alpha$ line (cyan stars) fluxes. The sampled light curves are not obviously correlated, although the parent light curves are.  }\label{fig:sims_lcs}	

\end{figure}

To demonstrate the effect of reflection on the light curves, we show in Fig.~\ref{fig:sims_pds} a typical power spectrum of continuum X-ray variations, which are damped by a reprocessor at a distance of $50$\,light days; for simplicity we assume a simple thin spherical shell reprocessor, but acknowledge that more realistic distributions could be thicker, clumpier, and have a toroidal or ionization cone-like structure seen at a specific orientation with respect to the line-of-sight, all of which can impact delay times. The corresponding continuum and reflected Fe K$\alpha$ light curves are shown in Fig.~\ref{fig:sims_lcs}. The reflected light curve is differentially delayed (by design), which results in a clear drop in variability amplitude at high frequencies (or short timescales).
If we observe these simulated light curves in a manner comparable to the real data, the sparse sampling would result in two effects: (1) the sampled light curves could appear uncorrelated, preventing a measurement of any lag, and (2) the reduction in variance will be less deterministic, as the reflected light curve can be sampled at times when it is more variable by chance. To illustrate this, Fig.~\ref{fig:sims_lcs} shows continuum (black dots) and Fe K$\alpha$ line (cyan stars) flux measurements for a random selection of observing epochs. The sampled light curves are not obviously correlated, although the parent light curves are. We note that observational noise has not been added to these simulations; all the scatter is produced by the intrinsic fluctuations of the source. 

The variance of the continuum light curve in the above example, after sampling, is still larger than the variance of the reflected light curve, although the ratio of the variances depends on the sampling pattern. Intuitively, the degree of apparent randomness in the reflected light curve, and the scatter between the continuum and reflected variance ratios, increases as the time lag increases compared to the characteristic timescale of fluctuations, as quantified by the bend timescale in the power spectrum. The delay between continuum and reflected light curves, together with the sparse sampling, can explain part or all of the scatter seen in the flux-flux plots in Appendix Figs.~\ref{fig:fluxes_vs_q}.

With this in mind, we examine the observed values of $var_{r}$ in Fig.~\ref{subfig:exvars_fe_cont}. Somewhat surprisingly, we find that 11\% (4/38) of sources lie $>$1-$\sigma$ above $d_{\rm r} = 1 \rm \: ld$ (black continuous line), although all of these appear consistent with unity  within 2-$\sigma$ uncertainties. Similarly, 21\% (8/38) of sources lie above the $d_{\rm r} = 10 \rm \: ld$ line (${\sim}0.009$\,pc, close to the lower grouping of sources in Fig.\ref{fig:rfe_rsub_rhb}), but at 2-$\sigma$, only Cen\,A remains confidently above this value. In total, more than 50\% have values consistent with unity, implying little damping of the continuum by the reflector (i.e., angle-averaged light-crossing timescales to the reflector are comparable to the continuum variability timescales), or alternatively that we are not observing the true continuum fluctuations (as is likely the case for the three Compton-thick AGN, which all have ratios consistent with unity). Notably, all the sources which show stronger than expected continuum excess variances in Fig.~\ref{subfig:exvars_cont_sim} consistently have $var_{r}$ values below unity.

Although it is straightforward to estimate $var_{r}$ for our fiducial reflector model as a function of the break frequency of the continuum light curve power spectrum and the light crossing time of the reflector, we expect a large scatter in measured values due to the small number of data points and the stochastic nature of the light curves. We therefore adopt a Monte Carlo approach, as described below, to place meaningful constraints on the size of the reflector, incorporating both the stochasticity of the intrinsic variations and the observational noise.

We employ the same setup as for simulating continuum light curves from a bending powerlaw underlying power spectrum, described in $\S$\ref{sec:NXS}.
To simulate reflected light curves, we multiply  both the real and imaginary parts of the Fourier transform of the simulated continuum light curve by a  $\sinc(\nu\tau)$ function, which corresponds to a convolution with a top hat function of the light curve in time space. We then shifted the inverse Fourier-transformed curve forward in time by $\tau/2$ days. This corresponds to reflection of the continuum by a spherical shell of radius $R=c\tau/2$, leading to an immediate response from the front end of the reflector, followed by reflection from the rest of the shell until the light from the back end finally reaches the observer $2R/c=\tau$ days after the start of the initial front-end response \citep[e.g., \S5.1 in][]{Arevalo2009reprocessing}. This particular response function is chosen for simplicity, and reproduces the main characteristics of a reflected light curve, that is suppressed variability amplitude and average time delay. This is sufficient to estimate the size of the reflector, but not its geometry. Replacing the idealized thin shell reprocessor above (i.e., the $\sinc^2$ multiplicative function on the power spectrum) with the average of three reprocessors of similar size (e.g., $\tau, \tau\times 1.5, \tau/1.5$) can approximate the response of a finite width reprocessor. We implemented this setup as well, noting that it made little difference on the resulting sizes. The sizes reported below correspond to the latter finite-width reprocessor case. 

Transfer functions for different reprocessors, such as a thick shells with different radial matter distributions, and flat or flared disks are presented in $\S 5.1$ of \citet[][]{Arevalo2009reprocessing} for a given average delay $\tau$ between the direct and reprocessed light curves. All these functions have in common a flat response for frequencies below 1$\tau$ and consistently drop above it. Since we can only attempt to measure the decrease in variance of the \kalfa{} line light curve compared to that of the continuum, our interest is to compare the integrals of the original and filtered power spectra, not their precise shape. Therefore, any response function that generally complies with the shape described above will suffice.
We note however that reprocessors that are strongly asymmetric, such as clouds lying along only one axis or a compact region far from the nucleus are only partially captured by our approach. These reprocessors can have a smoothing effect proportional to their diameters, but a delay proportional to the average distance to the source. In these cases the reduction in \kalfa{} line variance would give an estimate of the diameter of such reprocessors, but the delays, and therefore the allowed scatter between realizations of the variance ratio, would be underestimated.

For each source, we consider its spectral parameters and choose a sampling timescale $dt$ that is related to its bending timescale as $dt=T_b/100$ or $dt=0.1$, whichever was shortest, such that the sampling resolution is at least a 100th of the $T_b$ scale. We resample the simulated continuum and \kalfa{} light curves according to the observed epochs.  
Finally, we compute the ratio of the variances ($\rm var_{r,sim}=\sigma^2_{\rm Fe, sim}/\sigma^2_{c,sim}$). As expected, this ratio decreases as the light crossing time of the reflector increases. We run sets of 100 simulations and record the median and median-absolute-deviation (MAD) of $\rm var_{r,sim}$ for a range of values of the light crossing time. To determine the $\tau$ values which correspond to our estimated reprocessor size and its upper and lower limits, we adopt the following statistic:
\begin{equation} \label{eq:X}
X(\tau) = \frac{\rm var_{r,real} - \rm var_{r,sim}}{\sqrt{\rm MAD( var_{r,sim})^2+err(var_{r,real})^2}},
\end{equation}
\noindent 
where $\rm var_{r,real}$ is the median value of the $\rm var_{r}$ distribution and $\rm var_{r,sim}$ is the median value of the simulated ratios. 
 
The $\tau$ values that return $X=0,1,-1$ correspond to the reprocessor size and its $\pm1\sigma$ uncertainties in light days, respectively. In order to constrain the size of the reprocessor, we first explore a range of $\tau$ values between 0.01--10,000 days, and then numerically look for the roots $X=0,1,-1$ using Newton's method.

Table~\ref{t:summary} summarizes our estimates for the size of the reprocessor ($R_{\rm rep}$) derived from the values of $\tau$ at $X=0,1,-1$, as well as the $\nu_b$ value used in the simulations. The $R_{\rm rep}$ values are also shown in summary Figure~\ref{fig:summary_sizes} (black squares), for comparison against the other reprocessor size estimates or limits derived from the spectral and imaging (see $\S$\ref{sec:image}) analysis. The simulations provide estimates of the reprocessor size for 24 sources in our sample.

We note that this approach can only provide limits on the size of the reflector if a limit of the measured ratio is between 0 and 1. If the upper and lower limits of the measured ratio, considering errors from observational noise only, already cover the whole range of possible ratio values, then any reprocessor size is consistent with the data and no limits can be placed. This is the case for 14 of our sources, as can be seen in Fig.~\ref{fig:exvars}c.  Conversely, if one or both limits on the measured ratio fall between 0 and 1, it is in principle possible to place limits on the size of the reflector. The additional scatter in possible ratio values produced by the red-noise nature and sparse sampling of the light curves can, however, extend the error bars beyond the 0--1 range and prevent a limit to be placed. This happened in the case of Pictor A. In addition, for sources where the lower limit on the measured ratio was above 1, an upper limit on the reprocessor size could still be placed in the case of NGC1068 since the additional scatter in the ratio allowed a value of $\tau$ for which $X=1$. For sources MCG-6-30-15, MR2251-178 and 4C+74.26, the smallest value of $\tau$ explored still produced a value of $X>1$, so in these cases the upper limits on $\tau$ were set to the smallest value explored. 

\subsection{Correlations between the Fe $\rm K\alpha$ line and 2--10\,keV continuum} \label{sub:slope}

A strong, positive correlation between \kalfa{} line and 2--10 keV continuum fluxes, which is much simpler to assess compared to more complicated lag analyses, should indicate that the reflector lies in close proximity to the source of the X-ray continuum emission. 
Thus we explore potential correlations between the observed 2--10 keV continuum and Fe $\rm K\alpha$ line light curves for sources in our sample. As mentioned above, any lag in the \kalfa\ line light curves, which is expected from reprocessing travel time delays, can reduce the apparent correlation between the observed light curves. The degree of loss of correlation is a function of the ratio between the light crossing time of the reprocessor and the characteristic timescales of fluctuations, as well as the geometry of the reprocessor. 

\begin{figure}
% %\centering
\resizebox{\hsize}{!}{\includegraphics{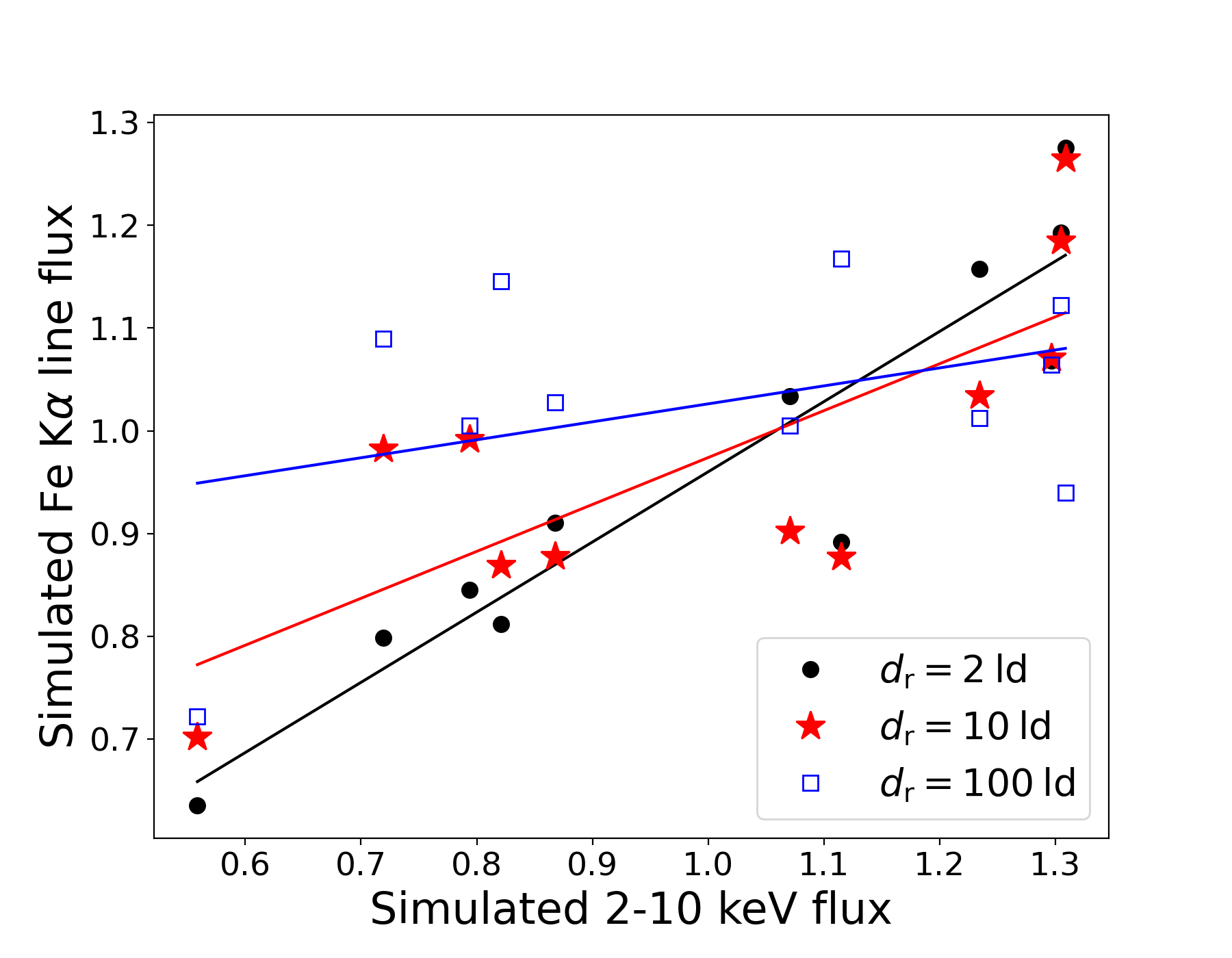}}
\vspace{-0.2cm}
\caption{Comparison of simulated \kalfa{} and 2--10 keV continuum mean-normalized fluxes. The continuum simulations use a power-spectral bend timescale of 10 days, while three different reprocessor diameters ($d_{\rm r}$) are chosen: 2 light days (black dots), 10 light days (red stars) and 100 light days (blue open squares). The simulated light curves are resampled using the observing epochs
shown in Fig.~\ref{fig:sims_lcs}). The different sets highlight how varying the reprocessor size can damp potential flux-flux correlations. Evidently, the larger the reprocessor, the larger the scatter in the flux-flux relation, and the flatter the best fitting linear regression (solid lines with corresponding colors).}
\label{fig:sims_flux_flux}
\end{figure}

For illustration, Fig. \ref{fig:sims_flux_flux} shows a comparison between the fluxes from the sampled observing epochs of the simulated light curves in Fig.~\ref{fig:sims_lcs} (i.e., the dots and stars). In that simulation, the power-spectral bend timescale is 10 days, while the reprocessor diameter is 100 light days. The corresponding flux-flux points, normalized to their respective means, are plotted in blue open squares. The correlation is weak, with a large scatter, and the best-fitting linear regression yields a slope of $0.17\pm 0.19$; the expected slope for a perfect correlation and equal variability amplitudes is 1. However, if the same continuum light curve is reprocessed by a smaller structure, the correlation improves and the slope approaches a value of 1. In the same figure we show the result of sampling the light curves on the same observing epochs but using a reprocessor diameter of 10 days (red stars) and 2 days (black dots). The best-fitting slopes for these smaller reprocessors are $0.46\pm 0.18$ and $0.68\pm 0.12$, respectively. We note that the loss of correlation in these sampled light curves is due entirely to the effect of the reflection, both smoothing and delaying the \kalfa\ light curve, since no observational noise is added to these simulations. A key point here is that, at least for a symmetric reflector observed with a relatively sparse cadence, its light travel distance only needs to be a $\sim$10 times larger than the continuum source variability timescale for any correlation to be almost completely washed out.

To assess if the \kalfa{} line and continuum light curves of our sources are correlated, we fit a line using the python package {\sc Linmix}.\footnote{see \url{https://github.com/jmeyers314/linmix}}  This package is based on the code of \citet{2007ApJ...665.1489K}, which describes a Bayesian method to perform linear regressions with measurement errors in both variables. The code runs between 5000 and 100000 steps of a Markov chain Monte Carlo to produce samples from the posterior distribution of the model parameters, given the data. We choose this code since it does not assume a particular distribution for the errors and incorporates nondetections, both of which are fundamental for this study since in several cases the fluxes of the Fe K$\alpha$ line are upper limits. The flux errors obtained in the spectral analysis are mildly asymmetric, as the data often straddle the division between Poisson and Gaussian distributions. We symmetrize them adopting the average between upper and lower errors, in order to use them as input to {\sc Linmix}, but this does not strongly impact the results for most sources; for sources with well-constrained fluxes, the positive and negative errors only differ by a few percent ($\lesssim 5\%$), while for poorly constrained fluxes, we adopt conservative (larger) errors or upper limits.

\begin{table*}
%\begin{adjustbox}{width=\textwidth}
\caption{Summary of fitted and computed quantities} 
\centering
\scalebox{0.63}{
\begin{tabular}{l c c c r r c c c c c c c l } 
 \hline
 \hline
\noalign{\smallskip}
\multicolumn{1}{c}{Source} & \multicolumn{1}{c}{$R_{\rm sub}$} & \multicolumn{1}{c}{$R_{\rm Fe\:K\alpha}$} & \multicolumn{1}{c}{$\upsilon_{\rm FWHM}$} & \multicolumn{1}{c}{{$ R_{\rm rep}$}} & \multicolumn{1}{c}{$\rm R_{IA}$} & \multicolumn{1}{c}{$m$} & \multicolumn{1}{c}{$P_{\rm var}$} & \multicolumn{1}{c}{$P_{\rm var}$}  & \multicolumn{1}{c}{$\sigma^2_{c,rms}$} & \multicolumn{1}{c}{$\sigma^2_{Fe,rms}$} & \multicolumn{1}{c}{$\sigma^2_{Fe,rms}/\sigma^2_{c,rms}$} & \multicolumn{1}{c}{$\sigma^2_{\rm c,sim}$} & \multicolumn{1}{c}{$\rm t_{\rm lc}$}  \\
& \multicolumn{1}{c}{(pc)} & \multicolumn{1}{c}{(pc)} & \multicolumn{1}{c}{(km\:s$^{-1}$)} & \multicolumn{1}{c}{(pc)} & \multicolumn{1}{c}{(pc)} &  & \multicolumn{1}{c}{$F_{\rm 2-10\: keV}$} & \multicolumn{1}{c}{{ $F_{\rm Fe\:K\alpha}$}} &  &  &  & & (years) \\

\multicolumn{1}{c}{(1)} & \multicolumn{1}{c}{(2)}& \multicolumn{1}{c}{(3)} & \multicolumn{1}{c}{(4)} & \multicolumn{1}{c}{(5)} & \multicolumn{1}{c}{(6)} & \multicolumn{1}{c}{(7)} & \multicolumn{1}{c}{(8)} & \multicolumn{1}{c}{(9)} & \multicolumn{1}{c}{(10)} & \multicolumn{1}{c}{(11)} & \multicolumn{1}{c}{(12)} & \multicolumn{1}{c}{(13)} & \multicolumn{1}{c}{(14)}   \\
\noalign{\smallskip}
\hline
1H0707-495       & ...             & ...                 & ...             & $>$0.00084 & $<2083$ &$-0.28 \pm 0.27$ & 1 & 0.37    &$0.212^{+0.006}_{-0.006}$&$-0.05^{+0.10}_{-0.13}$&$-0.15^{+0.41}_{-0.6}$ & $0.1 \pm 0.05$ & 18.98 \\ 
2MASXJ23444      & $5.85 \pm 3.04$ & ...                 & ...             & ---       & $<27579$ &$1.42 \pm 0.62$ & 1 & 0.91 &$0.057^{+0.010}_{-0.011}$&$0.21^{+0.15}_{-0.22}$&$3^{+3}_{-3}$  & $0.04\pm0.04$  & 6.36 \\ 
3C120            & $0.49 \pm 0.25$ & $0.083 \pm 0.053$   & $1954 \pm 625$  & ---       & $<1711$ &$-0.13 \pm 0.87$ & 1 & 0.99 &$0.020^{+0.001}_{-0.001}$&$-0.05^{+0.16}_{-0.43}$&$-3^{+9}_{-18}$ & $0.09 \pm 0.05$ & 13.37 \\ 
3C273            & $4.11 \pm 2.14$ & $0.54  \pm 1.24$    & $2700 \pm 3082$ & ---       & $<7096$  &$0.43 \pm 0.67$ & 1 & 1   &$0.076^{+0.005}_{-0.005}$&$1.5^{+1.4}_{-2.2}$&$20^{+16}_{-31}$ & $0.08 \pm 0.03$ & 18.07 \\ 
3C445            & $0.71 \pm 0.37$ & $0.057 \pm 0.026$   & $4990 \pm 1133$ & ---       & $<2820$ &--- & 1 & 0.53 &$0.007^{+0.004}_{-0.005}$&$0.004^{+0.063}_{-0.106}$&$-0.7^{+5}_{-14}$& $0.07\pm 0.06$  & 9.68 \\ 
4C+29.30         & $0.36 \pm 0.19$ & ...                 & ...             & ---       & $<3229$ &--- & 1 & 0.47 &$0.02^{+0.04}_{-0.05}$&$0.7^{+0.6}_{-0.9}$&$35^{+76}_{-98}$& $0.1 \pm 0.07$ & 8.89 \\ 
4C+74.26         & $1.29 \pm 0.67$ & ...                 & ...             & ---       & $<4789$ &--- & 1 & 0.99  &$0.096^{+0.007}_{-0.007}$&$0.7^{+0.3}_{-0.5}$&$7^{+4}_{-4}$& $0.04 \pm 0.03$ & 4.31 \\ 
Cen\,A             & $0.04 \pm 0.02$ & $0.10  \pm 0.05$    & $1816 \pm 461$  & $<$0.039  & $<157$  &$1.14 \pm 0.20$ & 1 & 1 &$0.60^{+0.01}_{-0.01}$&$0.56^{+0.11}_{-0.16}$&$0.86^{+0.23}_{-0.23}$ & $0.08 \pm 0.03$ & 18.03 \\ 
Circinus\,Galaxy & $0.08 \pm 0.04$ & $0.0050 \pm 0.0004$ & $1463 \pm 69$   & $<$0.15  & $95\pm 15$ &$0.32 \pm 0.18$ & 1 & 1  &$0.012^{+0.002}_{-0.002}$&$0.012^{+0.006}_{-0.009}$&$1.0^{+0.7}_{-0.7}$ & $0.11 \pm 0.05$ & 18.27 \\ 
CygnusA          & $0.76 \pm 0.40$ & ...                 & ...             & ---       & --- &$1.12 \pm 0.72$ & 1 & 1 &$0.029^{+0.005}_{-0.006}$&$-0.006^{+0.128}_{-0.152}$&$-0.4^{+5.4}_{-6.8}$ & $0.05 \pm 0.03$ & 17.01 \\ 
H1821+643        & $3.09 \pm 1.61$ & ...                 & ...             & ---   & $<8559$ &$2.32 \pm 0.76$ & 1 & 1  &$0.37^{+0.09}_{-0.09}$&$0.6^{+0.4}_{-0.7}$&$1.7^{-0.7}_{-1.9}$ & $0.06\pm0.04$  & 7.27 \\ 
IC4329A          & $0.42 \pm 0.22$ & $0.16  \pm 0.17$    & $1503 \pm 799$  & $<$ 0.08   & $<849$  &$1.09 \pm 0.28$ & 1 & 1  &$0.1208^{+0.001}_{-0.0009}$&$0.12^{+0.04}_{-0.06}$&$1.0^{+0.3}_{-0.5}$ & $0.1 \pm 0.05$ & 17.05 \\ 
M81              & $0.00 \pm 0.00$ & $0.26  \pm 0.63$    & $1323 \pm 1587$ & --- &$<6$&$0.78 \pm 0.34$ & 1 & 1  &$0.94^{+0.03}_{-0.03}$&$0.7^{+0.5}_{-0.9}$&$0.7^{+0.5}_{-0.9}$ & $0.05\pm0.03$ & 16.87 \\ 
MCG-6-30-15      & $0.11 \pm 0.06$ & ...                 & ...             & $<$0.00084       & $<414$  &$0.40 \pm 0.58$ & 1 & 1   &$0.0357^{+0.0003}_{-0.0002}$&$0.11^{+0.04}_{-0.06}$&$3.1^{+1.1}_{-1.7}$ & $0.1 \pm 0.05$ & 12.57 \\ 
MR2251-178       & $0.98 \pm 0.51$ & $0.16  \pm 0.19$    & $2401 \pm 1442$ & $<8.39\cdot 10^{-6}$       & $<3196$ &$0.17 \pm 0.73$ & 1 & 1 &$0.0478^{+0.0005}_{-0.0006}$&$0.26^{+0.13}_{-0.16}$&$5.40^{+2.6}_{-3.5}$ & $0.06 \pm 0.04$ & 15.04 \\ 
MRK\,1040          & $0.24 \pm 0.13$ & ...                 & ...             & $0.018^{+2.22}_{-0.017}$   & $<878$  &$0.83 \pm 0.56$ & 1 & 1 &$0.196^{+0.003}_{-0.003}$&$0.10^{+0.05}_{-0.10}$&$0.56^{+0.24}_{-0.44}$  & $0.1 \pm 0.07$ & 13.07 \\ 
MRK\,1210          & $0.18 \pm 0.09$ & $0.023 \pm 0.017$   & $3666 \pm 1291$ & $<8.39\cdot 10^{-5}$ & $<712$  &$0.67 \pm 1.35$ & 1 & 0.46  &$0.50^{+0.07}_{-0.08}$&$1.6^{+1.1}_{-1.2}$&$3.5^{+1.0}_{-2.6}$ & $0.14 \pm 0.11$ & 6.84 \\ 
MRK\,273           & $0.15 \pm 0.08$ & ...                 & ...             & $>$0.0059    & $<1948$ &$-0.76 \pm 1.13$ & 1 & 0.77 &$0.8^{+0.2}_{-0.4}$&$0.2^{+0.4}_{-1.3}$&$0.2^{+0.7}_{-1.3}$ & $0.05 \pm 0.03$ & 2.85 \\ 
MRK\,290           & $0.19 \pm 0.10$ & $0.0040 \pm 0.0045$ & $4973 \pm 2526$ & 4.2$^{\textrm{---}}_{\textrm{---}}$    & $<1539$ &$0.32 \pm 0.34$ & 1 & 1 &$0.127^{+0.004}_{-0.004}$&$0.04^{+0.13}_{-0.21}$&$0.3^{+1.0}_{-1.6}$ & $0.09 \pm 0.07$ & 15.10 \\ 
MRK\,3             & $0.33 \pm 0.17$ & $0.0040 \pm 0.0009$ & $2739 \pm 321$  & $<$0.84 & $<717$ &--- & 1 & 1  &$0.034^{+0.004}_{-0.004}$&$0.02^{+0.02}_{-0.02}$&$0.6^{+0.5}_{-0.5}$ & $0.11 \pm 0.05$ & 16.85 \\ 
MRK\,509           & $0.54 \pm 0.28$ & $0.054 \pm 0.021$   & $3442 \pm 661$  & 0.02$^{\textrm{---}}_{\textrm{---}}$  &$<1780$  &$0.39 \pm 0.23$ & 1 & 0.70  &$0.042^{+0.0004}_{-0.0003}$&$0.02^{+0.03}_{-0.04}$&$0.7^{+1.0}_{-1.7}$& $0.04 \pm 0.03$ & 11.88  \\ 
MRK\,766           & $0.11 \pm 0.06$ & ...                 & ...             & $>$0.016  &$<686$&$0.17 \pm 0.24$ & 1 & 0.96  &$0.175^{+0.002}_{-0.002}$&$0.02^{+0.05}_{-0.08}$&$0.1^{+0.3}_{-0.5}$ & $0.1 \pm 0.05$ & 14.17 \\ 
NGC\,1068          & $0.12 \pm 0.06$ & $0.0050 \pm 0.0014$ & $3217 \pm 491$  &  $<$0.318      & $795\pm176$ &$0.77 \pm 0.24$ & 1 & 1 &$0.014^{+0.003}_{-0.003}$&$0.03^{+0.01}_{-0.02}$&$2.2^{+1.1}_{-1.1}$ & $0.09 \pm 0.04$ & 14.53 \\ 
NGC\,1275          & $0.47 \pm 0.24$ & $0.019 \pm 0.031$   & $3237 \pm 2613$ & 0.003$^{\textrm{---}}_{\textrm{---}}$& --- &$0.07 \pm 0.31$ & 1 & 1  &$0.69^{+0.02}_{-0.02}$&$0.6^{+0.3}_{-0.5}$&$0.8^{+0.4}_{-0.9}$ & $0.1 \pm 0.04$ & 18.22 \\ 
NGC\,1365          & $0.07 \pm 0.03$ & $0.028 \pm 0.017$   & $3745 \pm 1111$ & $1.26^{+ 0.62}_{-1.16}$ &$<289$&$0.34 \pm 0.06$ & 1 & 1 &$1.220^{+0.007}_{-0.007}$&$0.27^{+0.03}_{-0.04}$&$0.22^{+0.03}_{-0.03}$& $0.08 \pm 0.04$ & 9.08 \\ 
NGC\,2992          & $0.05 \pm 0.02$ & $0.310 \pm 0.209$   & $1990 \pm 672$  & $5.87^{+1.3}_{-3.66}$    & $<412$  &$0.24 \pm 0.06$ & 1 & 1  &$3.59^{+0.02}_{-0.02}$&$0.11^{+0.03}_{-0.04}$&$0.03^{+0.01}_{-0.01}$& $0.07 \pm 0.05$ & 9.99 \\ 
NGC\,3393          & $0.10 \pm 0.05$ & $0.009 \pm 0.005$   & $4572 \pm 1343$ & ---       & $<664$  &--- & 1 & 1 &$0.10^{+0.04}_{-0.04}$&$0.020^{+0.11}_{-0.19}$&$0.3^{+1.8}_{-1.6}$ & $0.1 \pm 0.07$ & 8.77 \\ 
NGC\,3516          & $0.11 \pm 0.06$ & $0.009 \pm 0.003$   & $3861 \pm 598$  & ---    & $<471$  &$0.11 \pm 0.26$ & 1 & 0.31  &$0.29^{+0.01}_{-0.01}$&$0.3^{+0.3}_{-0.5}$&$1.2^{+1.0}_{-1.7}$ & $0.09 \pm 0.08$ & 6.04 \\ 
NGC\,3783          & $0.24 \pm 0.13$ & $0.029 \pm 0.006$   & $2146 \pm 238$  & 0.078$^{+0.5}_{-0.067}$ &$<518$&$-0.18 \pm 0.28$ & 1 & 1  &$0.079^{+0.001}_{-0.001}$&$0.04^{+0.01}_{-0.01}$&$0.42^{+0.15}_{-0.16}$ & $0.11 \pm 0.05$ & 16.93 \\ 
NGC\,4051          & $0.02 \pm 0.01$ & $0.0010 \pm 0.0003$ & $2984 \pm 573$  & ---       & $<124$  &$0.10 \pm 0.07$ & 1 & 1 &$0.14^{+0.01}_{-0.01}$&$0.2^{+0.4}_{-0.8}$&$1.4^{+2.8}_{-5.7}$ & $0.11 \pm 0.04$ & 16.07 \\ 
NGC\,4151          & $0.07 \pm 0.03$ & $0.034 \pm 0.008$   & $2487 \pm 279$  & 1.26$^{+0.71}_{-0.79}$ &$<178$&$0.15 \pm 0.09$ & 1 & 1 &$0.375^{+0.001}_{-0.001}$&$0.090^{+0.006}_{-0.005}$&$0.24^{+0.01}_{-0.02}$ & $0.06 \pm 0.03$ & 15.81 \\ 
NGC\,4388          & $0.16 \pm 0.08$ & $0.0050 \pm 0.0018$ & $3070 \pm 528$  & $0.725^{+ 1.12}_{-0.68}$  &$<449$&--- & 1 & 1 &$0.395^{+0.007}_{-0.006}$&$0.08^{+0.02}_{-0.03}$&$0.20^{+0.06}_{-0.08}$ & $0.12 \pm 0.08$ & 8.95 \\ 
NGC\,5548          & $0.19 \pm 0.10$ & $0.028 \pm 0.011$   & $3257 \pm 682$  & $>0.57$  & $<907$  &$0.07 \pm 0.11$ & 1 & 0.94  &$0.0624^{+0.0005}_{-0.0005}$&$0.004^{+0.006}_{-0.008}$&$0.07^{+0.09}_{-0.11}$ & $0.08 \pm 0.05$ & 15.96 \\ 
NGC\,6300          & $0.03 \pm 0.02$ & ...                 & ...             & $<0.0008$  & $<161$  &--- & 1 & 0.54  &$1.73^{+0.05}_{-0.04}$&$1.8^{+0.2}_{-0.1}$&$1.02^{+0.11}_{-0.03}$& $0.10\pm0.08$  & 8.29 \\ 
NGC\,7469          & $0.20 \pm 0.10$ & $0.0040 \pm 0.0034$ & $3431 \pm 1298$ & $>0.0008$       & $<862$  &$-0.13 \pm 0.32$ & 1 & 0.72 &$0.0094^{+0.0004}_{-0.0003}$&$0.000^{+0.007}_{-0.012}$&$0.1^{+0.7}_{-1.2}$ & $0.11 \pm 0.06$ & 15.01 \\ 
NGC\,7582          & $0.27 \pm 0.14$ & ...                 & ...             & $2.46^{+3.63}_{-1.2}$   & $<281$  &--- & 1 & 1 &$1.26^{+0.04}_{-0.05}$&$0.09^{+0.04}_{-0.07}$&$0.07^{+0.03}_{-0.06}$ & $0.08 \pm 0.05$ & 17.6 \\ 
PKS2153-69       & $0.16 \pm 0.09$ & ...                 & ...             & ---       & $<1470$ &--- & 1 & 0.43 &$0.12^{+0.03}_{-0.03}$&$-0.2^{+0.6}_{-1.9}$&$-1.7^{+5.0}_{-12.5}$ & $0.10\pm 0.06$  & 13 \\ 
PictorA          & $0.26 \pm 0.14$ & ...                 & ...             & ---       & $<1813$ &$0.63 \pm 0.40$ & 1 & 0.61 &$0.338^{0.015}_{-0.014}$&$-0.07^{+0.40}_{-0.94}$&$-0.4^{+1.2}_{-2.9}$ & $0.11 \pm 0.05$ & 14.99   \\ 

\hline
\end{tabular} }
\tablefoot{Col.(1): Object name; 
Col.(2): Dust sublimation radius computed with equation \ref{eq:rsub} for the sources with a $R_{\rm Fe\: K\alpha}$ measurement;
Col.(3): \kalfa{} radius estimated from $\upsilon_{\rm FWHM}$; Col.(4): deconvolved Full Width at Half Maximum velocity; 
Col.(5) Size of the \kalfa{} reprocessor computed in $\S$\ref{sec:rep_sims} with 1-$\sigma$ errors or 2-$\sigma$ limits; 
Col.(6) spatial extent of \kalfa{} emitting region obtained from imaging analysis, where values denote largest bin with a 3-$\sigma$ detection and with errors correspond to 25$\%$ of bin width, while upper limits are calculated for 2\farcs5 (4\farc{} for Cen\,A and 2MASXJ23444), below which the effects of pileup severely limit any estimate for this sample (see $\S$\ref{sec:image}); 
Col.(7) $\rm F_{\rm Fe\:K\alpha}$--$\rm F_{2-10\:keV}$ slope and 1-$\sigma$ errors; 
Col.(8) variability probability $P_{\rm var}$ for continuum light curve; 
Col.(9) variability probability $P_{\rm var}$ for \kalfa{} line light curve; 
Cols. (10-12):
$50\%$ percentile of the distributions of normalized excess variance, $\sigma^2_{rms}$, calculated through the Monte Carlo method described in $\S$\ref{sec:NXS}, for the continuum and Fe K$\alpha$ line light curves, as well as the ratio between them, for all AGN in the sample. Error bars  represent the ${\sim}1\sigma$ uncertainties due to different realizations of the observational Poisson noise in the light curves. Negative $\sigma^2_{rms}$ values can occur when the Poisson noise uncertainty $\sigma^2_{err}$ is larger than the intrinsic variance in Eq.~\ref{eq:NXS}. Col.(13): normalized excess variance of the simulated continuum light curve and ${\sim}1\sigma$ uncertainties (see $\S$\ref{sec:rep_sims}); Col.(14): maximum light curve timespan in years. Asymmetric upper and lower errors are calculated separately following $\S$1.73 of \citet{Lyons1991}.}
\label{t:summary}
\end{table*}

\begin{figure*} 
\centering
$\begin{subfigure}[b]{0.45\textwidth}
 \includegraphics[width=\textwidth]{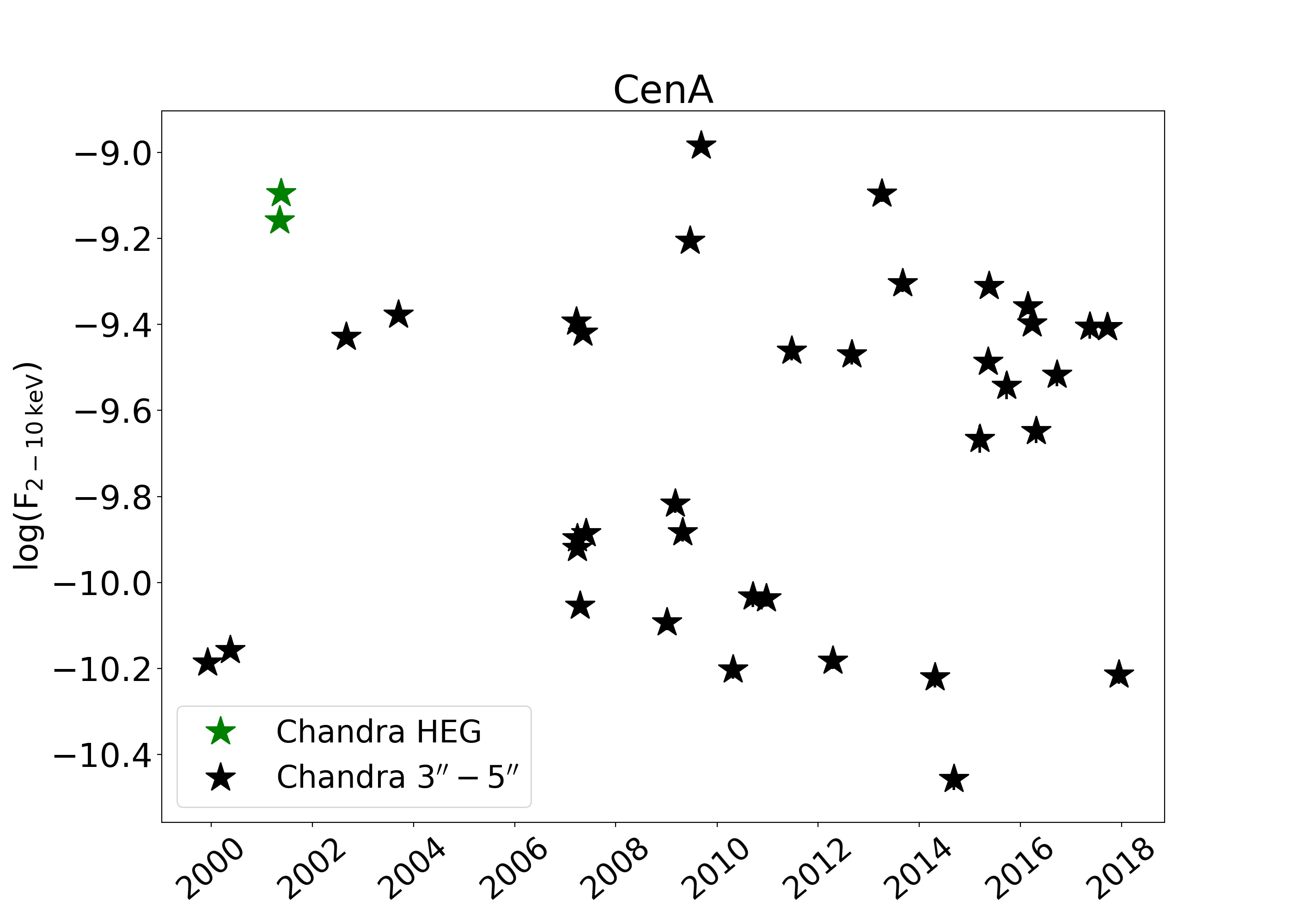}
 \end{subfigure}$
$\begin{subfigure}[b]{0.45\textwidth} 
 \includegraphics[width=\textwidth]{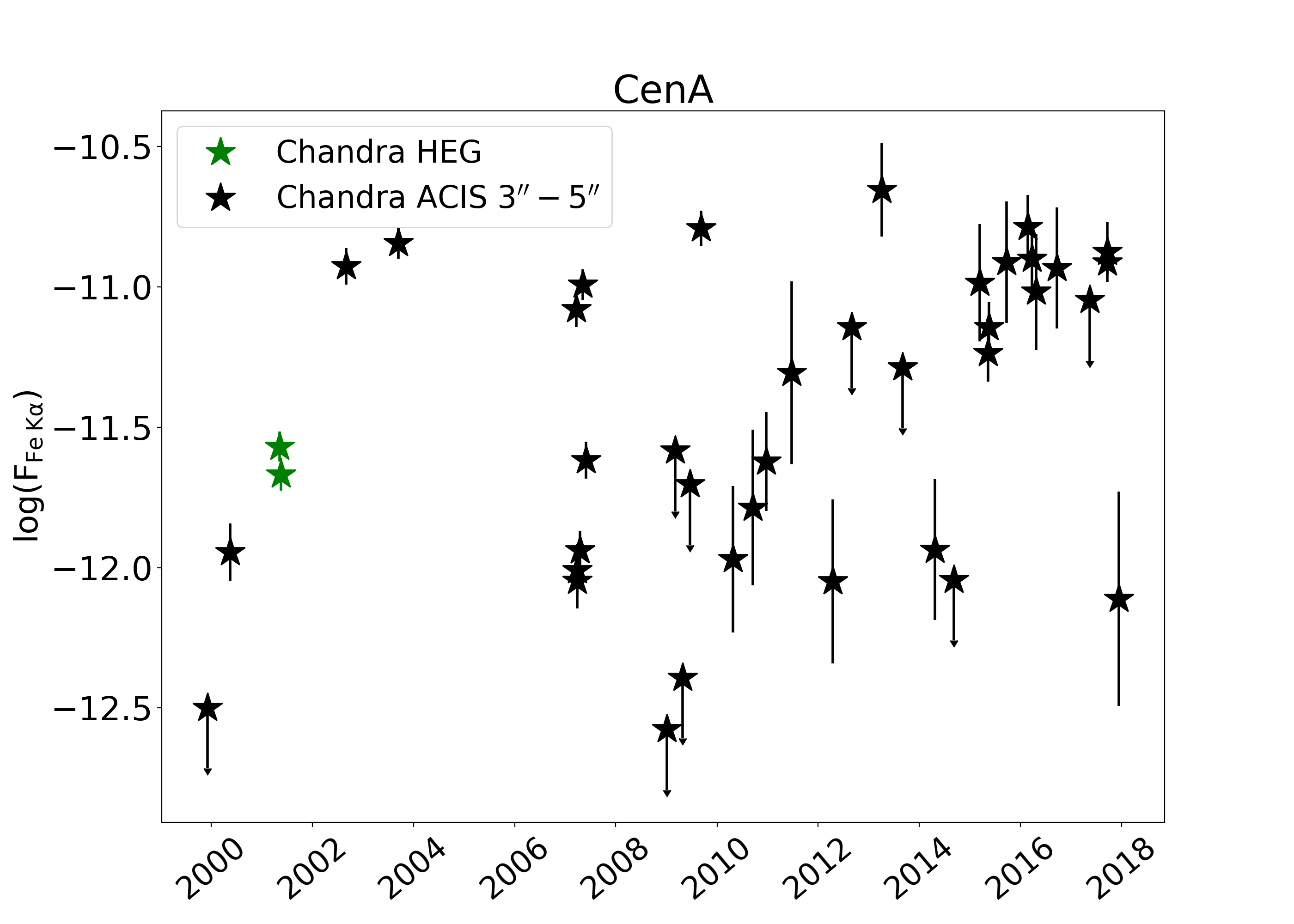}  
 \end{subfigure}$  
 
$\begin{subfigure}[b]{0.45\textwidth}
 \includegraphics[width=\textwidth]{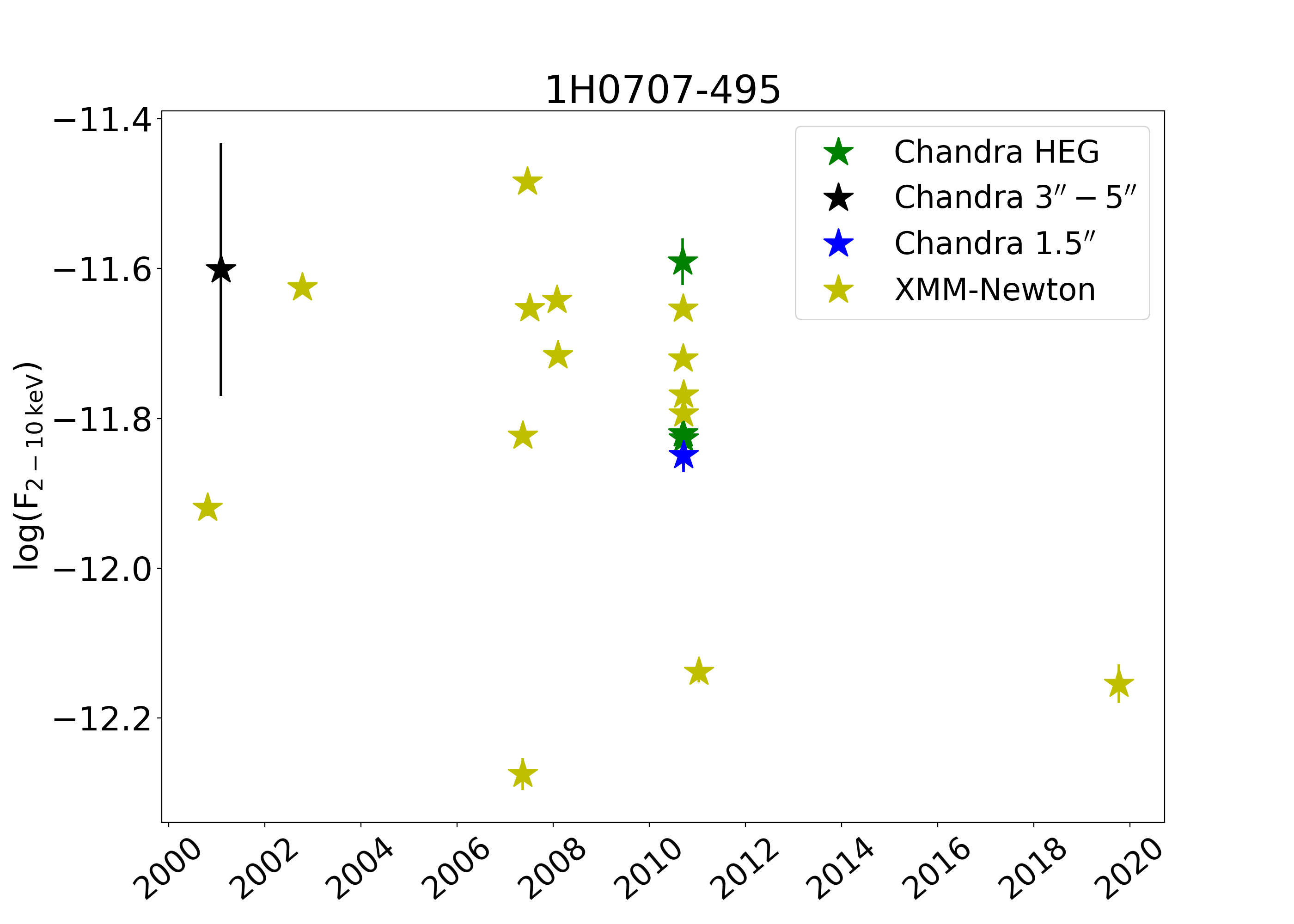}
 \end{subfigure}$
$\begin{subfigure}[b]{0.45\textwidth} 
 \includegraphics[width=\textwidth]{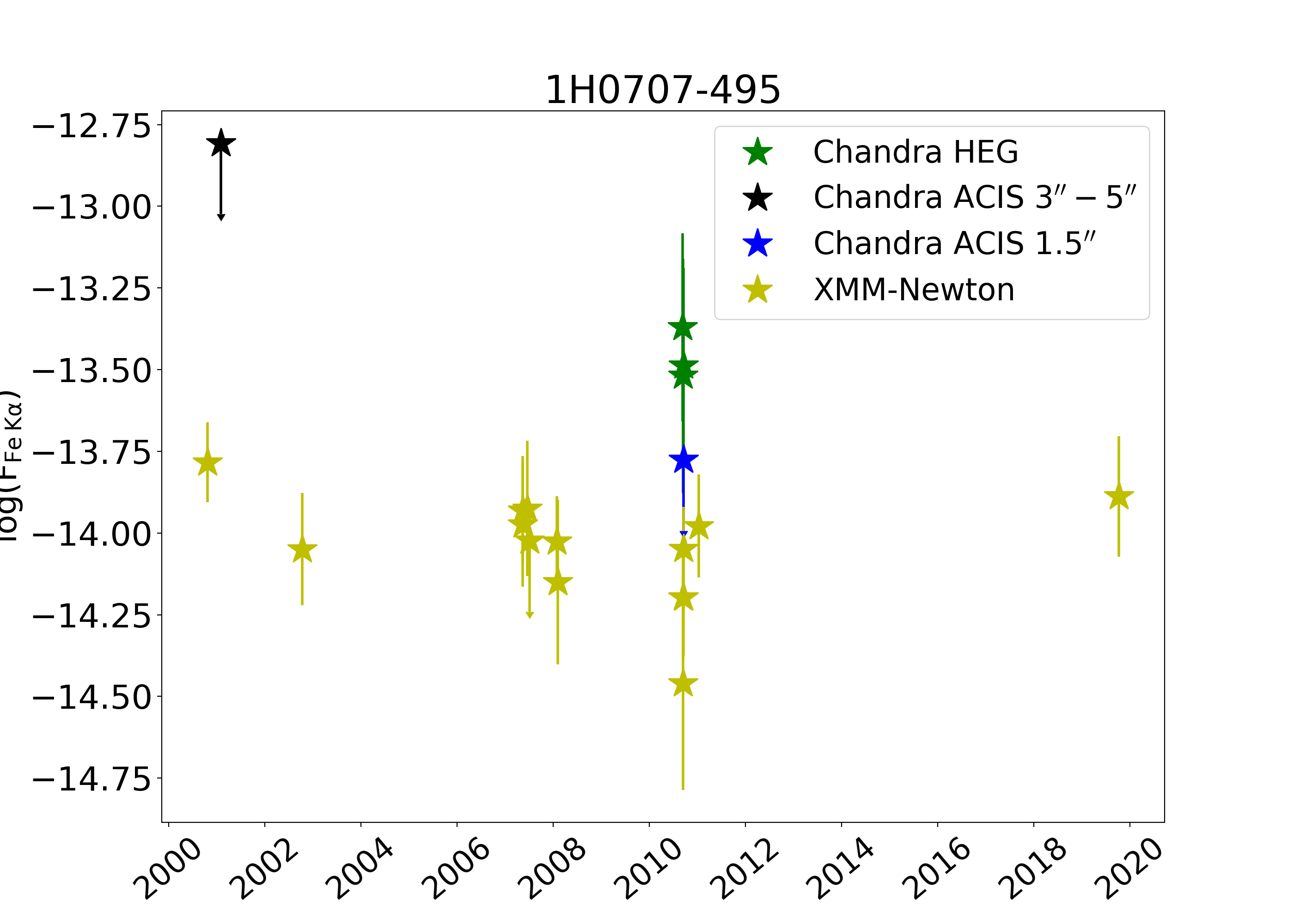}  
 \end{subfigure}$  

$\begin{subfigure}[b]{0.45\textwidth}
 \includegraphics[width=\textwidth]{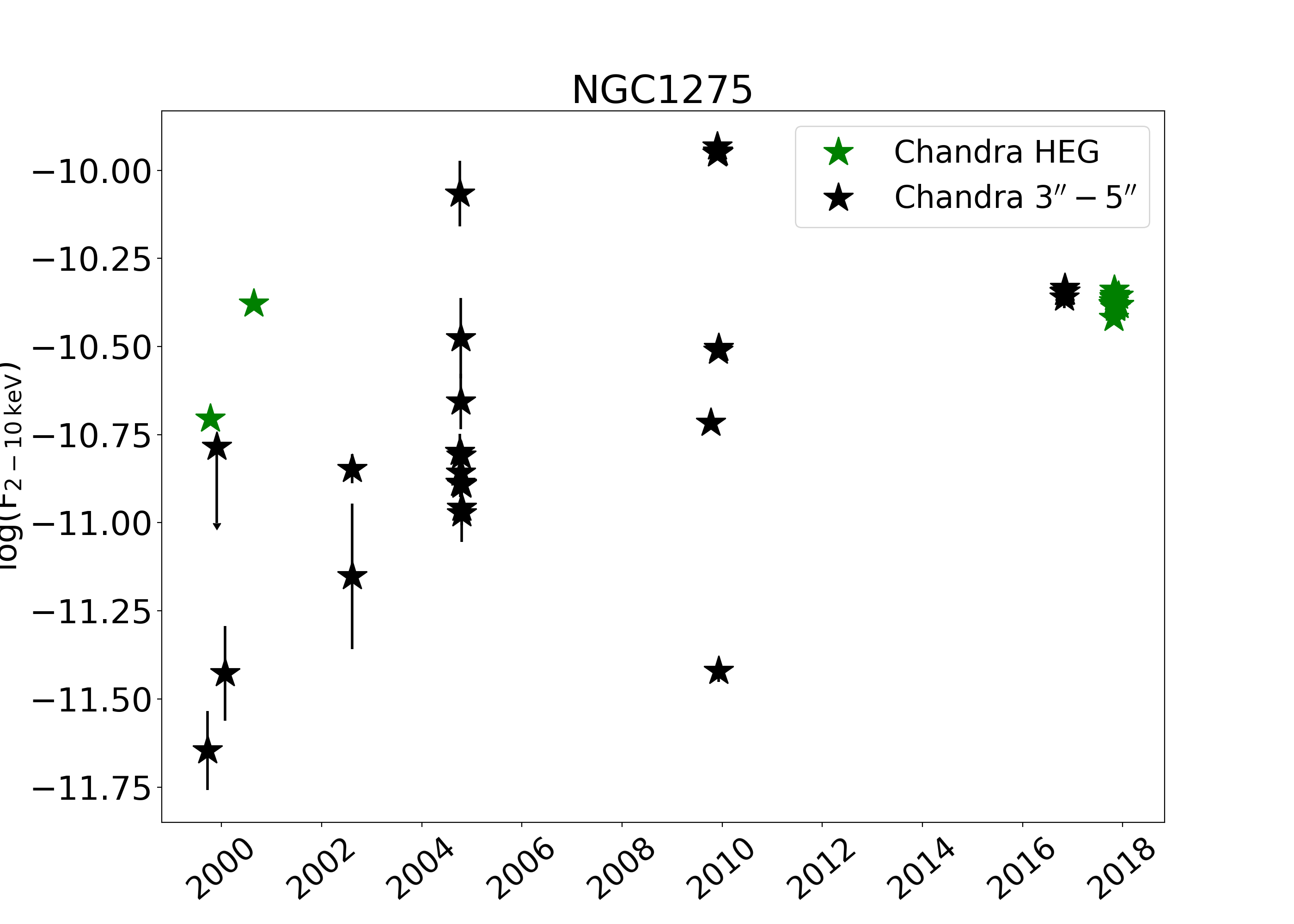}
 \end{subfigure}$
$\begin{subfigure}[b]{0.45\textwidth} 
 \includegraphics[width=\textwidth]{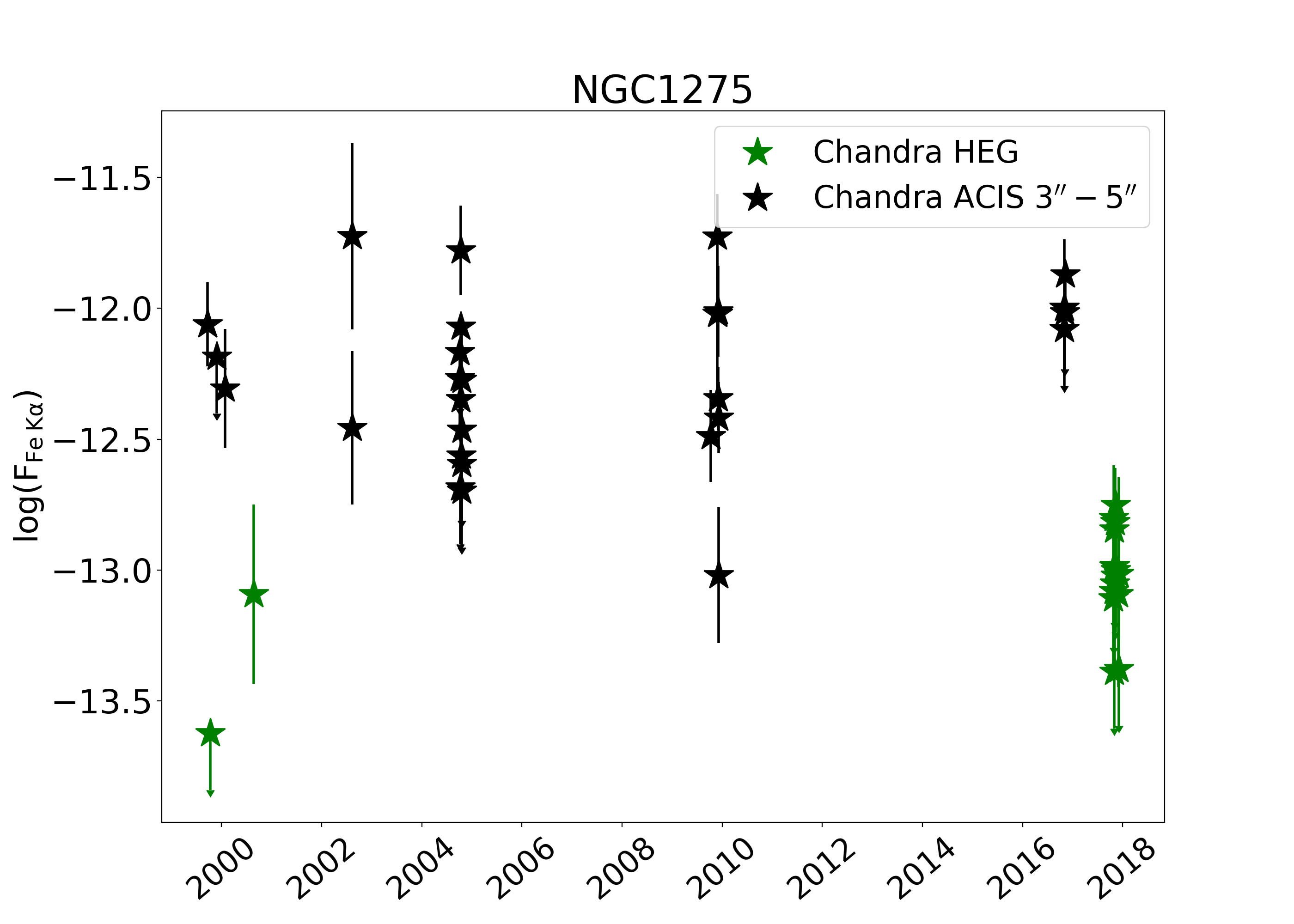}  
 \end{subfigure}$  

\caption{Light curves for the 2--10 keV continuum (left) and Fe K$\alpha$ line  (right) fluxes for the galaxy Centaurus A (upper panels), 1H0707-495 (middle panels) and NGC\,1275 (lower panels). The green and black stars denote fluxes measured from the \textit{Chandra} HEG first-order spectra and \textit{Chandra} ACIS 3\farc--5\farc{} annular spectra, respectively, while yellow stars are fluxes measured from \textit{XMM-Newton} pn observations. The flux units are $\rm erg\:cm^{-2}\: s^{-1}$ and the error plotted are 1$\sigma$ confidence.} \label{fig:lc_3ex}
 \end{figure*}

The {\sc Linmix} package requires at least five measurements, which all sources in our sample satisfy by definition. The best-fit linear regression results in 29 AGN with well-constrained slopes, and nine AGN with unconstrained slopes. Appendix Fig.~\ref{fig:fluxes_vs_q} compares the Fe K$\alpha$ line and continuum fluxes, with the best-fit slope shown in the label. In each plot, the thin red lines correspond to the samples of the posterior distribution and the red thick line is the average posterior. Table~\ref{t:summary} reports the values of the well-constrained slopes ($m$), the maximum light curve timespan ($\rm t_{lc}$), and the $P_{var}$ parameters obtained for the Fe K$\alpha$ line and 2--10 keV continuum light curves.

We find three sources with a $\rm F_{\rm Fe\:K\alpha}$--$\rm F_{2-10\:keV}$ slope consistent with one (i.e., $m{+}1\sigma{>}$1.0 and $m{-}2\sigma{>}$0.5). The most outstanding case is Cen\,A, with a slope $m=1.14 \pm 0.20$. The top plots of Fig.~\ref{fig:lc_3ex} show the light curves of Cen\,A for the continuum and Fe K$\alpha$ line. The \kalfa{} line light curve is strongly variable, with $P_{\rm var, Fe}{=}1$ and $\sigma_{\rm Fe}{=}0.53^{+0.11}_{-0.15}$, and appears to perfectly track the continuum, with variations of up to 1\,dex in five days. This suggests that Cen\,A has a compact reflector, very close to the source of the X-ray continuum. Similarly strong constraints are seen for H1821+643 and IC\,4329A.

On the other hand, six objects have slopes consistent with zero (i.e., $m{-}1\sigma{\approx}$0 and $m{+}2\sigma{<}$0.5); all have $\sigma_{\rm Fe}$ consistent with $\approx$0.0 as well, although only one shows no clear \kalfa{} line variability ($P_{\rm var,Fe}{<}0.5$). This latter source is 1H0707-495, with a slope of $m{=}{-}0.28{\pm}0.27$; its light curves are shown in the middle panels of Fig.~\ref{fig:lc_3ex}. While the continuum light curve is clearly variable ($P_{\rm var,c}{=}1$, $\sigma_{\rm c}{=}0.195{\pm}0.005$), the \kalfa{} line light curve has a $P_{\rm var}=0.37$ and $\sigma_{\rm Fe}{=}{-}$0.18--0.05, indicating no discernible line flux variability over a $\sim$19 year timescale. Such behavior suggests that the light crossing size between the X-ray continuum source and the reprocessor is substantially larger than the typical continuum variability timescale; based on the simulations carried out in the previous section, the reprocessor is ${\ga}$3 light days away, which is still consistent with BLR clouds. However, in a recent investigation, \citet{2021Boller} found that the X-ray spectrum of 1H0707-495 is dominated by relativistic reflection, and that the absorber is probably ionized, leading to weak \kalfa{} variability. The other five sources are NGC\,3783, NGC\,4051, NGC\,4151, NGC\,5548, and NGC\,7469.

A majority of the remaining AGN show some degree of variability of the \kalfa{} emission ($P_{\rm var,Fe}{>}0.9$), but span a wide range in terms of slope uncertainties; some have well-determined intermediate slopes, while others have completely unconstrained slopes. The lower panels of Fig.~\ref{fig:lc_3ex} show the light curve for NGC\,1275, which has a low but uncertain slope of $m{=}0.07{\pm}0.31$, although its \kalfa{} line light curve appears variable, with $P_{\rm var}{=}1$ and $\sigma_{\rm Fe}{=}$0.029--0.83. 

The analyses made in $\S$\ref{sec:NXS},~\ref{sec:rep_sims}, and~\ref{sub:slope} indicate that different AGN have distinct morphological and structural distributions of reflecting clouds. Some sources have a {\it compact} reprocessor, whereby the \kalfa{} flux tracks the continuum variations on timescales comparable to each observation exposure. For other sources the line varies but the correlation between the continuum and the line is weak due to the damping effects of a large or distant reprocessor. And finally, some sources show little variability in the line and do not have correlated fluxes, suggesting that the reflector is sufficiently large or distant, compared to the variability timescales of the continuum source, to wash out any reaction from the line.

\subsection{Potential influence of relativistically broadened \kalfa{}}

As noted in Table~\ref{t:properties}, nearly half of the objects in our sample have been argued to have relativistically broadened \kalfa{} emission in the literature. Thus, an important consideration for interpreting the above results is to understand the influence that any potential relativistically broadened \kalfa{} emission will have on the variability measurements of either the continuum or the narrow \kalfa{} line. We begin with a general caveat regarding the veracity of relativistically broadened \kalfa{} detections in the literature, which remain generally controversial for the majority of objects for which a detection has been claimed. This is in part due to a combination of limited photon statistics (e.g., \citealp{2009Brenneman} argue that spectra with $\sim$$10^6$ counts are generally required to confirm relativistically blurred components in $<$10\,keV spectra) and lack of good quality, simultaneous spectral constraints spanning both $\sim$0.5--8\,keV and $\sim$8--100\,keV, which are essential to lock down intrinisic spectral slopes and reflection fractions (these are often degenerate even in high-count $\sim$0.5--8\,keV spectra due to potential combinations of neutral and ionized absorption). As a consequence, the constraints on the contributions of such relativistically blurred Fe K$\alpha$ components remain poorly constrained \citep[e.g., reflection fractions ranging from $\lesssim$0.01 to $\sim$20][]{2009Brenneman}.

Importantly, the continuum as we model it will automatically absorb the bulk of any relativistically broad/blurred line flux and variability. Since both components are strongly correlated, with reported lags on the order of minutes to hours \citep{deMarco2009, Kara2016}, our observation by observation analysis should not be strongly affected, given the typical length of the observations. Another consideration is whether the relativistically broadened \kalfa{} profile peaks near 6.4\,keV, and thus if a portion of that component could be assigned to the narrow \kalfa{} lines that we fit, and lead to an enhancement in correlations. \citet{Hu2019} performed a stacking analysis of 193 RQ and 97 RL AGN, arguing that average broad line components are detected in both subsamples,
%, although the fit significance compared to 3 narrow lines of 6.4 keV (Fe I K$\alpha$), 6.7 keV (Fe XXV), 6.96 keV (Fe XXVI), and 7.06 keV (Fe I K$\beta$) keV is not directly demonstrated. 
although the average broad-line \kalfa{} components are subdominant compared to the narrow component by factors of $\sim$4 for RQ and $\sim$2 for RL AGN in the 6.2--6.6\,keV regime, respectively. Similarly, \citet{Falocco2014} fit the stacked spectra from 263 X-ray unabsorbed AGN, finding that narrow ("unresolved") \kalfa{} emission accounts for $\gtrsim$70\% of the total 6.2--6.6\,keV line equivalent width. Given that our simple continuum model already incorporates for some of the broad \kalfa{} in the 6.2--6.6\,keV regime, we assume that the narrow line as we measure it suffers from  $<$10--20\% contamination at most. Further considering the arguments presented in Appendix~\ref{ap:mult_ref}, the \kalfa{} line variances that we report should be related to the square of the aforementioned fractional contributions, and hence strongly dominated by the narrow component in all cases.

\subsection{Variability properties compared with AGN and host galaxy properties}\label{sec:correlations}

To gain further insight into the different behaviors of the Fe K$\alpha$ line in our sample, we compare the $\sigma_{\rm Fe}/\sigma_c$ values and $\rm F_{\rm Fe\:K\alpha}$--$\rm F_{2-10\:keV}$ slopes (for 29 sources in the sample with firm $\sigma_{\rm Fe}/\sigma_c$ and slopes measurements) with several AGN properties (see Figs.~\ref{fig:ratio_rel} and \ref{fig:slope_rel} of Appendix\,\ref{ap:var_prop}). The line-of-sight column density and radio-loudness are taken from \citet{2017ApJS..233...17R}, who derived the X-ray properties of the sources through X-ray spectral fitting. The SMBH mass, Eddington ratio, and Seyfert type are taken from \citet{2017ApJ...850...74K} and BASS DR2 (\citeauthor{2022Koss} (submitted)). The SMBH masses were estimated using broad-line measurements assuming virial motion, or stellar velocity dispersion, assuming the $M_{\rm BH}-\sigma_{*}$ relation. To test possible correlations, we compute the Spearman correlation coefficient $\rho$ with its p-value and also perform a linear regression between the variability features and the properties mentioned above. The Spearman coefficient varies between $-1$ and $1$, with $\rho=0$ and $\rho=\pm 1$ implying no correlation and a perfect correlation, respectively, and the p-value represents the probability of that the values are uncorrelated.

\begin{figure}
%\centering
\resizebox{\hsize}{!}{\includegraphics{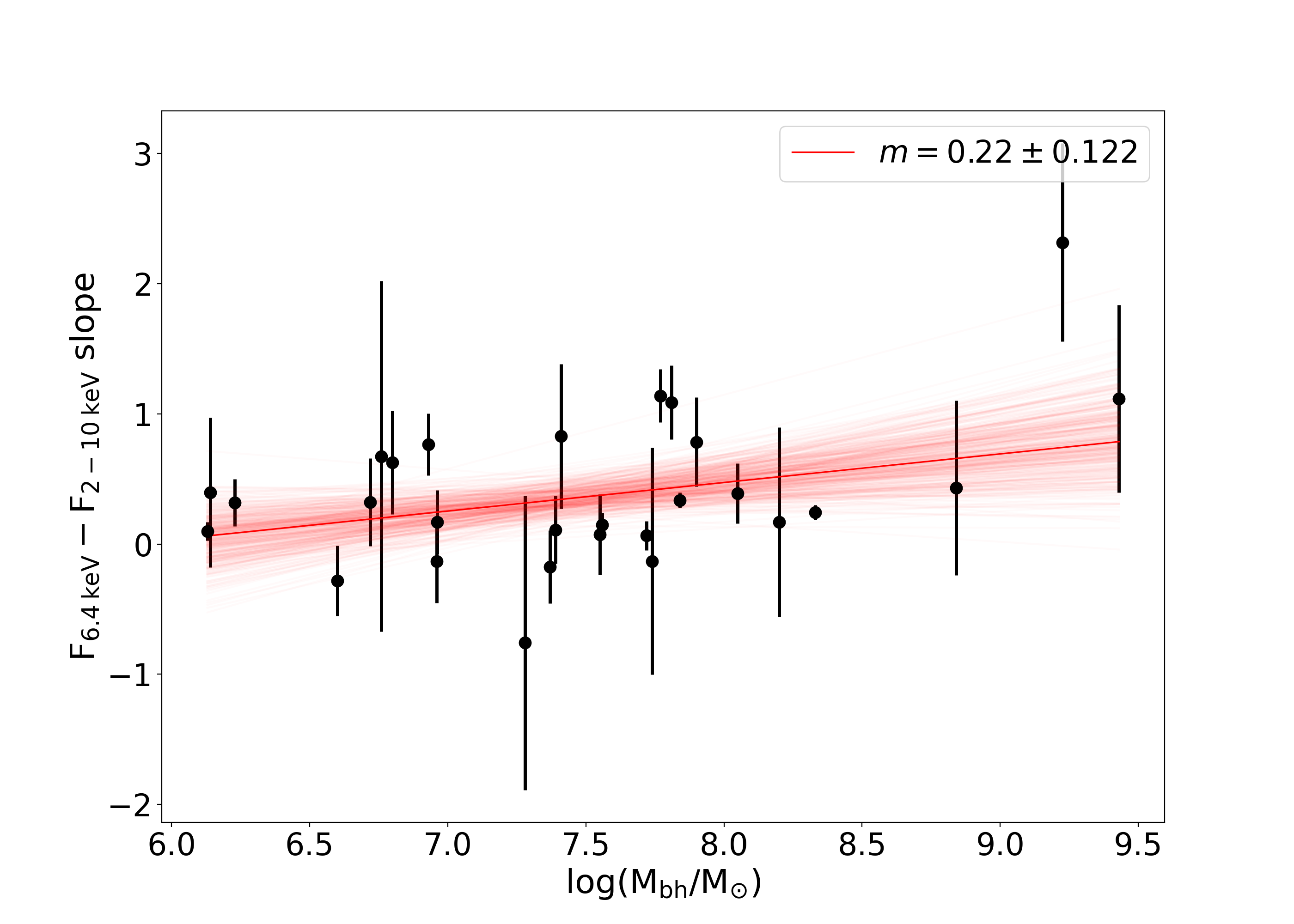}} % [trim=0 0 0 80,clip,width=0.5\textwidth]
\caption{Comparison between the $\rm F_{\rm Fe\:K\alpha}$--$\rm F_{2-10\:keV}$ slopes and the SMBH mass ($M_{\rm BH}$). }\label{fig:slope-MBH}	
% Agrandar ejes y labels
\end{figure}

\begin{table}
\caption{Correlation coefficients between the $\rm F_{\rm Fe\:K\alpha}$--$\rm F_{2-10\:keV}$ and $\sigma_{\rm Fe}/\sigma_c$ ratios and some AGN and host galaxy properties.  } 
%\begin{adjustbox}{width=\textwidth}
\scriptsize
% \centering
\scalebox{0.95}{
\begin{tabular}{c c c c c} 
 \hline
 \hline
\noalign{\smallskip}
 Variability feature& Property & linear regression slope & $\rho$ & p-value\\
\noalign{\smallskip}
\hline
\noalign{\smallskip}
$\rm F_{\rm Fe\:K\alpha}$--$\rm F_{2-10\:keV}$ slope & $ M_{\rm BH}$ & $0.22\pm 0.12$ & 0.36& 0.05 \\
 & $L_{\rm bol}/L_{\rm Edd}$ &$-0.06\pm 0.07$ & -0.1 & 0.63\\
 & $N_{\rm H}$ &$0.07\pm 0.06$ & 0.2 & 0.3 \\
 &radio-loudness& $0.13\pm 0.09$ & 0.24 & 0.21 \\

\noalign{\smallskip}

$\sigma_{\rm Fe}/\sigma_c$ & $ M_{\rm BH}$ & $-0.25 \pm 0.17$ & -0.03& 0.89 \\
 & $L_{\rm bol}/L_{\rm Edd}$ &  $0.04\pm 0.1$ & 0.13 & 0.51\\
 & $N_{\rm H}$ & $0.05\pm 0.08$ & 0.2 & 0.33 \\
 &radio-loudness& $0.12\pm 0.07$ & 0.008& 0.966 \\

\noalign{\smallskip}

\hline
 \end{tabular} }

\label{t:var-prop}
\end{table} 

Fig.~\ref{fig:slope-MBH} compares the slope and the central SMBH mass, which is the only set of properties where we find a weak correlation ($\rho =0.36$, p-value$=$0.05, and linear regression slope $m{=}0.22{\pm}0.13$). If the continuum variations are produced by the corona, we expect somewhat longer bend timescales for larger SMBHs following Eq.\ref{eq:Tbreak}, presumably with some scatter due to the effect of spin on the innermost stable circular orbit. To further test whether the weak correlation is real, we perform a bootstrap analysis to estimate the confidence interval of the Spearman coefficient, finding that $\rho$ ranges between 0.01 and 0.65 at a $95\%$ confidence level. The wide confidence interval suggests that our results are compatible with both a positive and nonexistent correlation between the slope and SMBH mass, which does not allow us to draw a definitive conclusion. In contrast, Figure~\ref{subfig:ratio-mbh} in the Appendix compares $\sigma_{\rm Fe}/\sigma_c$ with central SMBH mass, where the linear regression and the Spearman test suggest no clear correlation ($\rho=0.21$, p-value$=0.29$, and slope $m{=}{-}0.12{\pm}0.2$), implying that the observed damping does not depend on the SMBH mass. The disagreement between these two indicators, combined with the low strength of the relationship with the slope, suggests that the weak correlation in the former may not be real.

Similarly, we search for correlations between both $\sigma_{\rm Fe}/\sigma_c$ (Figs.~\ref{subfig:ratio-er}, \ref{subfig:ratio-nh} \ref{subfig:ratio-RL}, and \ref{subfig:ratio-type}) and $\rm F_{\rm Fe\:K\alpha}$--$\rm F_{2-10\:keV}$ slopes 
(Figs.~\ref{subfig:rel-er}, \ref{subfig:rel-nh}, \ref{subfig:rel-RL} and \ref{subfig:rel-type})
versus the Eddington ratio, 
the line-of-sight column density, 
radio loudness 
and AGN type, respectively. Table\,\ref{t:var-prop} reports the linear regression slope, $\rho$, and p-value for each property. We might expect potential relations with any one of these parameters. For instance, the $N_{\rm H}$ and AGN type should trace the overall reflector geometry, which should imprint itself on the \kalfa{} line properties. Likewise, the Eddington ratio has been linked to variations in $N_{\rm H}$, potentially sculpting the inner few 10s of pc via radiative feedback \citep[e.g.,][]{2017Natur.549..488R}. Finally, whether an AGN is radio-loud or not may be associated with certain physical conditions related to black hole spin and magnetic threading of the accretion disk, which could extend to the broader local environment. Among all of these possibilities, however, we find no clear trends.

Summarizing, we find a weak correlation between the $\rm F_{\rm Fe\:K\alpha}$--$\rm F_{2-10\:keV}$ slopes with the SMBH mass, and no correlations with the Eddington ratio, column density, radio-loudness, or AGN type. Therefore, the only property that might affect the reaction of the Fe K$\alpha$ line to the continuum variations appears to be the SMBH mass.

\section{Imaging analysis with \textit{Chandra}} \label{sec:image}

To complement the spectral results above, we analyze the \textit{Chandra} images to look for possible spatially extended \kalfa{} emission in our sample. Extended \kalfa{} emission has been observed in a handful of nearby, typically Compton-thick, AGN, and we want to examine whether sources where the \kalfa{} flux reacts quickly to continuum variations are point-like, and sources with no statistical line variability may be more spatially extended. 

To quantify this, we investigate the radial profiles of the AGN in our sample, using both the azimuthally averaged profile in comparison to the nominal {\rm Chandra} PSF and by comparing averaged quadrant profiles to look for strong asymmetries. We analyze a rest-frame continuum-subtracted \kalfa{}-only image, as well as the high-energy continuum in the rest-frame 5--6\,keV band. The \kalfa{}-only image is created by taking the rest-frame 6.2--6.5\,keV band, which should capture the vast majority of 6.405\,keV photons,\footnote{Based on the calibration information in the {\rm Chandra} Proposer's Observatory Guide, the spectral resolution at $\approx$6.0\,keV after CTI-correction should be $\lesssim$\,eV for ACIS-S3, assuming chipy$<$512, and $\lesssim$270\,eV for ACIS-I, assuming chipy$<$950.} and subtracting the average between the rest-frame 5.9--6.2 and 6.5--6.8\,keV bands, which should remove the underlying 6.2--6.5\,keV continuum given that the ACIS effective area smoothly and linearly changes between these two bands. This scheme is not perfect, as the faint wings of the \kalfa{} spectral profile will extend into the 5.9--6.2 and 6.5--6.8\,keV bands, but should nonetheless yield the approximate spatial distribution of  \kalfa{} photons, is relatively straightforward to implement and most importantly does not suffer from the PSF calibration uncertainties at large annuli (see below). The rest-frame 5--6\,keV band is adopted as a proxy for the broader 2--10\,keV, because it will not be as strongly affected by absorption as lower energies, it is free of strong emission lines, and {\rm Chandra}'s sensitivity is still relatively high here.

The \textit{Chandra} PSF is predominantly a function of the energy and off-axis angle, being compact and roughly symmetric on-axis but broadening substantially both radially and azimuthally beyond off-axis angles $\ga$2\farcmin{}, with complex structure due to mirror (mis)alignment and aberrations, shadowing by mirror support struts, and increased scattering as a function of energy.\footnote{ \url{https://cxc.harvard.edu/ciao/PSFs/psf\_central.html}} The observed PSF shape further depends on the number of counts, with pileup potentially impacting the innermost pixels and at least a few counts/pixel needed to fully sample the PSF wings and complex structure. This combination makes it extremely difficult to calibrate the full shape of \textit{Chandra}'s PSF using in-flight observations of single bright targets.\footnote{\url{https://cxc.harvard.edu/ciao/PSFs/chart2/caveats.html}} As such, the PSF is typically simulated with numerical ray-trace calculations based upon a fiducial mirror model developed using preflight measurements. These models, however strongly underestimate the flux in the PSF wings at energies higher than 2 keV by factors of $\sim$2--3 beyond $>$3\farc{} (as we show in $\S$\ref{subsec:emp-PSF}). Given such limitations, we describe our approach to analyze the \textit{Chandra} images below.

\subsection{SAOTRACE and ${\mathtt{MARX}}$ simulated PSFs} \label{sec:psf}

The {\it Chandra} PSF can be simulated using either SAOTrace\footnote{\url{https://cxc.cfa.harvard.edu/cal/Hrma/SAOTrace.html}} or ${\mathtt{MARX}}$ \citep{2012SPIE.8443E..1AD}, both of which model the on-orbit performance of the various operationals modes of \textit{Chandra} for input source shapes and spectra. SAOTrace uses a more detailed physical model of the mirror geometry, which can be important for off-axis sources or to study the wings of the PSF out to several arcseconds, but is computationally very expensive for bright sources. ${\mathtt{MARX}}$ has as its default a slightly simplified description of the mirror, which is much faster to run and produces very similar overall results (differing by only a few percent in radial profiles). For these reasons, we use ${\mathtt{MARX}}$ (v5.4.0) for the simulations below. ${\mathtt{MARX}}$ takes as input the source spectrum, as well as parameters such as the sky position of the source AGN, exposure time of the observation, and grating type, to generate a simulated event list for the source. 

To understand ability of ${\mathtt{MARX}}$ (and by extension SAOTrace) to reproduce faithfully the wings of the PSF, we compare it to a bright, point-like X-ray source which has been extensively observed by {\it Chandra}. For this purpose, we analyze the archive ACIS-S HETG data for the X-ray binary Hercules X-1 (HERX1, hereafter), following the procedures outlined in $\S$\ref{subsec:chandra}. Importantly, HERX1 lies well above the Galactic Plane \citep[$N_{\rm H, Gal}{=}1.5\times10^{20}$\,cm$^{-2}$;][]{2016A&A...594A.116H}, and thus should not show strong contamination at large angles due to dust scattering \citep[e.g.,][]{1998ApJ...503..831S}, which could complicate interpretation of extended features at soft energies. Specifically, we merge the images and grating spectra for ObsIDs 2749, 3821, 3822, 4585, 6149, and 6150, resulting in total exposure time of 166.6\,ks. We model the first-order HEG and MEG spectra from each ObsID with a simple unabsorbed powerlaw model [$\Gamma=-0.125$, $F_{\rm 2-10\,keV}{=}(2.5)\: \times \: 10^{-10}$\,erg\,cm$^{-2}$\,s$^{-1}$], which we provide as input to MARX. HERX1 is relatively bright, with a comparable number of counts to the brightest AGN in our sample, and therefore has a well-defined radial profile out to $\sim$200\farc{} in the combined image. This high count rate, however, means the central few pixels are affected by pileup. For this reason, we restrict comparisons to outer radii, renormalizing the PSFs at 2\farcs5; at this radius, the radial profile is always found to be declining, with counts per frame is $<$0.01, well below the regime where pileup begins to occur, while the radial profile of grade 0 and 6 events remains flat and close to expected values.\footnote{We investigated using fainter sources to model the inner 2\farcs5, but the poor statistics of the individual objects are insufficient to map out the complex PSF structure.} We ran ${\mathtt{MARX}}$ using an aspect blur of 0\farcs2,\footnote{\url{https://cxc.cfa.harvard.edu/ciao/why/aspectblur.html}} and leaving pileup turned off, such that the ${\mathtt{MARX}}$ simulated images and radial profile yield intrinsic images and profiles.

\begin{figure}

\resizebox{\hsize}{!}{\includegraphics[trim=0 0 0 0, clip,width=\textwidth]{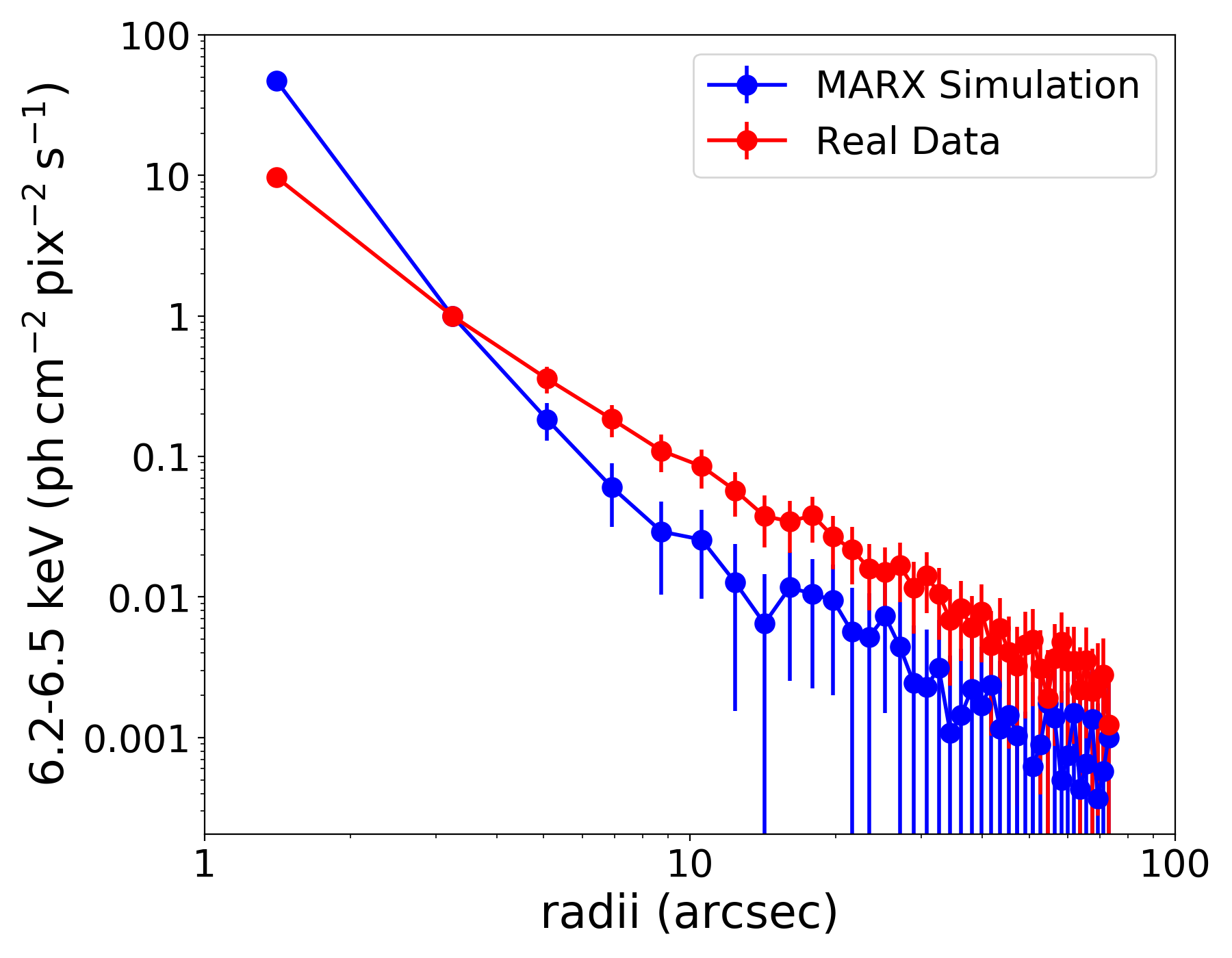}}
\resizebox{\hsize}{!}{\includegraphics[trim=0 0 0 0, clip,width=\textwidth]{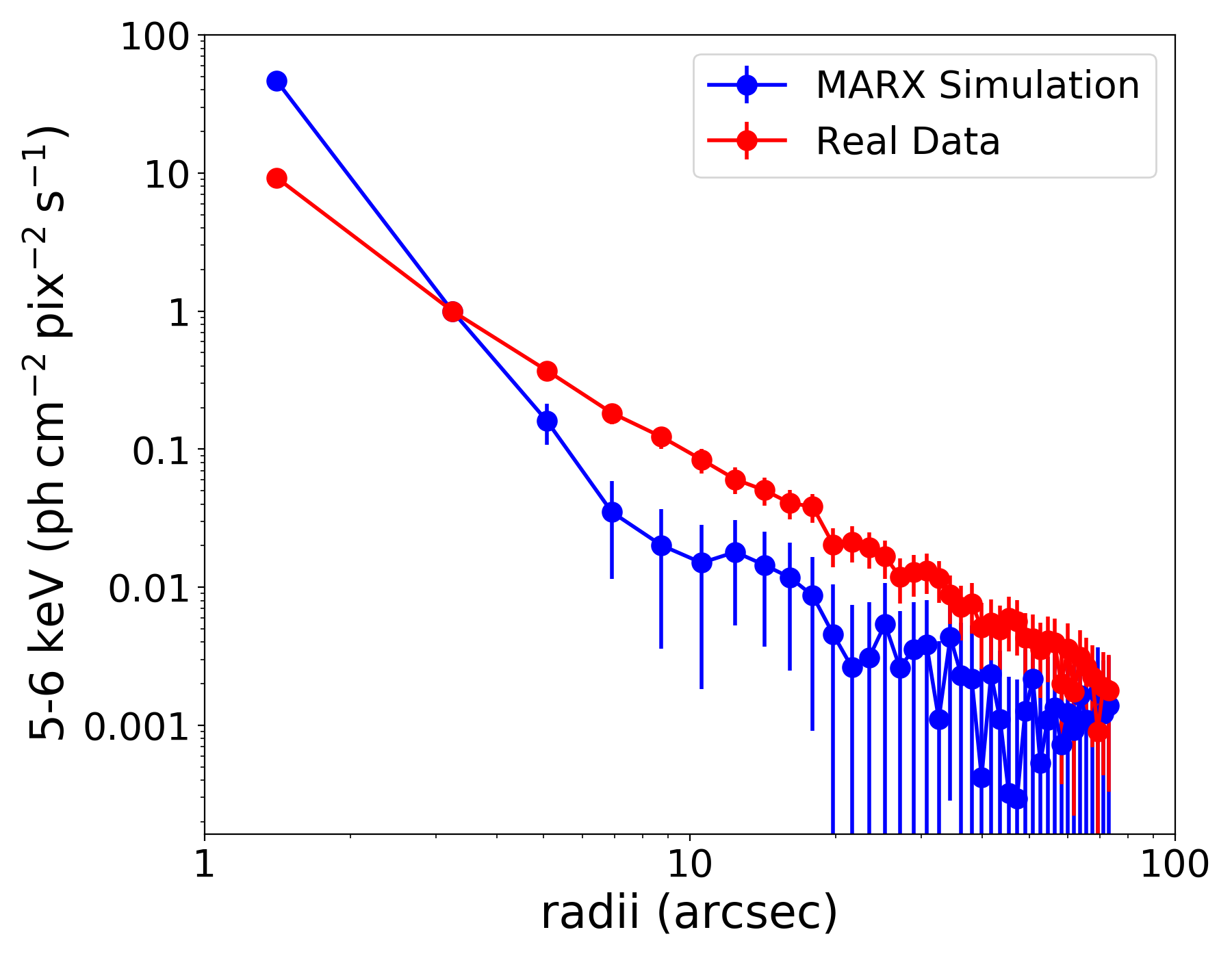}}

 \caption{Comparison between real and simulated radial profiles for the X-ray binary HERX1.  Panels (a) and (b) show radial profiles for the 6.2--6.5 keV and 5--6 keV ranges, respectively. The profiles are renormalized at 2\farcs5, to highlight the regime interior to which pileup effects may occur in the real data. The errors of profiles denote $99\%$ confidence. }{\label{fig:psf_marx} }  
\end{figure}

After generating empirical and simulated images of HERX1, we extract radial profiles for both, which we compare in Fig.~\ref{fig:psf_marx}, in two energy ranges characteristic of the continuum and \kalfa{} line. Both radial profiles show rough powerlaw declines. After normalizing the profiles at 2\farcs5, we see that the real data fall below the simulation at small radii (as expected due to pileup), while at large radii the real data are systematically higher. 

We expect that pileup should be a roughly symmetric effect, and hence we test the symmetry of the PSF for HERX1 on all scales by comparing the radial profiles in Fig.~\ref{fig:herx1} for four distinct quadrants (q1, q2, q3, and q4) as shown in the NGC\,1068 coordinate plane of Appendix Fig.~\ref{fig:Circinus1068}. Importantly, the quadrant profiles are consistent between 1\farc{}--80\farc{} to within errors. We see significant differences in the 0\farcs5 bin, which may be due to the known PSF artifact/spur at $\sim$0\farcs6--0\farcs8 caused by a misalignment in the HRMA mirrors\footnote{see \url{https://cxc.cfa.harvard.edu/ciao/caveats/psf\_artifact.html}}. We also see differences beyond 80\farc{}, which are the result of the dispersed order spectra.

A fundamental point here is that if we model sources based on simulated PSFs, which is the only method provided by the {\it Chandra} X-ray Center, we would misinterpret point-like emission as extended emission. Thus, for the rest of our analysis, we adopt the radial profile of HERX1 as our empirical {\it Chandra} PSF, which we compare to our AGN sample. 

\begin{figure}
\centering
\resizebox{\hsize}{!}{\includegraphics[width=0.55\textwidth]{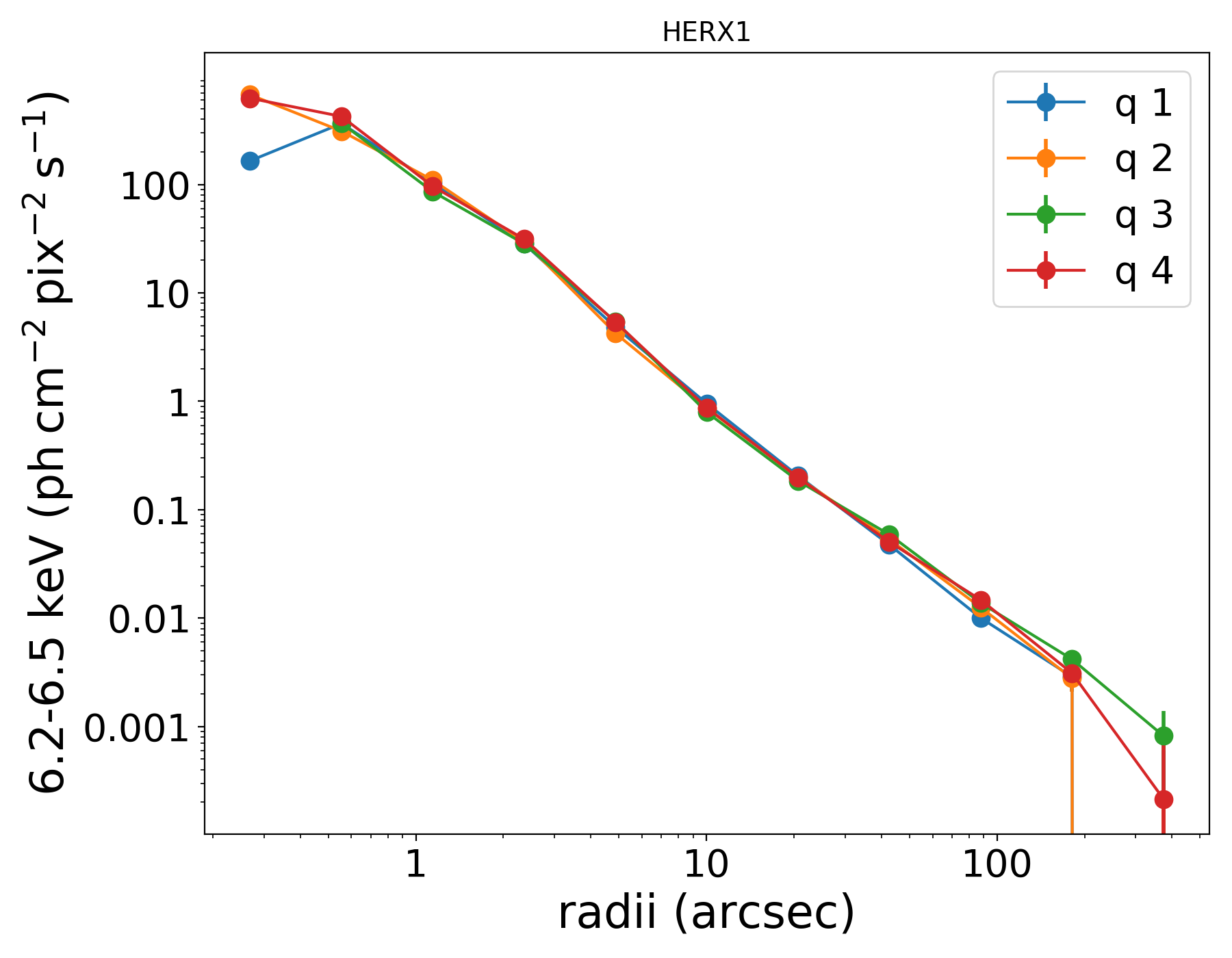}}
\caption{Comparison of 6.2--6.5 keV radial profiles per quadrant (q1, q2, q3, and q4, as defined in Appendix Fig.~\ref{fig:Circinus1068}) for HERX1. The profiles are statistically consistent, indicating that the PSF appears to be approximately symmetric out to at least $\sim$100\farc{} in the energy range which includes \kalfa{} emission for our sample. The errors plotted are $99\%$ confidence.}\label{fig:herx1}	
% Agrandar ejes y labels
\end{figure}

\subsection{Empirical PSF comparison}\label{subsec:emp-PSF}

We first compare the 6.2--6.5 keV radial profiles of HERX1 to the 6.2--6.5 keV radial profiles for all the sources in our sample in Fig.~\ref{fig:rprof_62-65}, to estimate whether any extended (\kalfa{}+continuum) emission is observed. We restrict our radial profile comparisons with HERX1 to beyond 2\farcs5; this only has a small impact in our analysis since the observations of most AGN in the sample are also affected by pileup in the central $\sim2$\farc{}, and thus are not reliable anyway. To make a proper comparison, we normalized each profile to the flux at 2\farcs5, except for Cen\,A and 2MASXJ23444 where we normalized at 4\farc{} to avoid their more severe pileup. Most of the sources have a profile consistent with that of HERX1, except for Cygnus A, H1821+643 and NGC\,1275. For these latter sources, a visual inspection suggests contamination by jets, diffuse emission from the galaxy cluster which the galaxy are centrally embedded, or starburst emission.

Next, we compare the \kalfa{}-only radial profiles (i.e., continuum-substracted) of our sample to that of HERX1 in Fig.\,\ref{fig:rprof_fe}.\footnote{HERX1 has a notable \kalfa{} line.} The azimuthally averaged radial profiles of all the sources are formally consistent with that of HERX1 at 3-$\sigma$ confidence, although some sources show $\approx$2-$\sigma$ excesses beyond 2\farcs5 in one or more bins compared with HERX1 (e.g., MRK\,1040, NGC\,4151, NGC\,1068). The profiles of Cygnus A, H1821+643 and NGC\,1275 have no counts after 2\farcs5, since we could not estimate the underlying \kalfa{} continuum. We exclude these sources from the remaining analysis since we cannot draw conclusions about the spatial extension of those sources. 

To assess whether the \kalfa{} emission is asymmetric in some cases, we compare the radial profile of each quadrant of the images. Figure~\ref{fig:all_quarters_line} shows the profiles divided by 25\% of the renormalized HERX1 radial profile used in Fig.\,\ref{fig:rprof_fe}, and the error bars are plotted at a 3-$\sigma$ level. In several cases, when the Fe K$\alpha$ emission is not very bright, we only obtain an upper limit for the flux at a given radius. The plots indicate that in most sources, the radial profiles are consistent between the quadrants, suggesting that the \kalfa{} emission is symmetric. However, we clearly see extended emission in the Circinus Galaxy and NGC\,1068. The Circinus Galaxy shows asymmetry around $\sim$1--5\farc{}, particularly in the E-W direction. As such, we rotate the quadrants by 45$^\circ$ to obtain a more robust measurement of the extent (see bottom panel of Figure\,\ref{fig:Circinus1068}). We find that quadrant 2 has more flux than the others, followed by quadrant 4. In the case of NGC\,1068, quadrants 2 and 4 also have more flux than quadrants 1 and 3 extending out to $\sim 10$\farc{}. This can be clearly appreciated in the \kalfa{}-only images of Fig.\,\ref{fig:Circinus1068}.   

We estimate the physical extent of the \kalfa{} emission for both sources by converting the angular size to physical distance. For the Circinus Galaxy ($z=0.0014$), the \kalfa{} emission extends up to $95\pm 15$ pc, and for NGC\,1068 ($z=0.0037$) up to $795 \pm 176$ pc. The errors correspond to the $25\%$ of the radii bin width. The rest of the sources do not show extended emission outside 2\farcs5 (4\farc{} in the cases of Cen\,A and 2MASXJ23444), which we can use as an upper limit for the \kalfa{} emission. A summary of the constraints calculated from the imaging analysis are listed in Table\,\ref{t:summary}.

We finally estimate the fractional contribution of the extended \kalfa{} emission in Circinus and NGC\,1068, by subtracting the \kalfa{} counts in quadrants 1 and 3 (which show little or no extended emission) from quadrants 2 and 4 (which show extended emission), and compare this to the overall \kalfa{} counts. We find that the extended emission contributes ${<}8.0 \pm 0.9 \%$ and ${<}16.7 \pm 7.7 \%$ of the total \kalfa{} flux in Circinus Galaxy and NGC\,1068, respectively. These values are strong upper limits to the real ones, since both images are affected by pileup. Although the value for NGC\,1068 is higher, for both AGN, the extended emission represents a small portion of the total \kalfa{} emission.

It is important to understand whether the extended emission seen in Circinus Galaxy and NGC\,1068 is a result of some observational bias, rather than an intrinsic difference. In particular, the Circinus Galaxy and NGC\,1068 have ${\sim}$1\,Ms and $\sim$0.5\,Ms of imaging data, respectively, which rank among the highest in the sample, compared to the majority of sources which only have ${\sim}$100\,ks of data. Comparing the total number of \kalfa{} counts among sources, these objects have factors of $>$60 and $>$5 more counts than the rest of the sample. 
They are also among the closest sources in the sample, such that only 15\% and 30\% of the sample would be able to resolve extended \kalfa{} emission on similar angular scales.
Finally, Circinus Galaxy and NGC\,1068 are both Compton-thick AGN (two of the three in the sample), meaning we likely observe far less direct emission from the BLR, accretion disk and corona, and hence achieve better contrast with which to observe extended features.  All of these factors likely play a part in producing radial profiles which exhibit extended emission for these two targets, and thus we do not rule out that higher quality data could find faint \kalfa{} extension in other sample sources.

\section{Constraints on the origin of the Fe K$\alpha$ line}\label{sec:discussion}

\begin{figure*}
\centering
% \hbox{\hspace{-2.5em}}
\hspace*{-0.5cm}\includegraphics[width=1.0\textwidth]{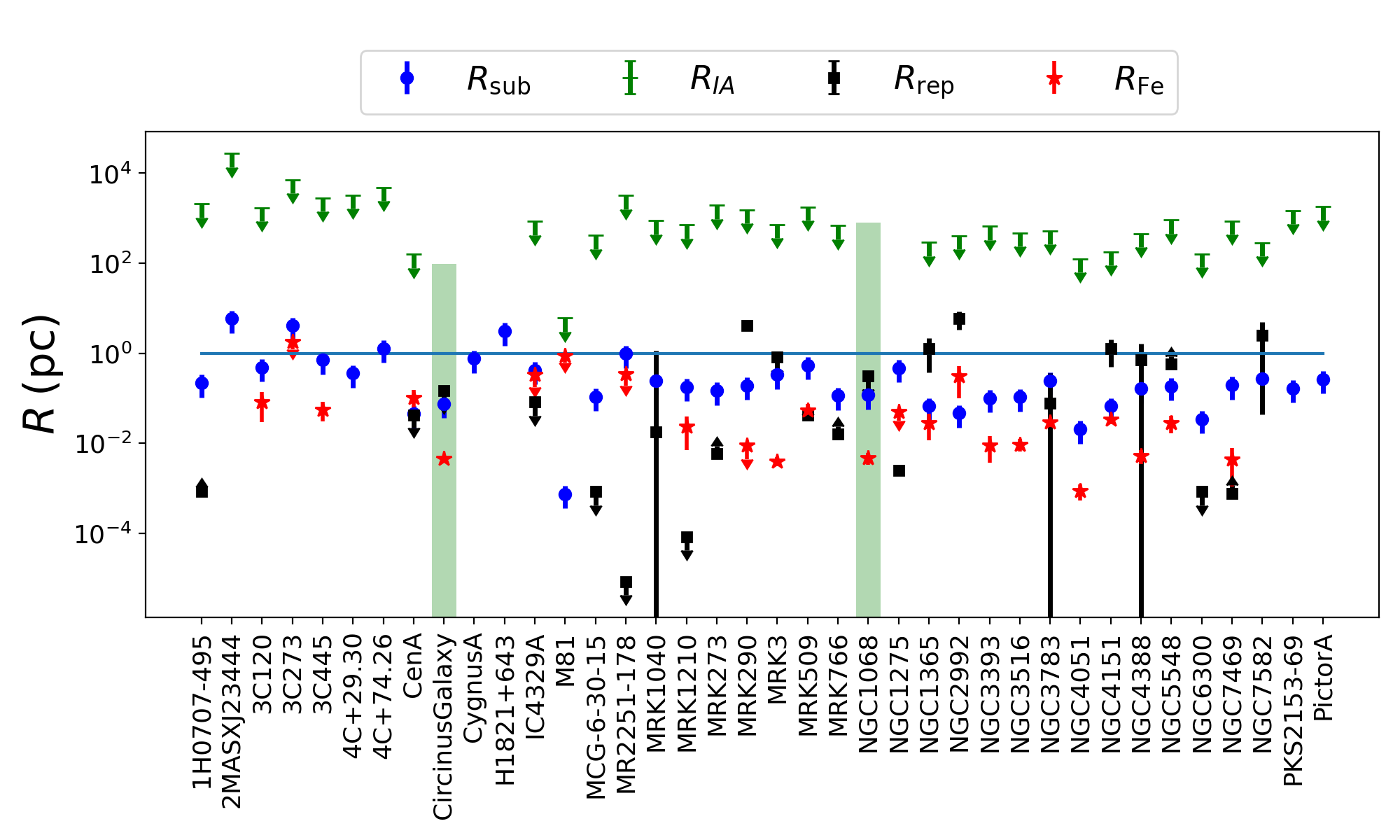}
\caption{Summary of the reprocessor size estimations computed through the spectral analysis (red stars/limits), the timing analysis (black squares/limits), and the imaging analysis (green lines/limits). The dust sublimation radius $R_{\rm sub}$ is plotted in blue as a reference. Black squares without errorbars are sources where the simulations could not provide upper or lower limits for the size of the reflector. The green shaded areas in Circinus and NGC\,1068 represent the extent of the extended emission that we observe in the imaging analysis. }\label{fig:summary_sizes}

\end{figure*}

Comparing the spectral, timing and imaging results for a representative sample of 38 AGN with at least five \textit{Chandra} observations, complemented by existing \textit{XMM-Newton} observations, we find multiple \kalfa{} location estimates for 37 of them.  Figure~\ref{fig:summary_sizes} summarizes the different \kalfa{} location constraints for our sample. Although there is a large dispersion among the constraints (with imaging being the least constraining due to fundamental instrument limitations), these three independent measurements show strong consistency for the majority of the sample, indicating that the bulk of the \kalfa{} emission typically originates inside the dust sublimation radius, and thus is presumably associated with the BLR or outer accretion disk. As we discuss below, only a handful of AGN show clear discrepancies between the indicators (Circinus Galaxy, NGC\,1068, NGC\,4151, and NGC\,5548) and only NGC\,2992 exhibits indications that the bulk of \kalfa{} arises from beyond the dust sublimation radius. Notably, several of these are among the best studied AGN in the local universe.

{The spectral analysis of $\S$\ref{sec:spec_fitting} implies that the bulk of the \kalfa{} emission for 21 out of 24 sources arises from inside the dust sublimation radius, and that for eight out of 11 sources arises from outside the radius of the optical BLR, suggesting that most of the time the \kalfa{} emitting clouds are associated with the BLR, and in a few cases with the outer accretion disk.} This is consistent with several previous studies, which all rely on a subset of the HETG spectra used here. For instance, \citet{2010ApJS..187..581S,2011ApJ...738..147S} and \citet{2011SCPMA..54.1354J} studied the \kalfa{} line core for several type 1 and type 2 AGN, finding that a fraction of the \kalfa{} line originates at factors of $\sim$0.7--2 and $\sim$0.7--11 times the radius of the optical BLR, for type 1 and type 2 AGN, respectively. \citet{2015ApJ...802...98M} similarly conclude that the \kalfa{} line lies between that of the broad Balmer emission lines and the dust reverberation radius for a similar sample of AGN. Alternatively, \citet{2015Gandhi} found that the \kalfa{} radii are similar or smaller than the radius of the optical BLR, and concluded that the dust sublimation radius is an outer envelope to the \kalfa{} bulk for type 1 AGN. Notably, all of the above studies, including ours, are based on the first-order HEG spectra, where a small fraction of the FWHM estimates are of the same order as the nominal spectral resolution. \citet{2016MNRAS.463L.108L} demonstrated that the best-fit line widths from the second- and third-order spectra are systematically lower then the first-order spectral fits by $\approx$30\%, although, in general, they remain consistent within the errors. In the case of NGC\,1275, \citet{2018PASJ...70...13H} measured a line width of 500--1600 $\rm km \: s^{-1}$ (90\% confidence) with {\it Hitomi}, which a factor of 2--6 lower than the value we found, although consistent within errors. Even with such additional systematic errors, the values remain firmly inside of the dust sublimation radius, under the assumption of virial motion.

The timing analysis of $\S$\ref{sec:lc_analysis} provides complementary information for the sample. The simple model considered in $\S$\ref{sec:NXS}, that invokes a universal power-spectral shape with a single break time-scale scaling with mass and accretion rate, accounts for much but not all of the differences in continuum variance between the sources. This can be observed in Fig.~\ref{fig:exvars}d, which shows that for some sources the continuum is more quiescent than expected ($\sigma_c/ \sigma_{c,sim}\ll1$), while in others it is significantly more variable ($\sigma_c/ \sigma_{c,sim}\gg1$). Notable deviations are seen in the Compton-thick sources, which show lower variability than the simple model predicts, consistent with the fact that the direct coronal emission is fully obscured in the observed band and variations might arise from slight changes in the very heavy obscuration \citep[e.g.,][]{Marinucci2016}. On the other hand, the one changing-look AGN in the sample (NGC1365) varies more than expected probably due to well-documented large changes in obscuration which appear to act in addition to intrinsic variations of the continuum, which remains directly visible \citep[e.g.,][]{2015Rivers}. Additionally, radio-loud sources show more scatter in $\sigma_c/ \sigma_{c,sim}$ than radio-quiet ones, possibly due to different levels of jet contribution, none of which is included in the model.

The simulations described in $\S$\ref{sec:rep_sims} provide estimates of the X-ray reprocessor size for 24 sources. Ten have upper limits on the reprocessor size of $<$1 pc, suggesting that the whole \kalfa{} flux arises in regions smaller than $R_{\rm sub}$, the fiducial inner wall of the torus, and in a few cases perhaps even from regions smaller than the BLR. For five sources, the upper limit on the reprocessor size is further out, on parsec scales, leaving open the possibility that a portion of the \kalfa{} flux could arise in the dusty torus. For the other eight sources, the simulations only provide a lower limit or a single value (without lower and upper limits) of the reprocessor size. In some of the latter cases, this is due to the fact that the light curves have very few observations ($\lesssim$7--9) to detect any correlation between the continuum and \kalfa{} fluxes, or the sampling of the observations in not constraining. In other cases, the \kalfa{} line flux does not show variability during the observations, possibly because the reflector is too large compared to the continuum variability timescales, and the simulations are not able to find an upper limit for the reflector. 

For Circinus Galaxy, the simulations predict a small reprocessor, with an upper limit of 0.15 pc. This is contrary to several investigations \citep[e.g,][]{2013Marinucci, 2021Uematsu}, which have predicted a larger reflector. This inconsistency could arise from the fact that we are not be adequately modeling the X-ray continuum from this source, as it is heavily obscured. In the cases of NGC\,4151 and NGC\,5548, the estimated reprocessor sizes appear to be inconsistent with the FWHM-based sizes, with the latter being smaller. This may suggest that the \kalfa{} emission has a more extended configuration than the bulk of the sample or that the FWHM sizes (and hence velocity profiles) are not purely virial in nature. 

Based on our variability analyses, we single out Cen A and IC 4329A, where significant \kalfa{} variability is detected on timescales of days and estimate upper limits for the reprocessor size of 50 and 98 light days, respectively. \citet{2016ApJ...821...15F} studied the narrow \kalfa{} emission of both sources using a sample of multiple observations of \textit{Suzaku} (six observations per source) and \textit{XMM-Newton} (two observations for Cen A) over periods of 4 and 9 years, respectively, and they only detected significant variability on timescales of 1000-2000 days. The main difference between these results and our could be related to the number of observations used in each study. We analyze nine observations of IC\,4329A and 39 observations of Cen A over a timespan of $\sim$18 years, which are sensitive to significant variability on shorter (and longer) timescales. 

\citet{2016ApJ...821...15F} also analyzed four observations of NGC\,4151, finding significant variability on scales of $\sim$1000 days. By contrast, \citet{2019ApJ...884...26Z} studied two decades of \textit{XMM-Newton} observations of NGC\,4151 and found a time delay between the \kalfa{} and continuum variability of $3.3^{+1.8}_{-0.7}$ days. In $\S$\ref{sec:lc_analysis}, we find a lower limit for the reprocessor of $\sim 550$ light days for NGC\,4151, suggesting that the bulk of the \kalfa{} emission mainly arises from or  inside the torus, although the velocity of the line suggests that a portion of the line originates on scales of 40 light days from the center. Our study, by design (i.e., only analyzing spectroscopy on observational exposure timescales), will not be very sensitive to rapid variability of the continuum or \kalfa{} and avoids highly complex spectral fits, which may be why we do not more robustly confirm the results of \citet{2019ApJ...884...26Z}.

Many of our results appear consistent with past findings in smaller samples. One notable exception is the high incidence of radio-loud AGN showing stronger excess variance compared to expectations for a standard disk+corona model. Relativistic beaming coupled with variations in jet or corona structure as a function of time could lead to a high level of associated continuum variability \citep[e.g.,][]{1997Ulrich,2008Chatterjee,2020Weaver}, even if the jet component is not completely dominant in the X-rays \citep[e.g.,][]{2012Cowperthwaite,2013Lohfink}. In large samples of radio-loud AGN \citep[e.g.,][]{2010MacLeod,2021Zhu}, the X-ray emission still appears to be dominated by the corona, although the variability is lower compared to radio-quiet AGNs due to contaminated from a jetted component. This could explain why four radio-loud sources in our sample are less variable than expected by the standard PSD model, but does not explain the stronger excess variance seen in others.

In $\S$\ref{sec:correlations}, we examine possible correlations between the observed X-ray continuum and \kalfa{} fluxes. Overall, in most of the cases the $\rm F_{\rm Fe\:K\alpha}$--$\rm F_{2-10\:keV}$ slope is consistent with zero or has a large scatter. The two sources for which we were able to detect strong correlations between the \kalfa{} to continuum variations are IC4329A and Cen A. These two outstanding sources show \kalfa{} flux variations on timescales of a few days, again suggesting a small reflector, which is confirmed by the simulations, as mentioned above. The case of Cen\,A, shown in Fig.~\ref{fig:lc_3ex}, is particularly striking, with $\approx$1\,dex continuum variations on timescales of weeks to months, and $\approx$1.5\,dex \kalfa{} variations on similar timescales. This level of variation is significantly higher than predicted by the simulations, implying a nonstandard source of variability (perhaps associated with a radio-jet launching region or a highly variable and luminous X-ray emitting knot in the jet at larger distances). IC4329A potentially demonstrates similar levels of variation, although this source only exhibits a strong downward variation  during a single epoch, and otherwise shows fairly typical variability behavior. 

An intriguing result is the case of NGC\,2992, which is the only source in the sample where the spectral and timing analyses estimate a \kalfa{} emission size larger than the dust sublimation radius. The spectral analysis finds that for this source $\upsilon_{\rm FWHM}=1990\pm 672 \rm \: km \: s^{-1}$, which is consistent with the value found by \citet{2017Murphy}, who also analyzed the {\it Chandra}-HETG spectrum of NGC\,2992. Although we find correlated variability between the intrinsic X-ray continuum and the \kalfa{} line ($0.24\pm 0.06$), our timing analysis locates the reflector in parsec scales ($2.2 \: {\rm pc}<R_{rep}<7.1 \: {\rm pc}$), which is expected since the \kalfa{} line responds to the continuum changes on timescales of years. \citet{1996Weaver} reached a similar conclusion by performing a timing analysis of 16 years of ASCA observations and estimated that the reprocessor is $\sim 3.2\rm \: pc$ away from the X-ray source.

We perform imaging analyses on the \textit{Chandra} ACIS images for 35 sources in our sample with sufficiently clean images to allow the removal of off-nuclear point-like sources and analyze their radial profiles. We find that two of the sources with the highest signal-to-noise data of the sample show extended emission outside of 2.5\farc{}, while we can only provide upper limits on the reflecting region for the rest of the sample. We cannot discard the possibility that faint extended emission might be detected in the rest of the sources if comparably high signal-to-noise data could be obtained. For the Circinus Galaxy and NGC\,1068, we find that the \kalfa{} emission extends up to 95 and 795 pc, respectively. Extended emission in the Circinus Galaxy was already reported by \citet{2013Marinucci} on scales of tens of parsec, based on radial profile analyses; by \citet{2014ApJ...791...81A} who found an overall broadening of the iron line in the \textit{Chandra} HEG spectrum due to spatial extension along the dispersion direction; and by \citet{2022Andonie} who found that the \kalfa{} line width systematically increases
as the spectral aperture increases. Similarly, \citet{2015ApJ...812..116B} found that a portion of the \kalfa{} emission is extended up to hundreds of parsec scales in NGC\,1068. 

Recently, \citet{2021ApJ...908..156Y} found that NGC\,4388 shows kpc-scale \kalfa{} extended emission by comparing the radial profiles of the \textit{Chandra} ACIS-S PSF with the one of NGC\,4388 in the 6.2--6.7 keV band. On the contrary, our analysis finds that NGC\,4388 does not present \kalfa{} emission outside of $\sim 450$ pc. Our analysis takes a very different approach, as we construct an empirical PSF that we compare to the radial profiles, as well as consider excesses for both continuum+\kalfa{} and \kalfa{}-only emission (i.e., after subtracting the continuum). On the other hand, \citet{2012Guainazzi} concluded that the \kalfa{} emission in Mrk\,3 extends up to $\simeq 300$\,pc by comparing a {\it Chandra} image with the PSF. Due to the nature of our imaging study, and in particular the inherent pileup of our empirical PSF, we only probe extended emission beyond 2\farc{}5, which for Mrk\,3 ($z=0.0135$) is equivalent to 717\,pc; thus cannot confirm nor refute the possibility of extended emission for this object.

The spectral, timing, and imaging analysis suggest that the \kalfa{} emission predominantly arises from a region inside the dust sublimation radius, probably in the BLR, or perhaps in some cases in the outer accretion disk. Typical spectral energy distributions of AGN indicate that a large fraction of the primary X-ray and UV emission is reprocessed by gas in the molecular torus, with the dust sublimation radius only setting a lower limit for the location of the dust. Thus, we might naively expect that the X-ray reflection would occur in the molecular torus. However, significant cold X-ray absorption is found to occur inside this radius, implying that neutral gas torus coexists with the BLR \citep[e.g.,][]{2015Davies}. Moreover, \citet{2019Ichikawa} studied a sample of 606 AGN in the BASS survey and demonstrated that an extra neutral gas component, in addition to the gas in the dusty torus, is needed to reproduce the X-ray reflection in AGN. Our results, in addition to the studies mentioned above plus many others \citep[e.g.,][]{2007Risaliti,2011Risaliti,2010Maiolino}, provide robust evidence for the presence of abundant dust-free gas inside the dust sublimation radius, which appears to be giving rise to the bulk of reflection in most local AGN. On the other hand, our imaging analysis finds modest \kalfa{} extended emission out to $\sim$0.1-kpc scales for two Compton-thick sources in our sample, suggesting that heavily-obscured reflection; probably in the Compton-thin ($N_{\rm H} \sim 10^{23} \: \rm cm^{-2}$) or Compton-thick regime, has a complex structure, with secondary/sub dominant reflection likely arising from the molecular torus, the ionization cone, and potentially broader scale gas in the host galaxy.

\section{Conclusions} \label{sec:conclusion}

In this work, we analyzed observations for 38 objects from \textit{Chandra} and \textit{XMM-Newton} observatories to constrain the neutral reflecting material in AGN, invoking spectral, timing and imaging analyses of Fe K$\alpha$ line emission. We fit the X-ray spectra of all the observations available of each source, measuring the \kalfa{} line FWHM and flux, and the 2-10 keV continuum flux. Then, we created light curves for the Fe K$\alpha$ line and the continuum and used them to provide an estimate of the reflector size through simulations that compute the delay between the Fe K$\alpha$ line and continuum variability. Finally, we perform an imaging analysis using \textit{Chandra}-ACIS images, to assess whether the sources of our sample are spatially extended. 

In total, we are able to provide estimates of the reflector size for 37 out of 38 AGN. We find:

\begin{enumerate}
    \item For 21 out of 24 sources with a \kalfa{} FWHM measurement, under the assumption of virialized motion, we estimate that the bulk of the \kalfa{} originates inside the dust sublimation radius, probably in the BLR or outer accretion disk.
    
    \item 37 sources show significant variability in the continuum, but only 18 AGN show variability in the \kalfa{} flux. The light curve simulations provide reflector size estimates for 24 sources. Of these, six have continuum excess variances that are $\gtrsim$2-$\sigma$ above expectations based on simulated power-spectral models, hinting at a different origin for the variability for those sources; meanwhile four lie $\lesssim$2-$\sigma$ below expectations, implying significant continuum damping. The simple model invoking a universal power-spectral shape accounts for much but not all of the differences in continuum variance between the sources.

    \item Nearly half of the sample show observed \kalfa{} excess variances equal to or higher (possibly by factors of up to $\sim$3--10 in a few cases) than the continuum ones, implying very little damping by any reprocessor geometry. Further specialized simulations, analyses, and possibly dedicated observations are required to determine plausible geometries. 
    
    \item Despite the wide range of variability properties, our constraints on the \kalfa{} photon reprocessor sizes confirm the picture from the FWHMs, whereby for $\approx$83\% of the systems the \kalfa{} emitting regions are consistent with being inside $R_{\rm sub}$, albeit with looser size limits bounded between $\sim$10$^{-3}$--10 pc. Also, for eight out of 11 sources, the \kalfa{} emission appears to originate from a radius consistent with or larger than $R_{\rm H\beta}$. One interesting outlier is NGC 2992, whose FWHM and timing constraints place the \kalfa{}-emitting region a factor of $\gtrsim$5-$\sim$10 times that of $R_{\rm sub}$.
    
    \item For most of the sources, we do not detect a clear correlation between the observed \kalfa{} line and continuum fluxes, although two outstanding objects, Cen\,A and IC\,4329A, show strong $\approx$0.5--1.5\,dex \kalfa{} variability on timescales of weeks to months. We find no significant trends between the variability features of the \kalfa{} line and AGN/host galaxy properties (e.g., SMBH mass, Eddington ratio, AGN type, radio-loudness and column density).

    \item Finally, we find that the \kalfa{} radial profiles for 33 of 35 AGN are consistent with being point-like beyond 2.5\arcsec{}, and derive loose upper limits for the reflecting clouds using that radius ($\sim$ 6 to 7096 pc). In the cases of the Circinus Galaxy and NGC\,1068, we observe \kalfa{} emission that extends out to 95$\pm$15 and 795$\pm$176 pc, respectively, suggesting the existence of reflecting clouds on scales of hundreds of parsecs. The extended emission in these sources is by no means dominant, but demonstrates that overall heavily-obscured reflection in AGN has a broad (physical) range of secondary contributions to the total \kalfa{} emission. 
    
    \end{enumerate}

As a main conclusion, we confirm that the \kalfa{} emission in the vast majority of AGN appears to arise from regions smaller than and presumably inside $R_{\rm sub}$, yet equal to or beyond $R_{\rm H\beta}$, and thus is associated either with either the outer BLR or accretion disk. That being said, the wide variety of continuum and \kalfa{} variability properties are not easily accommodated by a universal scenario for the production of the bulk of the \kalfa{} emission among local AGNs. More detailed analyses and future observational campaigns are likely required to make further sense of the variability properties hinted at in this study.  

In particular, future observations dedicated to modeling the resolved \kalfa{} line profiles of bright AGN with the calorimeters on {\it XRISM} \citep{2020XRISM} and the variability properties thereof, should thus improve our understanding of how reflection arises from the BLR, outer accretion disk or jet base regions. Likewise, observations being acquired over the next several years with {\it eROSITA} \citep{2021Predehl} should provide more systematic constraints on intermediate timescales albeit at relatively low signal-to-noise for bright AGN, while dedicated observing campaigns and archival studies of outliers such as Cen\,A and IC\,4329A should be pursued. Ultimately, dedicated campaigns with {\it Athena} \citep{2018SPIE10699E..1GB}, given its high sensitivity and spectral resolution, will place the best constraints on the location and origin of the ubiquitous \kalfa{} emission which emanates from AGN.

As part of this work, we additionally detected an inconsistency between the observed and simulated \textit{Chandra} PSFs beyond $\sim$ 3\farc{}--20\farc{}, where the simulations systematically underpredict the real data. This is notable both in direct PSF comparisons, normalized at 2\farc5, between observed and simulated sources, as well in indirect comparisons of fluxes derived from HEG and ACIS annular spectra. We derived a calibration constant between the HEG and 3\farc{}--5\farc{} annular spectra of $\rm 0.424\pm0.00507$, where we expect a value of 1 if the nominal PSF correction is accurate. Our findings imply that any aperture corrections and flux limits based on simulated \textit{Chandra} PSFs are likely to be systematically underestimated by at least a few to several percent.

\begin{acknowledgements}

This project has received funding from 
The European Union’s Horizon 2020 research and innovation programme under the Marie Sklodowska-Curie grant agreement No 860744 (CA),
ANID - Millennium Science Initiative Program - ICN12\_009 (CA, FEB), 
CATA-Basal - AFB-170002 (CA, FEB, ET), 
CATA-Puente - ACE210002 (FEB, ET),
CATA2 - FB210003 (FEB, ET), 
FONDECYT Regular - 1190818 (ET, FEB) and 1200495 (CA, FEB, ET), PIA ACT172033 (ET, PA), NCN$19\_058$ TITANS (PA, ET), the Max-Planck Society through a Max-Planck Partner Group with the university of Valparaíso (PA, RC),
Fondecyt Iniciacion grant 11190831 (CR),
ANID BASAL project FB210003 (CR),
the Science and Technology Facilities Council for support through grant code ST/T000244/1 (DMA),
and FONDECYT Postdoctorado for the project n. 3190213 (AT).

\end{acknowledgements}

% WARNING
%-------------------------------------------------------------------
% Please note that we have included the references to the file aa.dem in
% order to compile it, but we ask you to:
%
% - use BibTeX with the regular commands:
%   \bibliographystyle{aa} % style aa.bst
%   \bibliography{Yourfile} % your references Yourfile.bib
%
% - join the .bib files when you upload your source files
%-------------------------------------------------------------------

% \begin{thebibliography}{}
\bibliographystyle{aa}
\bibliography{biblio}

\begin{thebibliography}{142}
\expandafter\ifx\csname natexlab\endcsname\relax\def\natexlab#1{#1}\fi

\bibitem[{{Andonie} {et~al.}(2022){Andonie}, {Ricci}, {Paltani}, {Ar{\'e}valo},
  {Treister}, {Bauer}, \& {Stalevski}}]{2022Andonie}
{Andonie}, C., {Ricci}, C., {Paltani}, S., {et~al.} 2022, \mnras, 511, 5768

\bibitem[{{Ar{\'e}valo} {et~al.}(2014){Ar{\'e}valo}, {Bauer}, {Puccetti},
  {Walton}, {Koss}, {Boggs}, {Brandt}, {Brightman}, {Christensen}, {Comastri},
  {Craig}, {Fuerst}, {Gandhi}, {Grefenstette}, {Hailey}, {Harrison}, {Luo},
  {Madejski}, {Madsen}, {Marinucci}, {Matt}, {Saez}, {Stern}, {Stuhlinger},
  {Treister}, {Urry}, \& {Zhang}}]{2014ApJ...791...81A}
{Ar{\'e}valo}, P., {Bauer}, F.~E., {Puccetti}, S., {et~al.} 2014, \apj, 791, 81

\bibitem[{{Ar{\'e}valo} {et~al.}(2008){Ar{\'e}valo}, {McHardy}, \&
  {Summons}}]{Arevalo2008}
{Ar{\'e}valo}, P., {McHardy}, I.~M., \& {Summons}, D.~P. 2008, \mnras, 388, 211

\bibitem[{{Ar{\'e}valo} {et~al.}(2009){Ar{\'e}valo}, {Uttley}, {Lira},
  {Breedt}, {McHardy}, \& {Churazov}}]{Arevalo2009reprocessing}
{Ar{\'e}valo}, P., {Uttley}, P., {Lira}, P., {et~al.} 2009, \mnras, 397, 2004

\bibitem[{{Arnaud}(1996)}]{1996ASPC..101...17A}
{Arnaud}, K.~A. 1996, in Astronomical Society of the Pacific Conference Series,
  Vol. 101, Astronomical Data Analysis Software and Systems V, ed. G.~H.
  {Jacoby} \& J.~{Barnes}, 17

\bibitem[{{Barret} {et~al.}(2018){Barret}, {Lam Trong}, {den Herder}, {Piro},
  {Cappi}, {Houvelin}, {Kelley}, {Mas-Hesse}, {Mitsuda}, {Paltani}, {Rauw},
  {Rozanska}, {Wilms}, {Bandler}, {Barbera}, {Barcons}, {Bozzo}, {Ceballos},
  {Charles}, {Costantini}, {Decourchelle}, {den Hartog}, {Duband}, {Duval},
  {Fiore}, {Gatti}, {Goldwurm}, {Jackson}, {Jonker}, {Kilbourne}, {Macculi},
  {Mendez}, {Molendi}, {Orleanski}, {Pajot}, {Pointecouteau}, {Porter},
  {Pratt}, {Pr{\^e}le}, {Ravera}, {Sato}, {Schaye}, {Shinozaki}, {Thibert},
  {Valenziano}, {Valette}, {Vink}, {Webb}, {Wise}, {Yamasaki}, {Douchin},
  {Mesnager}, {Pontet}, {Pradines}, {Branduardi-Raymont}, {Bulbul}, {Dadina},
  {Ettori}, {Finoguenov}, {Fukazawa}, {Janiuk}, {Kaastra}, {Mazzotta},
  {Miller}, {Miniutti}, {Naze}, {Nicastro}, {Scioritino}, {Simonescu},
  {Torrejon}, {Frezouls}, {Geoffray}, {Peille}, {Aicardi}, {Andr{\'e}},
  {Daniel}, {Cl{\'e}net}, {Etcheverry}, {Gloaguen}, {Hervet}, {Jolly}, {Ledot},
  {Paillet}, {Schmisser}, {Vella}, {Damery}, {Boyce}, {Dipirro}, {Lotti},
  {Schwander}, {Smith}, {Van Leeuwen}, {van Weers}, {Clerc}, {Cobo}, {Dauser},
  {Kirsch}, {Cucchetti}, {Eckart}, {Ferrando}, \&
  {Natalucci}}]{2018SPIE10699E..1GB}
{Barret}, D., {Lam Trong}, T., {den Herder}, J.-W., {et~al.} 2018, in Society
  of Photo-Optical Instrumentation Engineers (SPIE) Conference Series, Vol.
  10699, Space Telescopes and Instrumentation 2018: Ultraviolet to Gamma Ray,
  ed. J.-W.~A. {den Herder}, S.~{Nikzad}, \& K.~{Nakazawa}, 106991G

\bibitem[{{Bauer} {et~al.}(2015){Bauer}, {Ar{\'e}valo}, {Walton}, {Koss},
  {Puccetti}, {Gandhi}, {Stern}, {Alexander}, {Balokovi{\'c}}, {Boggs},
  {Brandt}, {Brightman}, {Christensen}, {Comastri}, {Craig}, {Del Moro},
  {Hailey}, {Harrison}, {Hickox}, {Luo}, {Markwardt}, {Marinucci}, {Matt},
  {Rigby}, {Rivers}, {Saez}, {Treister}, {Urry}, \&
  {Zhang}}]{2015ApJ...812..116B}
{Bauer}, F.~E., {Ar{\'e}valo}, P., {Walton}, D.~J., {et~al.} 2015, \apj, 812,
  116

\bibitem[{{Behar} {et~al.}(2020){Behar}, {Kaspi}, {Paubert}, {Billot},
  {Peretz}, {Baldi}, {Laor}, {Kaastra}, \& {Mehdipour}}]{2020Behar}
{Behar}, E., {Kaspi}, S., {Paubert}, G., {et~al.} 2020, \mnras, 491, 3523

\bibitem[{{Bentz} {et~al.}(2009){Bentz}, {Peterson}, {Netzer}, {Pogge}, \&
  {Vestergaard}}]{2009Bentz}
{Bentz}, M.~C., {Peterson}, B.~M., {Netzer}, H., {Pogge}, R.~W., \&
  {Vestergaard}, M. 2009, \apj, 697, 160

\bibitem[{{Bischetti} {et~al.}(2017){Bischetti}, {Piconcelli}, {Vietri},
  {Bongiorno}, {Fiore}, {Sani}, {Marconi}, {Duras}, {Zappacosta}, {Brusa},
  {Comastri}, {Cresci}, {Feruglio}, {Giallongo}, {La Franca}, {Mainieri},
  {Mannucci}, {Martocchia}, {Ricci}, {Schneider}, {Testa}, \&
  {Vignali}}]{2017A&A...598A.122B}
{Bischetti}, M., {Piconcelli}, E., {Vietri}, G., {et~al.} 2017, \aap, 598, A122

\bibitem[{{Boller} {et~al.}(2021){Boller}, {Liu}, {Weber}, {Arcodia}, {Dauser},
  {Wilms}, {Nandra}, {Buchner}, {Merloni}, {Freyberg}, {Krumpe}, \&
  {Waddell}}]{2021Boller}
{Boller}, T., {Liu}, T., {Weber}, P., {et~al.} 2021, \aap, 647, A6

\bibitem[{{Brenneman} \& {Reynolds}(2009)}]{2009Brenneman}
{Brenneman}, L.~W. \& {Reynolds}, C.~S. 2009, \apj, 702, 1367

\bibitem[{{Cackett} {et~al.}(2014){Cackett}, {Zoghbi}, {Reynolds}, {Fabian},
  {Kara}, {Uttley}, \& {Wilkins}}]{2014MNRAS.438.2980C}
{Cackett}, E.~M., {Zoghbi}, A., {Reynolds}, C., {et~al.} 2014, \mnras, 438,
  2980

\bibitem[{{Canizares} {et~al.}(2000){Canizares}, {Huenemoerder}, {Davis},
  {Dewey}, {Flanagan}, {Houck}, {Markert}, {Marshall}, {Schattenburg},
  {Schulz}, {Wise}, {Drake}, \& {Brickhouse}}]{2000ApJ...539L..41C}
{Canizares}, C.~R., {Huenemoerder}, D.~P., {Davis}, D.~S., {et~al.} 2000,
  \apjl, 539, L41

\bibitem[{{Cash}(1979)}]{1979Cash}
{Cash}, W. 1979, \apj, 228, 939

\bibitem[{{Chatterjee} {et~al.}(2008){Chatterjee}, {Jorstad}, {Marscher}, {Oh},
  {McHardy}, {Aller}, {Aller}, {Balonek}, {Miller}, {Ryle}, {Tosti},
  {Kurtanidze}, {Nikolashvili}, {Larionov}, \& {Hagen-Thorn}}]{2008Chatterjee}
{Chatterjee}, R., {Jorstad}, S.~G., {Marscher}, A.~P., {et~al.} 2008, \apj,
  689, 79

\bibitem[{{Corral} {et~al.}(2011){Corral}, {Della Ceca}, {Caccianiga},
  {Severgnini}, {Brunner}, {Carrera}, {Page}, \&
  {Schwope}}]{2011A&A...530A..42C}
{Corral}, A., {Della Ceca}, R., {Caccianiga}, A., {et~al.} 2011, \aap, 530, A42

\bibitem[{{Cowperthwaite} \& {Reynolds}(2012)}]{2012Cowperthwaite}
{Cowperthwaite}, P.~S. \& {Reynolds}, C.~S. 2012, \apjl, 752, L21

\bibitem[{{Davies} {et~al.}(2015){Davies}, {Burtscher}, {Rosario},
  {Storchi-Bergmann}, {Contursi}, {Genzel}, {Graci{\'a}-Carpio}, {Hicks},
  {Janssen}, {Koss}, {Lin}, {Lutz}, {Maciejewski}, {M{\"u}ller-S{\'a}nchez},
  {Orban de Xivry}, {Ricci}, {Riffel}, {Riffel}, {Schartmann},
  {Schnorr-M{\"u}ller}, {Sternberg}, {Sturm}, {Tacconi}, \&
  {Veilleux}}]{2015Davies}
{Davies}, R.~I., {Burtscher}, L., {Rosario}, D., {et~al.} 2015, \apj, 806, 127

\bibitem[{{Davis} {et~al.}(2012){Davis}, {Bautz}, {Dewey}, {Heilmann}, {Houck},
  {Huenemoerder}, {Marshall}, {Nowak}, {Schattenburg}, {Schulz}, \&
  {Smith}}]{2012SPIE.8443E..1AD}
{Davis}, J.~E., {Bautz}, M.~W., {Dewey}, D., {et~al.} 2012, in Society of
  Photo-Optical Instrumentation Engineers (SPIE) Conference Series, Vol. 8443,
  \procspie, 84431A

\bibitem[{{de La Calle P{\'e}rez} {et~al.}(2010){de La Calle P{\'e}rez},
  {Longinotti}, {Guainazzi}, {Bianchi}, {Dov{\v{c}}iak}, {Cappi}, {Matt},
  {Miniutti}, {Petrucci}, {Piconcelli}, {Ponti}, {Porquet}, \&
  {Santos-Lle{\'o}}}]{deLaCalle2010}
{de La Calle P{\'e}rez}, I., {Longinotti}, A.~L., {Guainazzi}, M., {et~al.}
  2010, \aap, 524, A50

\bibitem[{{de Marco} {et~al.}(2009){de Marco}, {Iwasawa}, {Cappi}, {Dadina},
  {Tombesi}, {Ponti}, {Celotti}, \& {Miniutti}}]{deMarco2009}
{de Marco}, B., {Iwasawa}, K., {Cappi}, M., {et~al.} 2009, \aap, 507, 159

\bibitem[{{De Marco} {et~al.}(2013){De Marco}, {Ponti}, {Cappi}, {Dadina},
  {Uttley}, {Cackett}, {Fabian}, \& {Miniutti}}]{2013MNRAS.431.2441D}
{De Marco}, B., {Ponti}, G., {Cappi}, M., {et~al.} 2013, \mnras, 431, 2441

\bibitem[{{Done} {et~al.}(2012){Done}, {Davis}, {Jin}, {Blaes}, \&
  {Ward}}]{Done2012}
{Done}, C., {Davis}, S.~W., {Jin}, C., {Blaes}, O., \& {Ward}, M. 2012, \mnras,
  420, 1848

\bibitem[{{Done} {et~al.}(2007){Done}, {Sobolewska}, {Gierli{\'n}ski}, \&
  {Schurch}}]{2007DoneC}
{Done}, C., {Sobolewska}, M.~A., {Gierli{\'n}ski}, M., \& {Schurch}, N.~J.
  2007, \mnras, 374, L15

\bibitem[{{Du} {et~al.}(2016){Du}, {Lu}, {Hu}, {Qiu}, {Li}, {Huang}, {Wang},
  {Bai}, {Bian}, {Yuan}, {Ho}, {Wang}, \& {SEAMBH Collaboration}}]{2016Du}
{Du}, P., {Lu}, K.-X., {Hu}, C., {et~al.} 2016, \apj, 820, 27

\bibitem[{{Edelson} \& {Nandra}(1999)}]{1999ApJ...514..682E}
{Edelson}, R. \& {Nandra}, K. 1999, \apj, 514, 682

\bibitem[{{Edelson} {et~al.}(1990){Edelson}, {Krolik}, \& {Pike}}]{1990Edelson}
{Edelson}, R.~A., {Krolik}, J.~H., \& {Pike}, G.~F. 1990, \apj, 359, 86

\bibitem[{{Fabbiano}(2006)}]{Fabbiano2006}
{Fabbiano}, G. 2006, \araa, 44, 323

\bibitem[{{Fabbiano} {et~al.}(2017){Fabbiano}, {Elvis}, {Paggi}, {Karovska},
  {Maksym}, {Raymond}, {Risaliti}, \& {Wang}}]{2017ApJ...842L...4F}
{Fabbiano}, G., {Elvis}, M., {Paggi}, A., {et~al.} 2017, \apjl, 842, L4

\bibitem[{{Falocco} {et~al.}(2014){Falocco}, {Carrera}, {Barcons}, {Miniutti},
  \& {Corral}}]{Falocco2014}
{Falocco}, S., {Carrera}, F.~J., {Barcons}, X., {Miniutti}, G., \& {Corral}, A.
  2014, \aap, 568, A15

\bibitem[{{Fukazawa} {et~al.}(2016){Fukazawa}, {Furui}, {Hayashi}, {Ohno},
  {Hiragi}, \& {Noda}}]{2016ApJ...821...15F}
{Fukazawa}, Y., {Furui}, S., {Hayashi}, K., {et~al.} 2016, \apj, 821, 15

\bibitem[{{Gallimore} {et~al.}(2016){Gallimore}, {Elitzur}, {Maiolino},
  {Marconi}, {O'Dea}, {Lutz}, {Baum}, {Nikutta}, {Impellizzeri}, {Davies},
  {Kimball}, \& {Sani}}]{2016ApJ...829L...7G}
{Gallimore}, J.~F., {Elitzur}, M., {Maiolino}, R., {et~al.} 2016, \apjl, 829,
  L7

\bibitem[{{Gandhi} {et~al.}(2015){Gandhi}, {H{\"o}nig}, \&
  {Kishimoto}}]{2015Gandhi}
{Gandhi}, P., {H{\"o}nig}, S.~F., \& {Kishimoto}, M. 2015, \apj, 812, 113

\bibitem[{{Garc{\'\i}a-Burillo} {et~al.}(2016){Garc{\'\i}a-Burillo}, {Combes},
  {Ramos Almeida}, {Usero}, {Krips}, {Alonso-Herrero}, {Aalto}, {Casasola},
  {Hunt}, {Mart{\'\i}n}, {Viti}, {Colina}, {Costagliola}, {Eckart}, {Fuente},
  {Henkel}, {M{\'a}rquez}, {Neri}, {Schinnerer}, {Tacconi}, \& {van der
  Werf}}]{2016ApJ...823L..12G}
{Garc{\'\i}a-Burillo}, S., {Combes}, F., {Ramos Almeida}, C., {et~al.} 2016,
  \apjl, 823, L12

\bibitem[{{Ghisellini} {et~al.}(1994){Ghisellini}, {Haardt}, \&
  {Matt}}]{1994MNRAS.267..743G}
{Ghisellini}, G., {Haardt}, F., \& {Matt}, G. 1994, \mnras, 267, 743

\bibitem[{{Ghosh} \& {Pal}(2021)}]{2021Ghosh}
{Ghosh}, R. \& {Pal}, M. 2021, Research Notes of the American Astronomical
  Society, 5, 35

\bibitem[{{Gilli} {et~al.}(2000){Gilli}, {Maiolino}, {Marconi}, {Risaliti},
  {Dadina}, {Weaver}, \& {Colbert}}]{2000Gilli}
{Gilli}, R., {Maiolino}, R., {Marconi}, A., {et~al.} 2000, \aap, 355, 485

\bibitem[{{Gonz{\'a}lez-Mart{\'\i}n} \& {Vaughan}(2012)}]{2012A&A...544A..80G}
{Gonz{\'a}lez-Mart{\'\i}n}, O. \& {Vaughan}, S. 2012, \aap, 544, A80

\bibitem[{{Green} {et~al.}(1993){Green}, {McHardy}, \&
  {Lehto}}]{1993MNRAS.265..664G}
{Green}, A.~R., {McHardy}, I.~M., \& {Lehto}, H.~J. 1993, \mnras, 265, 664

\bibitem[{{Guainazzi} {et~al.}(2006){Guainazzi}, {Bianchi}, \&
  {Dov{\v{c}}iak}}]{2006AN....327.1032G}
{Guainazzi}, M., {Bianchi}, S., \& {Dov{\v{c}}iak}, M. 2006, Astronomische
  Nachrichten, 327, 1032

\bibitem[{{Guainazzi} {et~al.}(2012){Guainazzi}, {La Parola}, {Miniutti},
  {Segreto}, \& {Longinotti}}]{2012Guainazzi}
{Guainazzi}, M., {La Parola}, V., {Miniutti}, G., {Segreto}, A., \&
  {Longinotti}, A.~L. 2012, \aap, 547, A31

\bibitem[{{Guolo} {et~al.}(2021){Guolo}, {Ruschel-Dutra}, {Grupe}, {Peterson},
  {Storchi-Bergmann}, {Schimoia}, {Nemmen}, \& {Robinson}}]{2021Guolo}
{Guolo}, M., {Ruschel-Dutra}, D., {Grupe}, D., {et~al.} 2021, \mnras, 508, 144

\bibitem[{{Haardt} \& {Maraschi}(1991)}]{1991ApJ...380L..51H}
{Haardt}, F. \& {Maraschi}, L. 1991, \apjl, 380, L51

\bibitem[{{HI4PI Collaboration} {et~al.}(2016){HI4PI Collaboration}, {Ben
  Bekhti}, {Fl{\"o}er}, {Keller}, {Kerp}, {Lenz}, {Winkel}, {Bailin},
  {Calabretta}, {Dedes}, {Ford}, {Gibson}, {Haud}, {Janowiecki}, {Kalberla},
  {Lockman}, {McClure-Griffiths}, {Murphy}, {Nakanishi}, {Pisano}, \&
  {Staveley-Smith}}]{2016A&A...594A.116H}
{HI4PI Collaboration}, {Ben Bekhti}, N., {Fl{\"o}er}, L., {et~al.} 2016, \aap,
  594, A116

\bibitem[{{Hitomi Collaboration} {et~al.}(2018){Hitomi Collaboration},
  {Aharonian}, {Akamatsu}, {Akimoto}, {Allen}, {Angelini}, {Audard}, {Awaki},
  {Axelsson}, {Bamba}, {Bautz}, {Blandford}, {Brenneman}, {Brown}, {Bulbul},
  {Cackett}, {Chernyakova}, {Chiao}, {Coppi}, {Costantini}, {de Plaa}, {de
  Vries}, {den Herder}, {Done}, {Dotani}, {Ebisawa}, {Eckart}, {Enoto}, {Ezoe},
  {Fabian}, {Ferrigno}, {Foster}, {Fujimoto}, {Fukazawa}, {Furuzawa},
  {Galeazzi}, {Gallo}, {Gandhi}, {Giustini}, {Goldwurm}, {Gu}, {Guainazzi},
  {Haba}, {Hagino}, {Hamaguchi}, {Harrus}, {Hatsukade}, {Hayashi}, {Hayashi},
  {Hayashida}, {Hiraga}, {Hornschemeier}, {Hoshino}, {Hughes}, {Ichinohe},
  {Iizuka}, {Inoue}, {Inoue}, {Ishida}, {Ishikawa}, {Ishisaki}, {Iwai},
  {Kaastra}, {Kallman}, {Kamae}, {Kataoka}, {Katsuda}, {Kawai}, {Kelley},
  {Kilbourne}, {Kitaguchi}, {Kitamoto}, {Kitayama}, {Kohmura}, {Kokubun},
  {Koyama}, {Koyama}, {Kretschmar}, {Krimm}, {Kubota}, {Kunieda}, {Laurent},
  {Lee}, {Leutenegger}, {Limousin}, {Loewenstein}, {Long}, {Lumb}, {Madejski},
  {Maeda}, {Maier}, {Makishima}, {Markevitch}, {Matsumoto}, {Matsushita},
  {McCammon}, {McNamara}, {Mehdipour}, {Miller}, {Miller}, {Mineshige},
  {Mitsuda}, {Mitsuishi}, {Miyazawa}, {Mizuno}, {Mori}, {Mori}, {Mukai},
  {Murakami}, {Mushotzky}, {Nakagawa}, {Nakajima}, {Nakamori}, {Nakashima},
  {Nakazawa}, {Nobukawa}, {Nobukawa}, {Noda}, {Odaka}, {Ohashi}, {Ohno},
  {Okajima}, {Ota}, {Ozaki}, {Paerels}, {Paltani}, {Petre}, {Pinto}, {Porter},
  {Pottschmidt}, {Reynolds}, {Safi-Harb}, {Saito}, {Sakai}, {Sasaki}, {Sato},
  {Sato}, {Sato}, {Sawada}, {Schartel}, {Serlemitsos}, {Seta}, {Shidatsu},
  {Simionescu}, {Smith}, {Soong}, {Stawarz}, {Sugawara}, {Sugita},
  {Szymkowiak}, {Tajima}, {Takahashi}, {Takahashi}, {Takeda}, {Takei},
  {Tamagawa}, {Tamura}, {Tanaka}, {Tanaka}, {Tanaka}, {Tashiro}, {Tawara},
  {Terada}, {Terashima}, {Tombesi}, {Tomida}, {Tsuboi}, {Tsujimoto}, {Tsunemi},
  {Tsuru}, {Uchida}, {Uchiyama}, {Uchiyama}, {Ueda}, {Ueda}, {Uno}, {Urry},
  {Ursino}, {Watanabe}, {Werner}, {Wilkins}, {Williams}, {Yamada}, {Yamaguchi},
  {Yamaoka}, {Yamasaki}, {Yamauchi}, {Yamauchi}, {Yaqoob}, {Yatsu}, {Yonetoku},
  {Zhuravleva}, {Zoghbi}, \& {Kawamuro}}]{2018PASJ...70...13H}
{Hitomi Collaboration}, {Aharonian}, F., {Akamatsu}, H., {et~al.} 2018, \pasj,
  70, 13

\bibitem[{{Hu} {et~al.}(2019){Hu}, {Liu}, {Jin}, \& {Yuan}}]{Hu2019}
{Hu}, J., {Liu}, Z., {Jin}, C., \& {Yuan}, W. 2019, \mnras, 488, 4378

\bibitem[{{Ichikawa} {et~al.}(2019){Ichikawa}, {Ricci}, {Ueda}, {Bauer},
  {Kawamuro}, {Koss}, {Oh}, {Rosario}, {Shimizu}, {Stalevski}, {Fuller},
  {Packham}, \& {Trakhtenbrot}}]{2019Ichikawa}
{Ichikawa}, K., {Ricci}, C., {Ueda}, Y., {et~al.} 2019, \apj, 870, 31

\bibitem[{{Imanishi} {et~al.}(2016){Imanishi}, {Nakanishi}, \&
  {Izumi}}]{2016ApJ...822L..10I}
{Imanishi}, M., {Nakanishi}, K., \& {Izumi}, T. 2016, \apjl, 822, L10

\bibitem[{{Iwasawa} {et~al.}(1996){Iwasawa}, {Fabian}, {Reynolds}, {Nand ra},
  {Otani}, {Inoue}, {Hayashida}, {Brand t}, {Dotani}, {Kunieda}, {Matsuoka}, \&
  {Tanaka}}]{1996MNRAS.282.1038I}
{Iwasawa}, K., {Fabian}, A.~C., {Reynolds}, C.~S., {et~al.} 1996, \mnras, 282,
  1038

\bibitem[{{Iwasawa} {et~al.}(2018){Iwasawa}, {U}, {Mazzarella}, {Medling},
  {Sanders}, \& {Evans}}]{2018Iwasawa}
{Iwasawa}, K., {U}, V., {Mazzarella}, J.~M., {et~al.} 2018, \aap, 611, A71

\bibitem[{{Jansen} {et~al.}(2001){Jansen}, {Lumb}, {Altieri}, {Clavel}, {Ehle},
  {Erd}, {Gabriel}, {Guainazzi}, {Gondoin}, {Much}, {Munoz}, {Santos},
  {Schartel}, {Texier}, \& {Vacanti}}]{2001A&A...365L...1J}
{Jansen}, F., {Lumb}, D., {Altieri}, B., {et~al.} 2001, \aap, 365, L1

\bibitem[{{Jiang} {et~al.}(2011){Jiang}, {Wang}, \&
  {Shu}}]{2011SCPMA..54.1354J}
{Jiang}, P., {Wang}, J., \& {Shu}, X. 2011, Science China Physics, Mechanics,
  and Astronomy, 54, 1354

\bibitem[{{Kara} {et~al.}(2016){Kara}, {Alston}, {Fabian}, {Cackett}, {Uttley},
  {Reynolds}, \& {Zoghbi}}]{Kara2016}
{Kara}, E., {Alston}, W.~N., {Fabian}, A.~C., {et~al.} 2016, \mnras, 462, 511

\bibitem[{{Kelly}(2007)}]{2007ApJ...665.1489K}
{Kelly}, B.~C. 2007, \apj, 665, 1489

\bibitem[{{Kishimoto} {et~al.}(2007){Kishimoto}, {H{\"o}nig}, {Beckert}, \&
  {Weigelt}}]{2007A&A...476..713K}
{Kishimoto}, M., {H{\"o}nig}, S.~F., {Beckert}, T., \& {Weigelt}, G. 2007,
  \aap, 476, 713

\bibitem[{{Koss} {et~al.}(2017){Koss}, {Trakhtenbrot}, {Ricci}, {Lamperti},
  {Oh}, {Berney}, {Schawinski}, {Balokovi{\'c}}, {Baronchelli}, {Crenshaw},
  {Fischer}, {Gehrels}, {Harrison}, {Hashimoto}, {Hogg}, {Ichikawa}, {Masetti},
  {Mushotzky}, {Sartori}, {Stern}, {Treister}, {Ueda}, {Veilleux}, \&
  {Winter}}]{2017ApJ...850...74K}
{Koss}, M., {Trakhtenbrot}, B., {Ricci}, C., {et~al.} 2017, \apj, 850, 74

\bibitem[{{Koss} {et~al.}(2016){Koss}, {Assef}, {Balokovi{\'c}}, {Stern},
  {Gandhi}, {Lamperti}, {Alexander}, {Ballantyne}, {Bauer}, {Berney}, {Brandt},
  {Comastri}, {Gehrels}, {Harrison}, {Lansbury}, {Markwardt}, {Ricci},
  {Rivers}, {Schawinski}, {Trakhtenbrot}, {Treister}, \& {Urry}}]{2016Koss}
{Koss}, M.~J., {Assef}, R., {Balokovi{\'c}}, M., {et~al.} 2016, \apj, 825, 85

\bibitem[{{Koss} {et~al.}(2022, submitted){Koss}, {Trakhtenbrot}, \&
  {Ricci}}]{2022Koss}
{Koss}, M.~J., {Trakhtenbrot}, B., \& {Ricci}, C. 2022, submitted, \apj

\bibitem[{{Krolik} {et~al.}(1994){Krolik}, {Madau}, \&
  {Zycki}}]{1994ApJ...420L..57K}
{Krolik}, J.~H., {Madau}, P., \& {Zycki}, P.~T. 1994, \apjl, 420, L57

\bibitem[{{Lamer} {et~al.}(2003){Lamer}, {McHardy}, {Uttley}, \&
  {Jahoda}}]{2003MNRAS.338..323L}
{Lamer}, G., {McHardy}, I.~M., {Uttley}, P., \& {Jahoda}, K. 2003, \mnras, 338,
  323

\bibitem[{{Lanzuisi} {et~al.}(2014){Lanzuisi}, {Ponti}, {Salvato}, {Hasinger},
  {Cappelluti}, {Bongiorno}, {Brusa}, {Lusso}, {Nandra}, {Merloni},
  {Silverman}, {Trump}, {Vignali}, {Comastri}, {Gilli}, {Schramm},
  {Steinhardt}, {Sanders}, {Kartaltepe}, {Rosario}, \&
  {Trakhtenbrot}}]{2014ApJ...781..105L}
{Lanzuisi}, G., {Ponti}, G., {Salvato}, M., {et~al.} 2014, \apj, 781, 105

\bibitem[{{Liu}(2016)}]{2016MNRAS.463L.108L}
{Liu}, J. 2016, \mnras, 463, L108

\bibitem[{{Liu} {et~al.}(2019){Liu}, {Veilleux}, {Iwasawa}, {Rupke}, {Teng},
  {U}, {Tombesi}, {Sanders}, {Max}, \& {Mel{\'e}ndez}}]{2019Liu}
{Liu}, W., {Veilleux}, S., {Iwasawa}, K., {et~al.} 2019, \apj, 872, 39

\bibitem[{{Lohfink} {et~al.}(2013{\natexlab{a}}){Lohfink}, {Reynolds},
  {Jorstad}, {Marscher}, {Miller}, {Aller}, {Aller}, {Brenneman}, {Fabian},
  {Miller}, {Mushotzky}, {Nowak}, \& {Tombesi}}]{Lohfink2013}
{Lohfink}, A.~M., {Reynolds}, C.~S., {Jorstad}, S.~G., {et~al.}
  2013{\natexlab{a}}, \apj, 772, 83

\bibitem[{{Lohfink} {et~al.}(2013{\natexlab{b}}){Lohfink}, {Reynolds},
  {Jorstad}, {Marscher}, {Miller}, {Aller}, {Aller}, {Brenneman}, {Fabian},
  {Miller}, {Mushotzky}, {Nowak}, \& {Tombesi}}]{2013Lohfink}
{Lohfink}, A.~M., {Reynolds}, C.~S., {Jorstad}, S.~G., {et~al.}
  2013{\natexlab{b}}, \apj, 772, 83

\bibitem[{{Lu} {et~al.}(2016){Lu}, {Du}, {Hu}, {Li}, {Zhang}, {Wang}, {Huang},
  {Bi}, {Bai}, {Ho}, \& {Wang}}]{2016Lu}
{Lu}, K.-X., {Du}, P., {Hu}, C., {et~al.} 2016, \apj, 827, 118

\bibitem[{{Lubi{\'n}ski} \& {Zdziarski}(2001)}]{2001MNRAS.323L..37L}
{Lubi{\'n}ski}, P. \& {Zdziarski}, A.~A. 2001, \mnras, 323, L37

\bibitem[{{Lyons}(1991)}]{Lyons1991}
{Lyons}, L. 1991, {A Practical Guide to Data Analysis for Physical Science
  Students}

\bibitem[{{Lyu} \& {Rieke}(2021)}]{2021LyuJ}
{Lyu}, J. \& {Rieke}, G.~H. 2021, \apj, 912, 126

\bibitem[{{MacLeod} {et~al.}(2010){MacLeod}, {Ivezi{\'c}}, {Kochanek},
  {Koz{\l}owski}, {Kelly}, {Bullock}, {Kimball}, {Sesar}, {Westman}, {Brooks},
  {Gibson}, {Becker}, \& {de Vries}}]{2010MacLeod}
{MacLeod}, C.~L., {Ivezi{\'c}}, {\v{Z}}., {Kochanek}, C.~S., {et~al.} 2010,
  \apj, 721, 1014

\bibitem[{{Maiolino} {et~al.}(2010){Maiolino}, {Risaliti}, {Salvati},
  {Pietrini}, {Torricelli-Ciamponi}, {Elvis}, {Fabbiano}, {Braito}, \&
  {Reeves}}]{2010Maiolino}
{Maiolino}, R., {Risaliti}, G., {Salvati}, M., {et~al.} 2010, \aap, 517, A47

\bibitem[{{Marinucci} {et~al.}(2018){Marinucci}, {Bianchi}, {Braito}, {Matt},
  {Nardini}, \& {Reeves}}]{2018MNRAS.478.5638M}
{Marinucci}, A., {Bianchi}, S., {Braito}, V., {et~al.} 2018, \mnras, 478, 5638

\bibitem[{{Marinucci} {et~al.}(2017){Marinucci}, {Bianchi}, {Fabbiano}, {Matt},
  {Risaliti}, {Nardini}, \& {Wang}}]{2017MNRAS.470.4039M}
{Marinucci}, A., {Bianchi}, S., {Fabbiano}, G., {et~al.} 2017, \mnras, 470,
  4039

\bibitem[{{Marinucci} {et~al.}(2016){Marinucci}, {Bianchi}, {Matt},
  {Alexander}, {Balokovi{\'c}}, {Bauer}, {Brandt}, {Gandhi}, {Guainazzi},
  {Harrison}, {Iwasawa}, {Koss}, {Madsen}, {Nicastro}, {Puccetti}, {Ricci},
  {Stern}, \& {Walton}}]{Marinucci2016}
{Marinucci}, A., {Bianchi}, S., {Matt}, G., {et~al.} 2016, \mnras, 456, L94

\bibitem[{{Marinucci} {et~al.}(2014){Marinucci}, {Matt}, {Miniutti},
  {Guainazzi}, {Parker}, {Brenneman}, {Fabian}, {Kara}, {Arevalo},
  {Ballantyne}, {Boggs}, {Cappi}, {Christensen}, {Craig}, {Elvis}, {Hailey},
  {Harrison}, {Reynolds}, {Risaliti}, {Stern}, {Walton}, \&
  {Zhang}}]{2014ApJ...787...83M}
{Marinucci}, A., {Matt}, G., {Miniutti}, G., {et~al.} 2014, \apj, 787, 83

\bibitem[{{Marinucci} {et~al.}(2013){Marinucci}, {Miniutti}, {Bianchi}, {Matt},
  \& {Risaliti}}]{2013Marinucci}
{Marinucci}, A., {Miniutti}, G., {Bianchi}, S., {Matt}, G., \& {Risaliti}, G.
  2013, \mnras, 436, 2500

\bibitem[{{Marinucci} {et~al.}(2012){Marinucci}, {Risaliti}, {Wang}, {Nardini},
  {Elvis}, {Fabbiano}, {Bianchi}, \& {Matt}}]{2012MNRAS.423L...6M}
{Marinucci}, A., {Risaliti}, G., {Wang}, J., {et~al.} 2012, \mnras, 423, L6

\bibitem[{{Markowitz} {et~al.}(2003){Markowitz}, {Edelson}, {Vaughan},
  {Uttley}, {George}, {Griffiths}, {Kaspi}, {Lawrence}, {McHardy}, {Nandra},
  {Pounds}, {Reeves}, {Schurch}, \& {Warwick}}]{Markowitz2003}
{Markowitz}, A., {Edelson}, R., {Vaughan}, S., {et~al.} 2003, \apj, 593, 96

\bibitem[{{McHardy} {et~al.}(2005){McHardy}, {Gunn}, {Uttley}, \&
  {Goad}}]{Mchardy2005}
{McHardy}, I.~M., {Gunn}, K.~F., {Uttley}, P., \& {Goad}, M.~R. 2005, \mnras,
  359, 1469

\bibitem[{{McHardy} {et~al.}(2006){McHardy}, {Koerding}, {Knigge}, {Uttley}, \&
  {Fender}}]{2006Natur.444..730M}
{McHardy}, I.~M., {Koerding}, E., {Knigge}, C., {Uttley}, P., \& {Fender},
  R.~P. 2006, \nat, 444, 730

\bibitem[{{McHardy} {et~al.}(2004){McHardy}, {Papadakis}, {Uttley}, {Page}, \&
  {Mason}}]{Mchardy2004}
{McHardy}, I.~M., {Papadakis}, I.~E., {Uttley}, P., {Page}, M.~J., \& {Mason},
  K.~O. 2004, \mnras, 348, 783

\bibitem[{{Minezaki} \& {Matsushita}(2015)}]{2015ApJ...802...98M}
{Minezaki}, T. \& {Matsushita}, K. 2015, \apj, 802, 98

\bibitem[{{Murphy} {et~al.}(2017){Murphy}, {Nowak}, \& {Marshall}}]{2017Murphy}
{Murphy}, K.~D., {Nowak}, M.~A., \& {Marshall}, H.~L. 2017, \apj, 840, 120

\bibitem[{{Murphy} {et~al.}(2007){Murphy}, {Yaqoob}, \&
  {Terashima}}]{2007Murphy}
{Murphy}, K.~D., {Yaqoob}, T., \& {Terashima}, Y. 2007, \apj, 666, 96

\bibitem[{{Mushotzky} {et~al.}(1993){Mushotzky}, {Done}, \&
  {Pounds}}]{1993ARA&A..31..717M}
{Mushotzky}, R.~F., {Done}, C., \& {Pounds}, K.~A. 1993, \araa, 31, 717

\bibitem[{{Mushotzky} {et~al.}(1995){Mushotzky}, {Fabian}, {Iwasawa},
  {Kunieda}, {Matsuoka}, {Nandra}, \& {Tanaka}}]{1995MNRAS.272L...9M}
{Mushotzky}, R.~F., {Fabian}, A.~C., {Iwasawa}, K., {et~al.} 1995, \mnras, 272,
  L9

\bibitem[{{Nandra}(2006)}]{2006MNRAS.368L..62N}
{Nandra}, K. 2006, \mnras, 368, L62

\bibitem[{{Nandra} {et~al.}(1997){Nandra}, {George}, {Mushotzky}, {Turner}, \&
  {Yaqoob}}]{1997ApJ...476...70N}
{Nandra}, K., {George}, I.~M., {Mushotzky}, R.~F., {Turner}, T.~J., \&
  {Yaqoob}, T. 1997, \apj, 476, 70

\bibitem[{{Nandra} \& {Pounds}(1994)}]{1994MNRAS.268..405N}
{Nandra}, K. \& {Pounds}, K.~A. 1994, \mnras, 268, 405

\bibitem[{{Nenkova} {et~al.}(2008){Nenkova}, {Sirocky}, {Nikutta},
  {Ivezi{\'c}}, \& {Elitzur}}]{2008IINenkova}
{Nenkova}, M., {Sirocky}, M.~M., {Nikutta}, R., {Ivezi{\'c}}, {\v{Z}}., \&
  {Elitzur}, M. 2008, \apj, 685, 160

\bibitem[{Netzer(1990)}]{Netzer1990}
Netzer, H. 1990, AGN Emission Lines, ed. T.~J.-L. Courvoisier \& M.~Mayor
  (Berlin, Heidelberg: Springer Berlin Heidelberg), 57--158

\bibitem[{{Oh} {et~al.}(2018){Oh}, {Koss}, {Markwardt}, {Schawinski},
  {Baumgartner}, {Barthelmy}, {Cenko}, {Gehrels}, {Mushotzky}, {Petulante},
  {Ricci}, {Lien}, \& {Trakhtenbrot}}]{2018ApJS..235....4O}
{Oh}, K., {Koss}, M., {Markwardt}, C.~B., {et~al.} 2018, \apjs, 235, 4

\bibitem[{{Pahari} {et~al.}(2020){Pahari}, {McHardy}, {Vincentelli}, {Cackett},
  {Peterson}, {Goad}, {G{\"u}ltekin}, \& {Horne}}]{2020Pahari}
{Pahari}, M., {McHardy}, I.~M., {Vincentelli}, F., {et~al.} 2020, \mnras, 494,
  4057

\bibitem[{{Panessa} \& {Giroletti}(2013)}]{2013MNRAS.432.1138P}
{Panessa}, F. \& {Giroletti}, M. 2013, \mnras, 432, 1138

\bibitem[{{Panessa} {et~al.}(2015){Panessa}, {Tarchi}, {Castangia}, {Maiorano},
  {Bassani}, {Bicknell}, {Bazzano}, {Bird}, {Malizia}, \&
  {Ubertini}}]{2015MNRAS.447.1289P}
{Panessa}, F., {Tarchi}, A., {Castangia}, P., {et~al.} 2015, \mnras, 447, 1289

\bibitem[{{Paolillo} {et~al.}(2004){Paolillo}, {Schreier}, {Giacconi},
  {Koekemoer}, \& {Grogin}}]{2004ApJ...611...93P}
{Paolillo}, M., {Schreier}, E.~J., {Giacconi}, R., {Koekemoer}, A.~M., \&
  {Grogin}, N.~A. 2004, \apj, 611, 93

\bibitem[{{Papadakis} {et~al.}(2008){Papadakis}, {Chatzopoulos},
  {Athanasiadis}, {Markowitz}, \& {Georgantopoulos}}]{2008A&A...487..475P}
{Papadakis}, I.~E., {Chatzopoulos}, E., {Athanasiadis}, D., {Markowitz}, A., \&
  {Georgantopoulos}, I. 2008, \aap, 487, 475

\bibitem[{{Peterson} {et~al.}(2004){Peterson}, {Ferrarese}, {Gilbert}, {Kaspi},
  {Malkan}, {Maoz}, {Merritt}, {Netzer}, {Onken}, {Pogge}, {Vestergaard}, \&
  {Wandel}}]{2004ApJ...613..682P}
{Peterson}, B.~M., {Ferrarese}, L., {Gilbert}, K.~M., {et~al.} 2004, \apj, 613,
  682

\bibitem[{{Ponti} {et~al.}(2013){Ponti}, {Cappi}, {Costantini}, {Bianchi},
  {Kaastra}, {De Marco}, {Fender}, {Petrucci}, {Kriss}, {Steenbrugge}, {Arav},
  {Behar}, {Branduardi-Raymont}, {Dadina}, {Ebrero}, {Lubi{\'n}ski},
  {Mehdipour}, {Paltani}, {Pinto}, \& {Tombesi}}]{2013A&A...549A..72P}
{Ponti}, G., {Cappi}, M., {Costantini}, E., {et~al.} 2013, \aap, 549, A72

\bibitem[{{Predehl} {et~al.}(2021){Predehl}, {Andritschke}, {Arefiev},
  {Babyshkin}, {Batanov}, {Becker}, {B{\"o}hringer}, {Bogomolov}, {Boller},
  {Borm}, {Bornemann}, {Br{\"a}uninger}, {Br{\"u}ggen}, {Brunner}, {Brusa},
  {Bulbul}, {Buntov}, {Burwitz}, {Burkert}, {Clerc}, {Churazov}, {Coutinho},
  {Dauser}, {Dennerl}, {Doroshenko}, {Eder}, {Emberger}, {Eraerds},
  {Finoguenov}, {Freyberg}, {Friedrich}, {Friedrich}, {F{\"u}rmetz},
  {Georgakakis}, {Gilfanov}, {Granato}, {Grossberger}, {Gueguen}, {Gureev},
  {Haberl}, {H{\"a}lker}, {Hartner}, {Hasinger}, {Huber}, {Ji}, {Kienlin},
  {Kink}, {Korotkov}, {Kreykenbohm}, {Lamer}, {Lomakin}, {Lapshov}, {Liu},
  {Maitra}, {Meidinger}, {Menz}, {Merloni}, {Mernik}, {Mican}, {Mohr},
  {M{\"u}ller}, {Nandra}, {Nazarov}, {Pacaud}, {Pavlinsky}, {Perinati},
  {Pfeffermann}, {Pietschner}, {Ramos-Ceja}, {Rau}, {Reiffers}, {Reiprich},
  {Robrade}, {Salvato}, {Sanders}, {Santangelo}, {Sasaki}, {Scheuerle},
  {Schmid}, {Schmitt}, {Schwope}, {Shirshakov}, {Steinmetz}, {Stewart},
  {Str{\"u}der}, {Sunyaev}, {Tenzer}, {Tiedemann}, {Tr{\"u}mper}, {Voron},
  {Weber}, {Wilms}, \& {Yaroshenko}}]{2021Predehl}
{Predehl}, P., {Andritschke}, R., {Arefiev}, V., {et~al.} 2021, \aap, 647, A1

\bibitem[{{Puccetti} {et~al.}(2014){Puccetti}, {Comastri}, {Fiore},
  {Ar{\'e}valo}, {Risaliti}, {Bauer}, {Brandt}, {Stern}, {Harrison},
  {Alexander}, {Boggs}, {Christensen}, {Craig}, {Gandhi}, {Hailey}, {Koss},
  {Lansbury}, {Luo}, {Madejski}, {Matt}, {Walton}, \& {Zhang}}]{Puccetti2014}
{Puccetti}, S., {Comastri}, A., {Fiore}, F., {et~al.} 2014, \apj, 793, 26

\bibitem[{{Ricci} {et~al.}(2017{\natexlab{a}}){Ricci}, {Trakhtenbrot}, {Koss},
  {Ueda}, {Del Vecchio}, {Treister}, {Schawinski}, {Paltani}, {Oh}, {Lamperti},
  {Berney}, {Gand hi}, {Ichikawa}, {Bauer}, {Ho}, {Asmus}, {Beckmann}, {Soldi},
  {Balokovi{\'c}}, {Gehrels}, \& {Markwardt}}]{2017ApJS..233...17R}
{Ricci}, C., {Trakhtenbrot}, B., {Koss}, M.~J., {et~al.} 2017{\natexlab{a}},
  \apjs, 233, 17

\bibitem[{{Ricci} {et~al.}(2017{\natexlab{b}}){Ricci}, {Trakhtenbrot}, {Koss},
  {Ueda}, {Schawinski}, {Oh}, {Lamperti}, {Mushotzky}, {Treister}, {Ho},
  {Weigel}, {Bauer}, {Paltani}, {Fabian}, {Xie}, \&
  {Gehrels}}]{2017Natur.549..488R}
{Ricci}, C., {Trakhtenbrot}, B., {Koss}, M.~J., {et~al.} 2017{\natexlab{b}},
  \nat, 549, 488

\bibitem[{{Ricci} {et~al.}(2015){Ricci}, {Ueda}, {Koss}, {Trakhtenbrot},
  {Bauer}, \& {Gandhi}}]{2015ApJ...815L..13R}
{Ricci}, C., {Ueda}, Y., {Koss}, M.~J., {et~al.} 2015, \apjl, 815, L13

\bibitem[{{Risaliti} {et~al.}(2007){Risaliti}, {Elvis}, {Fabbiano}, {Baldi},
  {Zezas}, \& {Salvati}}]{2007Risaliti}
{Risaliti}, G., {Elvis}, M., {Fabbiano}, G., {et~al.} 2007, \apjl, 659, L111

\bibitem[{{Risaliti} {et~al.}(2011){Risaliti}, {Nardini}, {Salvati}, {Elvis},
  {Fabbiano}, {Maiolino}, {Pietrini}, \& {Torricelli-Ciamponi}}]{2011Risaliti}
{Risaliti}, G., {Nardini}, E., {Salvati}, M., {et~al.} 2011, \mnras, 410, 1027

\bibitem[{{Rivers} {et~al.}(2015){Rivers}, {Risaliti}, {Walton}, {Harrison},
  {Ar{\'e}valo}, {Baur}, {Boggs}, {Brenneman}, {Brightman}, {Christensen},
  {Craig}, {F{\"u}rst}, {Hailey}, {Hickox}, {Marinucci}, {Reeves}, {Stern}, \&
  {Zhang}}]{2015Rivers}
{Rivers}, E., {Risaliti}, G., {Walton}, D.~J., {et~al.} 2015, \apj, 804, 107

\bibitem[{{S{\'a}nchez} {et~al.}(2017){S{\'a}nchez}, {Lira}, {Cartier},
  {P{\'e}rez}, {Miranda}, {Yovaniniz}, {Ar{\'e}valo}, {Milvang-Jensen},
  {Fynbo}, {Dunlop}, {Coppi}, \& {Marchesi}}]{2017ApJ...849..110S}
{S{\'a}nchez}, P., {Lira}, P., {Cartier}, R., {et~al.} 2017, \apj, 849, 110

\bibitem[{{Shu} {et~al.}(2010){Shu}, {Yaqoob}, \& {Wang}}]{2010ApJS..187..581S}
{Shu}, X.~W., {Yaqoob}, T., \& {Wang}, J.~X. 2010, \apjs, 187, 581

\bibitem[{{Shu} {et~al.}(2011){Shu}, {Yaqoob}, \& {Wang}}]{2011ApJ...738..147S}
{Shu}, X.~W., {Yaqoob}, T., \& {Wang}, J.~X. 2011, \apj, 738, 147

\bibitem[{{Smith} \& {Dwek}(1998)}]{1998ApJ...503..831S}
{Smith}, R.~K. \& {Dwek}, E. 1998, \apj, 503, 831

\bibitem[{{Summons}(2007)}]{summons_thesis}
{Summons}, D.~P. 2007, PhD thesis, University of Southampton, give Publisher
  address, give URL

\bibitem[{{Summons} {et~al.}(2007){Summons}, {Ar{\'e}valo}, {McHardy},
  {Uttley}, \& {Bhaskar}}]{Summons2007}
{Summons}, D.~P., {Ar{\'e}valo}, P., {McHardy}, I.~M., {Uttley}, P., \&
  {Bhaskar}, A. 2007, \mnras, 378, 649

\bibitem[{{Tanaka} {et~al.}(1995){Tanaka}, {Nandra}, {Fabian}, {Inoue},
  {Otani}, {Dotani}, {Hayashida}, {Iwasawa}, {Kii}, {Kunieda}, {Makino}, \&
  {Matsuoka}}]{1995Natur.375..659T}
{Tanaka}, Y., {Nandra}, K., {Fabian}, A.~C., {et~al.} 1995, \nat, 375, 659

\bibitem[{{Terashima} \& {Wilson}(2003)}]{2003ApJ...583..145T}
{Terashima}, Y. \& {Wilson}, A.~S. 2003, \apj, 583, 145

\bibitem[{{Timmer} \& {Koenig}(1995)}]{1995A&A...300..707T}
{Timmer}, J. \& {Koenig}, M. 1995, \aap, 300, 707

\bibitem[{{Trippe} {et~al.}(2008){Trippe}, {Crenshaw}, {Deo}, \&
  {Dietrich}}]{2008Trippe}
{Trippe}, M.~L., {Crenshaw}, D.~M., {Deo}, R., \& {Dietrich}, M. 2008, \aj,
  135, 2048

\bibitem[{{Tzanavaris} {et~al.}(2019){Tzanavaris}, {Yaqoob}, {LaMassa},
  {Yukita}, \& {Ptak}}]{Tzanavaris2019}
{Tzanavaris}, P., {Yaqoob}, T., {LaMassa}, S., {Yukita}, M., \& {Ptak}, A.
  2019, \apj, 885, 62

\bibitem[{{Uematsu} {et~al.}(2021){Uematsu}, {Ueda}, {Tanimoto}, {Kawamuro},
  {Setoguchi}, {Ogawa}, {Yamada}, \& {Odaka}}]{2021Uematsu}
{Uematsu}, R., {Ueda}, Y., {Tanimoto}, A., {et~al.} 2021, \apj, 913, 17

\bibitem[{{Ulrich} {et~al.}(1997){Ulrich}, {Maraschi}, \& {Urry}}]{1997Ulrich}
{Ulrich}, M.-H., {Maraschi}, L., \& {Urry}, C.~M. 1997, \araa, 35, 445

\bibitem[{{Uttley} {et~al.}(2014){Uttley}, {Cackett}, {Fabian}, {Kara}, \&
  {Wilkins}}]{2014A&ARv..22...72U}
{Uttley}, P., {Cackett}, E.~M., {Fabian}, A.~C., {Kara}, E., \& {Wilkins},
  D.~R. 2014, \aapr, 22, 72

\bibitem[{{Uttley} \& {McHardy}(2005)}]{Uttley2005}
{Uttley}, P. \& {McHardy}, I.~M. 2005, \mnras, 363, 586

\bibitem[{{Vaughan} {et~al.}(2003{\natexlab{a}}){Vaughan}, {Edelson},
  {Warwick}, \& {Uttley}}]{Vaughan2003variance}
{Vaughan}, S., {Edelson}, R., {Warwick}, R.~S., \& {Uttley}, P.
  2003{\natexlab{a}}, \mnras, 345, 1271

\bibitem[{{Vaughan} {et~al.}(2003{\natexlab{b}}){Vaughan}, {Fabian}, \&
  {Nandra}}]{2003Vaughan}
{Vaughan}, S., {Fabian}, A.~C., \& {Nandra}, K. 2003{\natexlab{b}}, \mnras,
  339, 1237

\bibitem[{{Wang} {et~al.}(2011){Wang}, {Fabbiano}, {Elvis}, {Risaliti},
  {Mundell}, {Karovska}, \& {Zezas}}]{2011ApJ...736...62W}
{Wang}, J., {Fabbiano}, G., {Elvis}, M., {et~al.} 2011, \apj, 736, 62

\bibitem[{{Wang} {et~al.}(2014){Wang}, {Nardini}, {Fabbiano}, {Karovska},
  {Elvis}, {Pellegrini}, {Max}, {Risaliti}, {U}, \&
  {Zezas}}]{2014ApJ...781...55W}
{Wang}, J., {Nardini}, E., {Fabbiano}, G., {et~al.} 2014, \apj, 781, 55

\bibitem[{{Weaver} {et~al.}(1996){Weaver}, {Nousek}, {Yaqoob}, {Mushotzky},
  {Makino}, \& {Otani}}]{1996Weaver}
{Weaver}, K.~A., {Nousek}, J., {Yaqoob}, T., {et~al.} 1996, \apj, 458, 160

\bibitem[{{Weaver} {et~al.}(2020){Weaver}, {Williamson}, {Jorstad}, {Marscher},
  {Larionov}, {Raiteri}, {Villata}, {Acosta-Pulido}, {Bachev}, {Baida},
  {Balonek}, {Ben{\'\i}tez}, {Borman}, {Bozhilov}, {Carnerero}, {Carosati},
  {Chen}, {Damljanovic}, {Dhiman}, {Dougherty}, {Ehgamberdiev}, {Grishina},
  {Gupta}, {Hart}, {Hiriart}, {Hsiao}, {Ibryamov}, {Joner}, {Kimeridze},
  {Kopatskaya}, {Kurtanidze}, {Kurtanidze}, {Larionova}, {Matsumoto},
  {Matsumura}, {Minev}, {Mirzaqulov}, {Morozova}, {Nikiforova}, {Nikolashvili},
  {Ovcharov}, {Rizzi}, {Sadun}, {Savchenko}, {Semkov}, {Slater}, {Smith},
  {Stojanovic}, {Strigachev}, {Troitskaya}, {Troitsky}, {Tsai}, {Vince},
  {Valcheva}, {Vasilyev}, {Zaharieva}, \& {Zhovtan}}]{2020Weaver}
{Weaver}, Z.~R., {Williamson}, K.~E., {Jorstad}, S.~G., {et~al.} 2020, \apj,
  900, 137

\bibitem[{{Webb} {et~al.}(2020){Webb}, {Coriat}, {Traulsen}, {Ballet}, {Motch},
  {Carrera}, {Koliopanos}, {Authier}, {de la Calle}, {Ceballos}, {Colomo},
  {Chuard}, {Freyberg}, {Garcia}, {Kolehmainen}, {Lamer}, {Lin}, {Maggi},
  {Michel}, {Page}, {Page}, {Perea-Calderon}, {Pineau}, {Rodriguez}, {Rosen},
  {Santos Lleo}, {Saxton}, {Schwope}, {Tom{\'a}s}, {Watson}, \&
  {Zakardjian}}]{2020Webb}
{Webb}, N.~A., {Coriat}, M., {Traulsen}, I., {et~al.} 2020, \aap, 641, A136

\bibitem[{{Weisskopf} {et~al.}(2000){Weisskopf}, {Tananbaum}, {Van Speybroeck},
  \& {O'Dell}}]{2000SPIE.4012....2W}
{Weisskopf}, M.~C., {Tananbaum}, H.~D., {Van Speybroeck}, L.~P., \& {O'Dell},
  S.~L. 2000, Society of Photo-Optical Instrumentation Engineers (SPIE)
  Conference Series, Vol. 4012, {Chandra X-ray Observatory (CXO): overview},
  ed. J.~E. {Truemper} \& B.~{Aschenbach}, 2--16

\bibitem[{{Winter} {et~al.}(2009){Winter}, {Mushotzky}, {Reynolds}, \&
  {Tueller}}]{2009WinterL}
{Winter}, L.~M., {Mushotzky}, R.~F., {Reynolds}, C.~S., \& {Tueller}, J. 2009,
  \apj, 690, 1322

\bibitem[{{XRISM Science Team}(2020)}]{2020XRISM}
{XRISM Science Team}. 2020, arXiv e-prints, arXiv:2003.04962

\bibitem[{{Yaqoob} {et~al.}(1995){Yaqoob}, {Edelson}, {Weaver}, {Warwick},
  {Mushotzky}, {Serlemitsos}, \& {Holt}}]{1995ApJ...453L..81Y}
{Yaqoob}, T., {Edelson}, R., {Weaver}, K.~A., {et~al.} 1995, \apjl, 453, L81

\bibitem[{{Yaqoob} \& {Padmanabhan}(2004)}]{2004ApJ...604...63Y}
{Yaqoob}, T. \& {Padmanabhan}, U. 2004, \apj, 604, 63

\bibitem[{{Yaqoob} {et~al.}(1996){Yaqoob}, {Serlemitsos}, {Turner}, {George},
  \& {Nandra}}]{1996ApJ...470L..27Y}
{Yaqoob}, T., {Serlemitsos}, P.~J., {Turner}, T.~J., {George}, I.~M., \&
  {Nandra}, K. 1996, \apjl, 470, L27

\bibitem[{{Yi} {et~al.}(2021){Yi}, {Wang}, {Shu}, {Fabbiano}, {Pappalardo},
  {Wang}, \& {Yu}}]{2021ApJ...908..156Y}
{Yi}, H., {Wang}, J., {Shu}, X., {et~al.} 2021, \apj, 908, 156

\bibitem[{{Zhang} {et~al.}(2019){Zhang}, {Du}, {Smith}, {Zhao}, {Hu}, {Xiao},
  {Li}, {Huang}, {Wang}, {Bai}, {Ho}, \& {Wang}}]{2019Zhang}
{Zhang}, Z.-X., {Du}, P., {Smith}, P.~S., {et~al.} 2019, \apj, 876, 49

\bibitem[{{Zhu} {et~al.}(2021){Zhu}, {Timlin}, \& {Brandt}}]{2021Zhu}
{Zhu}, S.~F., {Timlin}, J.~D., \& {Brandt}, W.~N. 2021, \mnras, 505, 1954

\bibitem[{{Zoghbi} {et~al.}(2012){Zoghbi}, {Fabian}, {Reynolds}, \&
  {Cackett}}]{Zoghbi2012}
{Zoghbi}, A., {Fabian}, A.~C., {Reynolds}, C.~S., \& {Cackett}, E.~M. 2012,
  \mnras, 422, 129

\bibitem[{{Zoghbi} {et~al.}(2010){Zoghbi}, {Fabian}, {Uttley}, {Miniutti},
  {Gallo}, {Reynolds}, {Miller}, \& {Ponti}}]{Zoghbi2010}
{Zoghbi}, A., {Fabian}, A.~C., {Uttley}, P., {et~al.} 2010, \mnras, 401, 2419

\bibitem[{{Zoghbi} {et~al.}(2019){Zoghbi}, {Miller}, \&
  {Cackett}}]{2019ApJ...884...26Z}
{Zoghbi}, A., {Miller}, J.~M., \& {Cackett}, E. 2019, \apj, 884, 26

\end{thebibliography}
% \end{thebibliography}

% \newpage
\appendix \label{appendix}

\section{Observational details and best-fitting parameters} \label{ap:InfoObs}

Table\,\ref{t:ObsInfo} provides a summary of each observation used in this paper, together with the best-fitting parameters. Here we only show a random number of entries as an example. The full version of the table is available in electronic format at the CDS via http://cdsweb.u-strasbg.fr/cgi-bin/qcat?J/A+A/ and also reports the fluxes uncertainties.

\begin{table*}
\caption{Summary of the observations used in our work and its best-fitting parameters } 
%\begin{adjustbox}{width=\textwidth}
\scriptsize
\centering
\scalebox{0.66}{
\begin{tabular}{lllrlrlrrrrrrrrrrrrrr}
 \hline
 \hline
 \noalign{\smallskip}
 \multicolumn{1}{c}{Source} & \multicolumn{1}{c}{Observatory} &  \multicolumn{1}{c}{Instrument} &     \multicolumn{1}{c}{OBSID} &  \multicolumn{1}{c}{ Date} & \multicolumn{1}{c}{Exposure}& \multicolumn{1}{c}{Spectrum} & \multicolumn{1}{c}{z} &    \multicolumn{1}{c}{ $\Gamma$ }&  \multicolumn{1}{c}{ $\Gamma_{\rm lo}$} &  \multicolumn{1}{c}{ $\Gamma{\rm up}$} &       \multicolumn{1}{c}{ $N_{\rm H}$} &\multicolumn{1}{c}{ $N_{\rm H,lo}$} & \multicolumn{1}{c}{$N_{\rm H,up}$} & \multicolumn{1}{c}{$\sigma$} &\multicolumn{1}{c}{$\sigma_{lo}$ }& \multicolumn{1}{c}{$\sigma_{up}$}& \multicolumn{1}{c}{ $\log F_{\rm 2-10 \:keV}$} &      \multicolumn{1}{c}{ $\log F_{\rm Fe K \alpha}$ }& \multicolumn{1}{c}{Fit statistic }&  \multicolumn{1}{c}{d.o.f }\\
 
 \noalign{\smallskip}
   &  &   &      &    &   \multicolumn{1}{c}{(ks)} && &   &   &  & \multicolumn{1}{c}{$(\rm cm^{-2})$} & \multicolumn{1}{c}{$(\rm cm^{-2})$} &  \multicolumn{1}{c}{($\rm cm^{-2})$} & \multicolumn{1}{c}{(eV)} & \multicolumn{1}{c}{(eV)}  & \multicolumn{1}{c}{(eV)}& \multicolumn{1}{c}{$(\rm erg\:cm^{-2}\: s^{-1}$)} &       \multicolumn{1}{c}{$(\rm erg\:cm^{-2}\: s^{-1})$} &  &   \\
 \noalign{\smallskip}
\multicolumn{1}{c}{(1)} & \multicolumn{1}{c}{(2)}& \multicolumn{1}{c}{(3)} & \multicolumn{1}{c}{(4)} & \multicolumn{1}{c}{(5)} & \multicolumn{1}{c}{(6)} & \multicolumn{1}{c}{(7)} & \multicolumn{1}{c}{(8)} & \multicolumn{1}{c}{(9)} & \multicolumn{1}{c}{(10)} & \multicolumn{1}{c}{(11)} & \multicolumn{1}{c}{(12)} & \multicolumn{1}{c}{(13)} & \multicolumn{1}{c}{(14)}
& \multicolumn{1}{c}{(15)}& \multicolumn{1}{c}{(16)}
& \multicolumn{1}{c}{(17)}& \multicolumn{1}{c}{(18)}
& \multicolumn{1}{c}{(19)} & \multicolumn{1}{c}{(20)}
& \multicolumn{1}{c}{(21)}\\
\noalign{\smallskip}
\hline
\noalign{\smallskip}
      3C120 &     CHANDRA & ACIS-S HETG &      3015 & 2001-12-21 &     57.22 &  grating & 0.033 &  1.63 &     1.61 &     1.67 &   20.0 &    -99 &  20.81 &  20.73 &    12.54 &    31.34 &   -10.18 & -12.48 &        2235.21 &   1977 \\
      3C120 &     CHANDRA & ACIS-S HETG &     16221 & 2014-12-19 &     77.72 &  grating & 0.033 &  1.73 &      1.7 &     1.75 &   20.0 &   21.7 &    -99 &   80.0 &    64.98 &      -99 &   -10.08 & -13.17 &        2181.33 &   1977 \\
      3C120 &     CHANDRA & ACIS-S HETG &     17564 & 2014-12-22 &     30.29 &  grating & 0.033 &  1.68 &     1.64 &     1.71 &   20.0 &    -99 &  20.86 &  22.33 &    13.29 &    32.98 &   -10.09 & -12.37 &        2172.18 &   1977 \\
      3C120 &     CHANDRA & ACIS-S HETG &     17565 & 2014-12-27 &     43.31 &  grating & 0.033 &  1.61 &     1.56 &      1.7 &  20.56 &    -99 &  21.48 &   0.03 &     80.0 &      -99 &   -10.25 & -12.62 &        2205.12 &   1977 \\
      3C120 &     CHANDRA & ACIS-S HETG &     17576 & 2015-01-27 &     43.32 &  grating & 0.033 &  1.73 &      1.7 &     1.77 &   20.0 &    -99 &  21.04 &  10.02 &     0.43 &    20.73 &   -10.07 & -12.58 &        2233.64 &   1977 \\
      3C273 &     CHANDRA & ACIS-S HETG &      3457 & 2002-06-05 &     24.85 &  grating & 0.158 &  1.65 &     1.61 &     1.67 &   20.0 &   21.7 &    -99 &   0.12 &     80.0 &      -99 &    -9.99 & -13.11 &        2042.96 &   1977 \\
      3C273 &     CHANDRA & ACIS-S HETG &     19867 & 2017-06-26 &     26.91 &  grating & 0.158 &  1.64 &     1.56 &     1.73 &  21.76 &   21.5 &  21.92 &   80.0 &    50.88 &      -99 &   -10.01 & -13.35 &        2167.44 &   1977 \\
      3C273 &     CHANDRA & ACIS-S HETG &     20709 & 2018-07-04 &     29.57 &  grating & 0.158 &  1.56 &     1.53 &      1.6 &   20.0 &   21.7 &    -99 &   0.04 &     80.0 &      -99 &   -10.09 & -12.88 &        2240.57 &   1977 \\
   4C+74.26 &     CHANDRA & ACIS-S HETG &      4000 & 2003-10-06 &     37.18 &  grating & 0.100 &  1.57 &     1.44 &      1.7 &  21.45 &    -99 &  21.82 &   0.13 &     80.0 &      -99 &   -10.49 & -13.21 &        2145.55 &   1977 \\
       CenA &     CHANDRA & ACIS-S HETG &      1600 & 2001-05-09 &     46.85 &  grating & 0.002 &  1.53 &     1.49 &     1.57 &  22.98 &  22.97 &  22.98 &  22.27 &    17.29 &     28.0 &    -9.16 & -11.57 &        2143.59 &   1977 \\
       CenA &     CHANDRA & ACIS-S HETG &      1601 & 2001-05-21 &     51.51 &  grating & 0.002 &  1.51 &     1.48 &     1.55 &  22.99 &  22.98 &  22.99 &  10.63 &     4.13 &    16.97 &    -9.09 & -11.67 &        2046.16 &   1977 \\
    IC4329A &     CHANDRA & ACIS-S HETG &      2177 & 2001-08-26 &     59.09 &  grating & 0.016 &   1.6 &     1.56 &     1.64 &  21.18 &  20.65 &  21.42 &   80.0 &    52.79 &      -99 &    -9.62 & -12.24 &        1964.62 &   1977 \\
    IC4329A &     CHANDRA & ACIS-S HETG &     20070 & 2017-06-06 &     91.85 &  grating & 0.016 &  1.88 &     1.84 &     1.92 &  21.91 &  21.85 &  21.97 &   80.0 &    66.15 &      -99 &    -9.75 & -12.36 &        2124.94 &   1977 \\
    IC4329A &     CHANDRA & ACIS-S HETG &     20096 & 2017-06-14 &     19.76 &  grating & 0.016 &   1.8 &     1.72 &     1.88 &  21.77 &  21.56 &  21.92 &  15.21 &     7.61 &    28.42 &    -9.73 &  -12.2 &        2215.67 &   1977 \\
    IC4329A &     CHANDRA & ACIS-S HETG &     20097 & 2017-06-17 &     16.81 &  grating & 0.016 &  1.79 &     1.71 &     1.88 &  21.74 &  21.47 &   21.9 &  12.02 &     6.02 &    26.15 &    -9.74 & -12.27 &        2163.42 &   1977 \\
        M81 &     CHANDRA & ACIS-S HETG &      6346 & 2005-07-14 &     54.48 &  grating & 0.000 &   1.9 &     1.84 &     1.96 &   20.0 &   21.7 &    -99 &  23.91 &     5.23 &    57.47 &    -10.8 & -13.38 &        2156.21 &   1977 \\
        M81 &     CHANDRA & ACIS-S HETG &      6347 & 2005-07-14 &     63.87 &  grating & 0.000 &  1.69 &     1.63 &     1.75 &   20.0 &   21.7 &    -99 &   80.0 &    57.04 &      -99 &    -10.8 & -13.23 &        2166.87 &   1977 \\
        M81 &     CHANDRA & ACIS-S HETG &      5600 & 2005-08-14 &     35.96 &  grating & 0.000 &  1.79 &      1.7 &      1.9 &   20.0 &    -99 &  21.49 &  32.89 &    16.84 &      -99 &   -10.86 & -13.07 &        1942.80 &   1977 \\
        M81 &     CHANDRA & ACIS-S HETG &      6892 & 2006-02-08 &     14.76 &  grating & 0.000 &   1.9 &     1.81 &     2.08 &   20.0 &   21.7 &    -99 &   2.61 &     0.05 &     9.11 &   -10.91 & -12.65 &        1412.35 &   1977 \\
        M81 &     CHANDRA & ACIS-S HETG &      6893 & 2006-03-05 &     14.76 &  grating & 0.000 &  1.88 &     1.75 &     2.19 &   20.4 &    -99 &  21.95 &   0.01 &     80.0 &      -99 &   -10.82 & -13.12 &        1511.26 &   1977 \\
        M81 &     CHANDRA & ACIS-S HETG &      6895 & 2006-04-24 &     14.56 &  grating & 0.000 &  1.75 &     1.58 &     2.07 &  21.15 &    -99 &  22.02 &   0.01 &     80.0 &      -99 &   -10.83 & -13.27 &        1479.50 &   1977 \\
        M81 &     CHANDRA & ACIS-S HETG &      6897 & 2006-06-09 &     14.76 &  grating & 0.000 &  2.18 &     1.85 &     2.53 &  21.95 &    -99 &  22.27 &  38.72 &    12.06 &      -99 &   -10.82 & -12.82 &        1429.57 &   1977 \\
        M81 &     CHANDRA & ACIS-S HETG &      6901 & 2006-08-12 &     14.76 &  grating & 0.000 &  1.87 &     1.78 &     1.96 &   20.0 &    -99 &  21.06 &   9.43 &     0.01 &    68.06 &   -10.52 & -12.84 &        1987.40 &   1977 \\
MCG-6-30-15 &     CHANDRA & ACIS-S HETG &      4760 & 2004-05-19 &    169.49 &  grating & 0.008 &  1.76 &     1.72 &     1.81 &  21.64 &  21.48 &  21.75 &   80.0 &    75.21 &      -99 &   -10.19 & -12.84 &        2152.29 &   1977 \\
MCG-6-30-15 &     CHANDRA & ACIS-S HETG &      4759 & 2004-05-24 &    158.54 &  grating & 0.008 &  1.84 &      1.8 &     1.89 &  21.69 &  21.55 &  21.79 &   30.4 &    19.85 &      -99 &   -10.17 & -12.59 &        1911.14 &   1977 \\
MCG-6-30-15 &     CHANDRA & ACIS-S HETG &      4762 & 2004-05-27 &     37.55 &  grating & 0.008 &  1.89 &     1.79 &     1.98 &  21.53 &   20.9 &  21.77 &   80.0 &    33.05 &      -99 &   -10.16 & -13.08 &        2239.82 &   1977 \\
MCG-6-30-15 &  XMM-NEWTON &          PN & 693781401 & 2013-02-02 &     48.92 & circular & 0.008 &  1.79 &     1.77 &      1.8 &  21.54 &  21.46 &  21.61 &   30.0 &    23.48 &      -99 &   -10.12 & -12.51 &         227.86 &    164 \\
 MR2251-178 &     CHANDRA & ACIS-S HETG &     12828 & 2011-09-26 &    160.47 &  grating & 0.064 &  1.55 &      1.5 &      1.6 &  21.46 &  21.17 &  21.64 &  21.75 &     11.8 &    37.93 &   -10.28 & -13.07 &        2074.63 &   1977 \\
 MR2251-178 &     CHANDRA & ACIS-S HETG &     12829 & 2011-09-29 &    184.60 &  grating & 0.064 &  1.53 &     1.49 &     1.58 &  21.38 &  21.04 &  21.57 &    0.1 &     80.0 &      -99 &   -10.28 & -13.41 &        2189.60 &   1977 \\
 MR2251-178 &  XMM-NEWTON &          PN & 670120301 & 2011-11-13 &    128.31 & circular & 0.063 &  1.64 &     1.63 &     1.65 &  21.12 &  21.07 &  21.21 &   5.99 &     0.11 &      -99 &    -9.99 & -12.76 &         201.71 &    165 \\
 MR2251-178 &  XMM-NEWTON &          PN & 670120401 & 2011-11-15 &    132.98 & circular & 0.063 &  1.65 &     1.65 &     1.66 &  21.27 &  21.23 &  21.34 &   5.85 &     0.15 &      -99 &   -10.03 & -12.87 &         197.18 &    165 \\
    MRK1040 &     CHANDRA & ACIS-S HETG &     15075 & 2014-02-25 &     28.57 &  grating & 0.017 &  1.63 &     1.58 &     1.68 &   20.0 &   21.7 &    -99 &  55.77 &    35.58 &      -99 &    -10.3 & -12.52 &        2301.29 &   1977 \\
     MRK290 &     CHANDRA & ACIS-S HETG &      4399 & 2003-06-29 &     83.77 &  grating & 0.029 &  1.54 &      1.5 &      1.6 &   20.0 &    -99 &  21.14 &  54.87 &    25.52 &      -99 &   -10.77 & -13.14 &        2134.73 &   1977 \\
     MRK290 &     CHANDRA & ACIS-S HETG &      4442 & 2003-07-17 &     49.37 &  grating & 0.029 &   1.7 &     1.65 &      1.8 &   20.0 &    -99 &   21.4 &  45.05 &    29.53 &    75.29 &   -10.59 & -13.48 &        2208.43 &   1977 \\
     MRK509 &  XMM-NEWTON &          PN & 130720101 & 2000-10-25 &     31.64 & circular & 0.034 &  1.62 &     1.61 &     1.63 &   20.0 &   21.7 &    -99 &   30.0 &    18.27 &      -99 &   -10.13 &  -12.4 &         190.34 &    163 \\
     MRK509 &  XMM-NEWTON &          PN & 130720201 & 2001-04-20 &     44.41 & circular & 0.034 &  1.73 &     1.72 &     1.74 &   20.0 &   21.7 &    -99 &   0.27 &     3.33 &      -99 &    -10.0 & -12.43 &         240.60 &    165 \\
     MRK509 &  XMM-NEWTON &          PN & 306090401 & 2006-04-25 &     69.95 & circular & 0.034 &  1.78 &     1.78 &     1.79 &   20.0 &   21.7 &    -99 &   30.0 &    24.93 &      -99 &    -9.97 & -12.41 &         250.79 &    165 \\
     MRK766 &     CHANDRA & ACIS-S HETG &      1597 & 2001-05-07 &     89.08 &  grating & 0.013 &  1.96 &     1.93 &     1.99 &   20.0 &    -99 &  20.75 &  19.65 &    10.45 &    47.31 &   -10.46 & -13.06 &        2125.32 &   1977 \\
     MRK766 &     CHANDRA & ACIS-S HETG &     16310 & 2014-06-30 &     60.15 &  grating & 0.013 &  1.87 &     1.83 &     1.94 &   20.0 &    -99 &  21.31 &   0.12 &     80.0 &      -99 &   -10.38 & -13.28 &        2110.10 &   1977 \\
     MRK766 &     CHANDRA & ACIS-S HETG &     16311 & 2014-07-08 &    123.29 &  grating & 0.013 &  1.98 &     1.95 &     2.02 &   20.0 &    -99 &  21.02 &   4.46 &     0.29 &     13.8 &   -10.32 & -13.01 &        2093.02 &   1977 \\
   % NGC1275 &     CHANDRA & ACIS-S HETG &       333 & 1999-10-10 &     26.76 &  grating & 0.018 &  2.07 &     1.93 &     2.25 &  21.36 &    -99 &  21.85 &   80.0 &    48.23 &      -99 &    -10.7 & -23.77 &        2154.92 &   1977 \\
    NGC1275 &     CHANDRA & ACIS-S HETG &     20451 & 2017-10-30 &     36.47 &  grating & 0.018 &   2.0 &     1.94 &     2.09 &   20.0 &    -99 &  21.42 &   0.14 &     80.0 &      -99 &   -10.38 & -12.86 &        2166.28 &   1977 \\
    NGC1275 &     CHANDRA & ACIS-S HETG &     20449 & 2017-11-06 &     45.32 &  grating & 0.018 &  1.99 &     1.94 &     2.05 &   20.0 &    -99 &  21.25 &  29.33 &     5.94 &    53.29 &   -10.39 & -12.83 &        2312.66 &   1977 \\
    NGC1275 &     CHANDRA & ACIS-S HETG &     20824 & 2017-12-02 &     49.25 &  grating & 0.018 &  2.05 &     1.95 &     2.16 &  21.39 &    -99 &  21.73 &   80.0 &    39.11 &      -99 &   -10.36 & -13.12 &        2124.90 &   1977 \\
    NGC2992 &  XMM-NEWTON &          PN & 147920301 & 2003-05-19 &     28.92 & circular & 0.007 &  1.65 &     1.64 &     1.67 &  21.94 &  21.92 &  21.97 &   30.0 &     3.86 &      -99 &    -9.65 & -12.05 &         248.55 &    166 \\
    NGC2992 &  XMM-NEWTON &          PN & 654910501 & 2010-05-26 &     55.92 & circular & 0.007 &  1.76 &     1.73 &      1.8 &  22.11 &  22.07 &  22.14 &    0.1 &    24.15 &      -99 &   -10.48 & -12.33 &         209.93 &    156 \\
    NGC2992 &  XMM-NEWTON &          PN & 654910601 & 2010-06-05 &     55.92 & circular & 0.007 &  1.73 &     1.68 &     1.79 &  22.05 &  21.97 &  22.12 &   30.0 &    23.08 &      -99 &   -10.94 & -12.37 &         176.78 &    137 \\
    NGC2992 &  XMM-NEWTON &          PN & 654910701 & 2010-11-08 &     55.91 & circular & 0.007 &  1.47 &     1.43 &     1.52 &  21.81 &  21.69 &   21.9 &   30.0 &    24.84 &      -99 &   -10.91 & -12.38 &         154.64 &    152 \\
    NGC2992 &  XMM-NEWTON &          PN & 654911001 & 2010-12-08 &     60.52 & circular & 0.007 &  1.59 &     1.55 &     1.64 &  22.01 &  21.95 &  22.07 &   1.09 &     8.49 &      -99 &   -10.85 & -12.47 &         222.91 &    149 \\
    NGC3516 &     CHANDRA & ACIS-S HETG &      8452 & 2006-10-09 &     19.83 &  grating & 0.009 &  1.71 &      1.6 &     1.83 &  22.12 &  21.99 &  22.22 &  33.18 &    18.23 &    60.96 &   -10.04 & -12.29 &        2170.87 &   1977 \\
    NGC3516 &     CHANDRA & ACIS-S HETG &      7282 & 2006-10-10 &     41.41 &  grating & 0.009 &   1.6 &     1.51 &      1.7 &  22.01 &  21.88 &  22.12 &  63.83 &    40.05 &      -99 &   -10.19 & -12.38 &        2206.09 &   1977 \\
    NGC3516 &     CHANDRA & ACIS-S HETG &      8451 & 2006-10-11 &     47.36 &  grating & 0.009 &  1.77 &      1.7 &     1.84 &  22.01 &  21.92 &  22.09 &  24.71 &    16.51 &     39.4 &    -9.99 & -12.33 &        2265.66 &   1977 \\
    NGC3516 &     CHANDRA & ACIS-S HETG &      8450 & 2006-10-12 &     38.51 &  grating & 0.009 &  1.93 &     1.85 &     2.01 &  22.09 &   22.0 &  22.17 &  63.49 &    50.27 &    79.04 &    -9.99 & -12.16 &        2106.09 &   1977 \\
    NGC3516 &     CHANDRA & ACIS-S HETG &      7281 & 2006-10-14 &     42.44 &  grating & 0.009 &   1.5 &     1.41 &     1.58 &  22.06 &  21.94 &  22.15 &  31.72 &    23.55 &    42.27 &   -10.14 & -12.24 &        2158.91 &   1977 \\
    NGC3783 &     CHANDRA & ACIS-S HETG &       373 & 2000-01-20 &     56.43 &  grating & 0.010 &  1.53 &     1.46 &     1.59 &  21.57 &  21.26 &  21.74 &  26.48 &    16.31 &    37.03 &   -10.02 & -12.18 &        2114.13 &   1977 \\
    NGC3783 &     CHANDRA & ACIS-S HETG &      2090 & 2001-02-24 &    165.66 &  grating & 0.010 &  1.41 &     1.37 &     1.45 &  21.54 &  21.36 &  21.67 &  17.92 &    14.12 &    22.55 &    -10.1 & -12.27 &        2190.15 &   1977 \\
    NGC3783 &     CHANDRA & ACIS-S HETG &      2091 & 2001-02-27 &    168.86 &  grating & 0.010 &  1.42 &     1.38 &     1.46 &   21.6 &  21.44 &  21.71 &  24.13 &    19.89 &    29.11 &    -10.1 &  -12.2 &        2061.25 &   1977 \\
    NGC3783 &     CHANDRA & ACIS-S HETG &      2092 & 2001-03-10 &    165.45 &  grating & 0.010 &  1.46 &     1.42 &      1.5 &   21.5 &   21.3 &  21.64 &   15.0 &    11.53 &    18.76 &   -10.09 &  -12.2 &        2283.15 &   1977 \\
    NGC3783 &     CHANDRA & ACIS-S HETG &      2093 & 2001-03-31 &    166.13 &  grating & 0.010 &  1.49 &     1.47 &     1.52 &  20.19 &    -99 &  21.06 &  12.54 &     9.43 &    15.73 &    -9.95 & -12.19 &        2297.01 &   1977 \\
    NGC3783 &     CHANDRA & ACIS-S HETG &      2094 & 2001-06-26 &    166.18 &  grating & 0.010 &   1.5 &     1.46 &     1.53 &  21.65 &  21.54 &  21.75 &  14.85 &    11.15 &    18.76 &    -10.0 & -12.27 &        2154.12 &   1977 \\
    NGC4051 &     CHANDRA & ACIS-S HETG &       859 & 2000-03-24 &     79.77 &  grating & 0.002 &  1.58 &     1.54 &     1.63 &   20.0 &   21.7 &    -99 &  64.87 &    45.93 &      -99 &   -10.61 & -12.63 &        2241.61 &   1977 \\
    NGC4051 &     CHANDRA & ACIS-S HETG &     10777 & 2008-11-06 &     27.39 &  grating & 0.002 &  1.95 &     1.84 &     2.08 &   21.3 &    -99 &  21.75 &   22.6 &    15.54 &    32.18 &   -10.33 & -12.43 &        2165.96 &   1977 \\
    NGC4051 &     CHANDRA & ACIS-S HETG &     10775 & 2008-11-08 &     30.38 &  grating & 0.002 &  1.93 &     1.89 &     1.98 &   20.0 &   21.7 &    -99 &   80.0 &    62.98 &      -99 &   -10.29 & -12.52 &        2153.96 &   1977 \\
    NGC4051 &     CHANDRA & ACIS-S HETG &     10403 & 2008-11-09 &     37.53 &  grating & 0.002 &  1.89 &     1.76 &     2.01 &  21.61 &  20.65 &  21.89 &   61.1 &    38.38 &      -99 &   -10.44 & -12.92 &        2101.34 &   1977 \\
    NGC4051 &     CHANDRA & ACIS-S HETG &     10404 & 2008-11-12 &     19.76 &  grating & 0.002 &  1.72 &     1.63 &     1.81 &   20.0 &   21.7 &    -99 &  26.83 &    16.51 &    39.87 &   -10.68 & -12.42 &        1862.67 &   1977 \\
    NGC4051 &     CHANDRA & ACIS-S HETG &     10780 & 2008-11-25 &     26.00 &  grating & 0.002 &  1.71 &     1.66 &     1.77 &   20.0 &   21.7 &    -99 &   0.39 &     80.0 &      -99 &    -10.4 & -12.47 &        2073.48 &   1977 \\
    NGC4051 &     CHANDRA & ACIS-S HETG &     10824 & 2008-11-30 &      9.00 &  grating & 0.002 &  1.95 &     1.78 &      2.2 &   21.5 &    -99 &  22.01 &   0.01 &     80.0 &      -99 &   -10.41 & -12.51 &        1812.76 &   1977 \\
    NGC4051 &  XMM-NEWTON &          PN & 606321301 & 2009-05-15 &     32.64 & circular & 0.002 &  1.98 &     1.95 &     2.02 &  21.37 &  21.11 &  21.54 &   30.0 &     22.8 &      -99 &   -10.34 & -12.34 &         253.29 &    149 \\
    NGC4051 &  XMM-NEWTON &          PN & 606321701 & 2009-05-27 &     44.92 & circular & 0.002 &  1.73 &     1.69 &     1.77 &  21.91 &  21.83 &  21.97 &   30.0 &     8.24 &      -99 &   -10.56 & -12.52 &         200.18 &    150 \\
    NGC4051 &     CHANDRA & ACIS-S HETG &     18768 & 2016-02-11 &     91.88 &  grating & 0.002 &  1.99 &     1.91 &     2.06 &   21.3 &    -99 &   21.6 &  14.93 &     5.98 &    27.03 &   -10.34 & -12.68 &        2036.41 &   1977 \\
    NGC4051 &     CHANDRA & ACIS-S HETG &     17102 & 2016-02-26 &    112.88 &  grating & 0.002 &  1.86 &     1.78 &     1.93 &  21.63 &  21.32 &  21.81 &  43.77 &    35.64 &    57.03 &   -10.46 & -12.65 &        2090.69 &   1977 \\
    NGC4051 &     CHANDRA & ACIS-S HETG &     18785 & 2016-02-28 &     65.18 &  grating & 0.002 &  1.84 &      1.8 &     1.88 &   20.0 &    -99 &  20.81 &   80.0 &    60.99 &      -99 &   -10.51 & -12.72 &        2201.04 &   1977 \\
    NGC4051 &     CHANDRA & ACIS-S HETG &     17103 & 2016-03-01 &     62.42 &  grating & 0.002 &  1.82 &     1.69 &     1.94 &  21.64 &  20.95 &  21.89 &   80.0 &    58.77 &      -99 &   -10.61 & -12.76 &        2159.47 &   1977 \\
    NGC4051 &     CHANDRA & ACIS-S HETG &     18786 & 2016-03-02 &     64.93 &  grating & 0.002 &   1.7 &     1.65 &     1.74 &   20.0 &   21.7 &    -99 &   0.88 &     80.0 &      -99 &   -10.58 & -13.21 &        2179.06 &   1977 \\
    NGC4051 &     CHANDRA & ACIS-S HETG &     18787 & 2016-04-14 &     25.36 &  grating & 0.002 &  1.88 &     1.82 &     1.94 &   20.0 &   21.7 &    -99 &   80.0 &     40.1 &      -99 &   -10.43 & -13.23 &        2061.63 &   1977 \\
    NGC4151 &     CHANDRA & ACIS-S HETG &      7830 & 2007-07-21 &     49.34 &  grating & 0.003 &  1.42 &     1.38 &     1.47 &  22.49 &  22.47 &  22.51 &  29.88 &    18.74 &    48.33 &    -9.53 & -11.72 &        2127.47 &   1977 \\
    NGC5548 &  XMM-NEWTON &          PN &  89960301 & 2001-07-09 &     95.82 & circular & 0.017 &  1.65 &     1.65 &     1.66 &   20.0 &   21.7 &    -99 &   30.0 &    22.33 &      -99 &    -9.99 & -12.33 &         145.58 &    166 \\
    NGC5548 &  XMM-NEWTON &          PN & 720110401 & 2013-06-30 &     55.91 & circular & 0.017 &  1.53 &     1.52 &     1.55 &  22.16 &  22.15 &  22.18 &   30.0 &    18.41 &      -99 &   -10.06 & -12.37 &         228.03 &    165 \\
    NGC5548 &  XMM-NEWTON &          PN & 720110701 & 2013-07-15 &     55.91 & circular & 0.017 &  1.46 &     1.45 &     1.48 &  22.24 &  22.23 &  22.26 &   0.05 &    15.32 &      -99 &    -10.1 & -12.27 &         244.61 &    165 \\
    NGC5548 &  XMM-NEWTON &          PN & 720111201 & 2013-07-27 &     55.91 & circular & 0.017 &  1.47 &     1.46 &     1.49 &  22.24 &  22.23 &  22.26 &   0.47 &    22.96 &      -99 &   -10.06 & -12.27 &         251.65 &    166 \\
    NGC5548 &  XMM-NEWTON &          PN & 720111601 & 2014-02-04 &     55.92 & circular & 0.017 &  1.46 &     1.44 &     1.48 &  21.91 &  21.88 &  21.95 &   0.66 &    21.15 &      -99 &   -10.17 & -12.28 &         242.21 &    165 \\
    NGC5548 &  XMM-NEWTON &          PN & 771000101 & 2016-01-14 &     35.91 & circular & 0.017 &  1.41 &     1.39 &     1.43 &   22.0 &  21.95 &  22.03 &   30.0 &    22.52 &      -99 &   -10.24 &  -12.4 &         180.33 &    164 \\
    NGC5548 &  XMM-NEWTON &          PN & 771000201 & 2016-01-16 &     32.91 & circular & 0.017 &  1.45 &     1.43 &     1.48 &   22.0 &  21.97 &  22.04 &   0.02 &     15.0 &      -99 &   -10.21 & -12.32 &         202.22 &    164 \\
    NGC7469 &     CHANDRA & ACIS-S HETG &      2956 & 2002-12-12 &     78.59 &  grating & 0.016 &  1.81 &     1.78 &     1.84 &   20.0 &   21.7 &    -99 &  12.47 &     4.89 &    19.62 &    -10.4 & -12.57 &        2241.20 &   1977 \\
    NGC7469 &     CHANDRA & ACIS-S HETG &      3147 & 2002-12-13 &     68.63 &  grating & 0.016 &  1.74 &      1.7 &     1.83 &   20.0 &    -99 &  21.41 &   49.7 &    34.35 &    79.02 &   -10.46 & -12.52 &        2192.94 &   1977 \\
    NGC7469 &  XMM-NEWTON &          PN & 207090201 & 2004-12-03 &     79.11 & circular & 0.016 &  1.83 &     1.83 &     1.84 &   20.0 &   21.7 &    -99 &   0.33 &    23.79 &      -99 &   -10.11 & -12.33 &         228.97 &    165 \\
    NGC7469 &  XMM-NEWTON &          PN & 760350201 & 2015-06-12 &     89.72 & circular & 0.016 &  1.97 &     1.96 &     1.97 &   20.0 &   21.7 &    -99 &   30.0 &    23.21 &      -99 &   -10.14 &  -12.3 &         270.05 &    164 \\
    NGC7469 &  XMM-NEWTON &          PN & 760350301 & 2015-11-24 &     85.92 & circular & 0.016 &  1.83 &     1.83 &     1.84 &   20.0 &   21.7 &    -99 &   0.17 &    21.81 &      -99 &   -10.11 & -12.38 &         261.30 &    165 \\
    NGC7469 &  XMM-NEWTON &          PN & 760350401 & 2015-12-15 &     84.81 & circular & 0.016 &  1.92 &     1.91 &     1.93 &   20.0 &   21.7 &    -99 &   30.0 &     4.06 &      -99 &   -10.24 & -12.31 &         229.54 &    160 \\
    NGC7469 &  XMM-NEWTON &          PN & 760350801 & 2015-12-28 &    100.52 & circular & 0.016 &  1.87 &     1.86 &     1.87 &   20.0 &   21.7 &    -99 &   0.57 &    24.58 &      -99 &   -10.11 & -12.28 &         256.73 &    165 \\
    
\noalign{\smallskip}            
\hline
\end{tabular}}

\tablefoot{ This table shows a random number of entries of the full table as an example. The limits reported in the tables are at 1-$\sigma$. The full table also reports the fluxes uncertainties. Col.(1): Object name; Col.(2): Observatory; Col.(3) Instrument; Col.(4) observation ID;  Col(5): date observations were made; Col.(6): exposure time of the observation; Col.(7): shape of the extracted spectrum; Col.(8) redshift of the sources taken from \citet{2017ApJ...850...74K}; Col.(9): photon index; Col.(10): lower limit of the photon index; ; Col.(11): upper limit of the photon index; Col.(12): column density in logarithmic scale; Col.(13): lower limit of the column density in logarithmic scale; ; Col.(14): upper limit of the column density in logarithmic scale; Col.(15): \kalfa{} line width; Col.(16): lower limit of the \kalfa{} line width; Col.(17): upper limit of the \kalfa{} line width; Col.(18): intrinsic 2--10 keV flux; Col.(19): \kalfa{} flux; Col.(20): fit statistic; Col.(21): degrees of freedom of the fit.  }
\label{t:ObsInfo}
\end{table*}

\section{Pileup and annular spectra recalibration} \label{ap:pileup}

We cannot use saturation-affected data to study flux variability, and thus we need to know whether each ACIS observation is affected by pileup. A good way to estimate the amount of pileup of an observation is to compute the count rate per frame time of the observation, since it has a relation with the pileup fraction\footnote{see \url{https://cxc.harvard.edu/csc/memos/files/Davis\_pileup.pdf}}.
When the count rate per frame is less than 0.1, the pileup fraction is $<5\%$. We used this threshold to separate saturated from unsaturated observations.  

To compute the total expected count rate per frame time of each observation, we used the 3\farc{}--5\farc{} annular spectra for all the ACIS observations, and the following expression:

\begin{equation} \label{eq:cr}
    \rm count \:s^{-1}_{3^{\prime\prime}-5^{\prime\prime}}\:  \times (33.3) \times frame\:time,
\end{equation}

\noindent where the factor 33.3 is the effective aperture correction, which is the inverse of the fractional power encircled in a 3\farc{}--5\farc{} annulus, which is approximated as 0.03 based on the digitization and integration of the 1\,keV encircled energy profile of Fig. 6.1 from the \textit{Chandra} Proposer's Observatory Guide.\footnote{\url{https://asc.harvard.edu/proposer/POG/html/chap6.html}} We assume that the 1 keV value provides a rough average over the entire relevant {\it Chandra} energy range.

If the observation is affected by pileup, we adopted the aperture-corrected 3\farc{}--5\farc{} annular spectrum, if not, we used the aperture-corrected 1.5\farc{} circular spectrum. 

We discovered an inconsistency when comparing fluxes given by the 1.5\farc{} aperture and 3\farc{}--5\farc{} annular spectra, suggesting an inconsistency in the aperture correction for the annular spectrum. Given that the central aperture spectra could be affected by saturation, we performed a consistency check between the 3\farc{}--5\farc{} annular spectrum with the HEG spectrum in the 2-10 keV energy range, for all the observations performed in grating mode. To do this, we used PyXspec to fit the following simple model:

\begin{equation} \label{eq:model_cte}
\rm  \textsc{constant} \times \textsc{phabs\_1}  \times \textsc{phabs\_2}  \times (\textsc{powerlaw}+\textsc{zgauss}).
\end{equation}

\noindent The description of each component is the same as in Section \ref{sec:spec_fitting}, aside from the \textsc{constant} component, which is added to compute the calibration constant between the spectra. The constant value was fixed to 1 for the grating spectrum and let free to vary for the annular spectrum.

\begin{figure} 
$\begin{subfigure}[b]{0.3\textwidth}
\includegraphics[trim= 105 120 0 20, clip, width=1.6\textwidth]{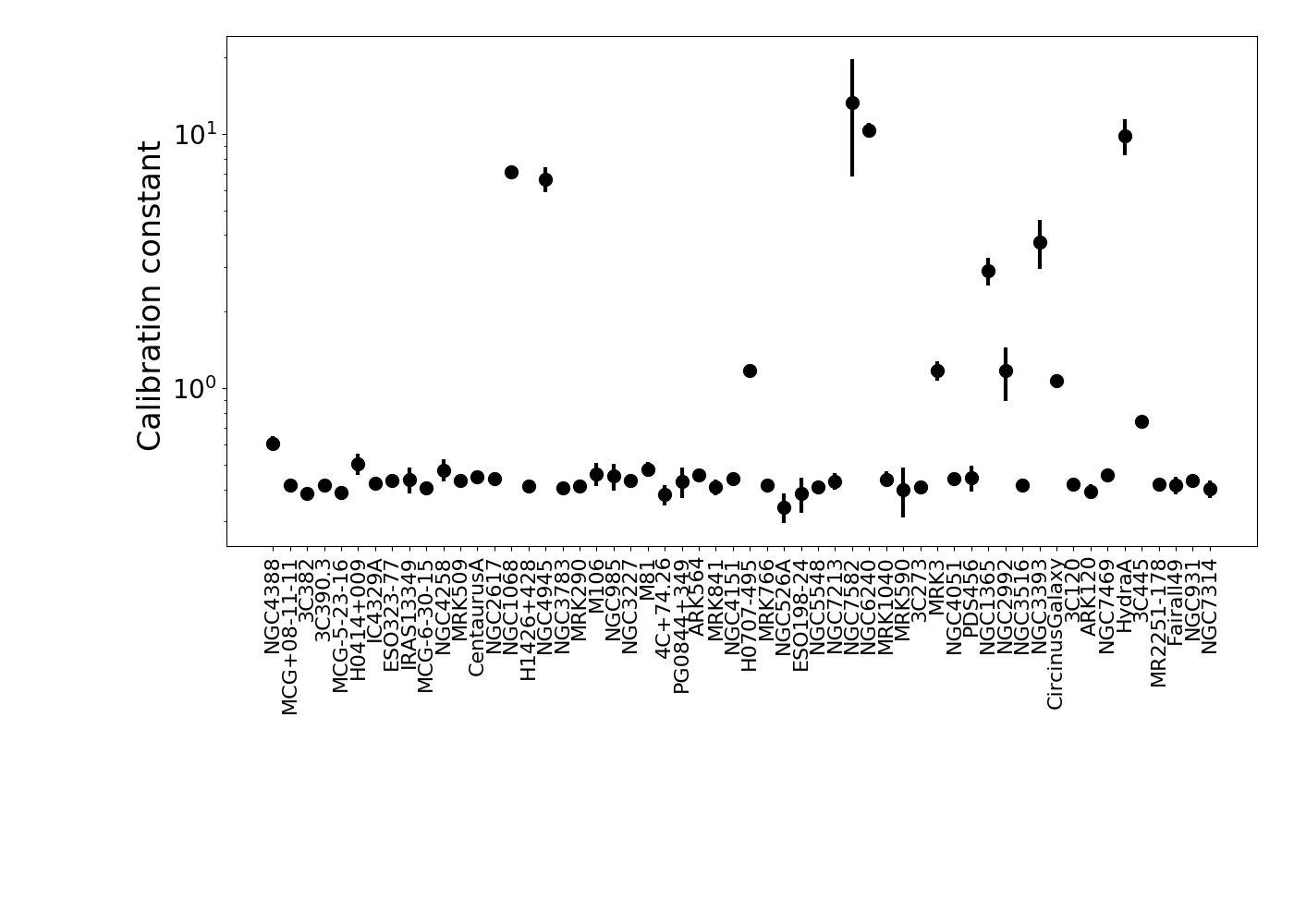} 
%\caption{}
%\label{}
 \end{subfigure}$

 $\begin{subfigure}[b]{0.3\textwidth}
 \includegraphics[trim= 105 120 0 20, clip, width=1.6\textwidth]{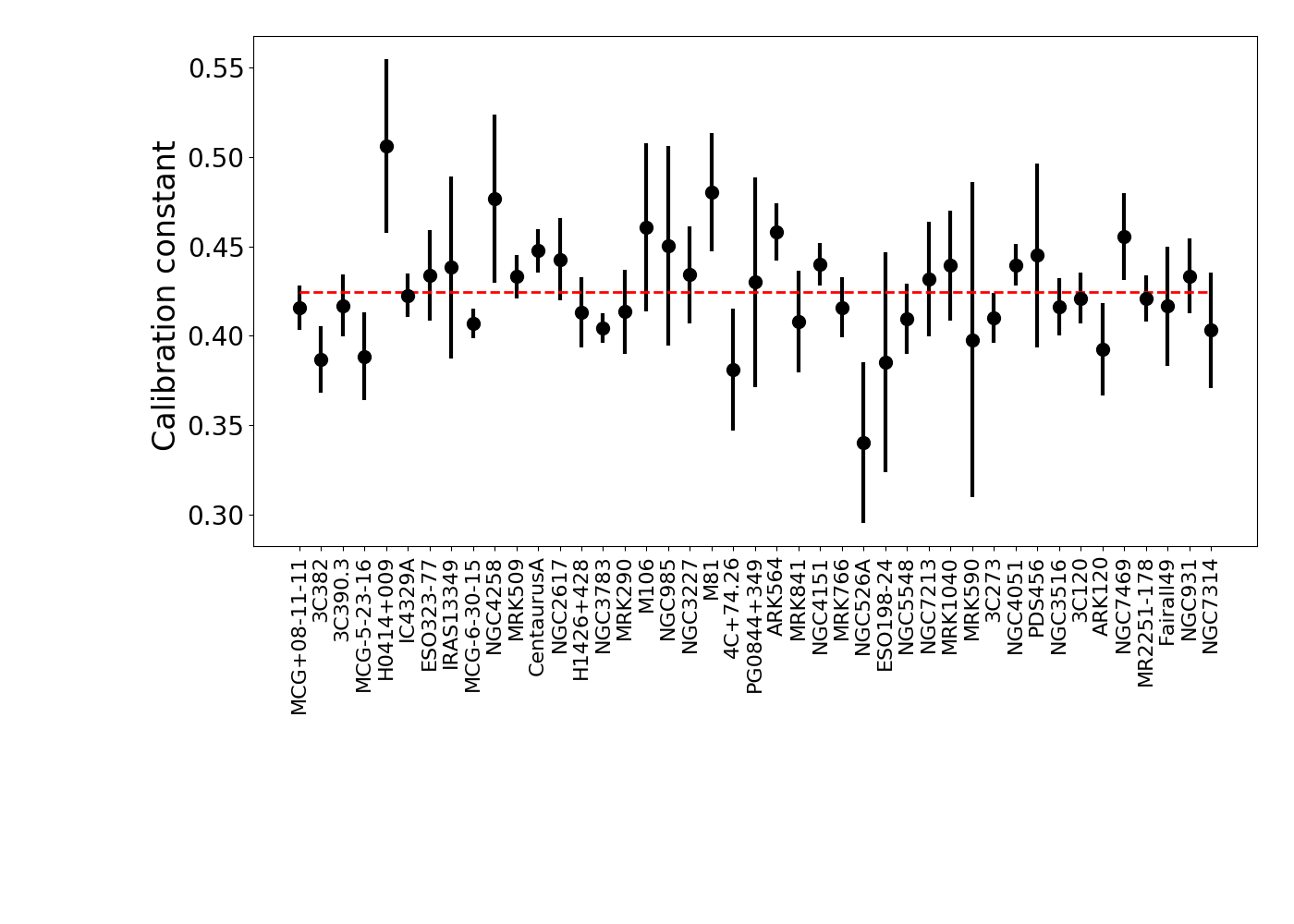}
 %\caption{}
 %\label{}
 \end{subfigure}$

 %\captionsetup{width=1.5\linewidth}  
 \caption{Calibration constants between the \textit{Chandra} HEG first-order and ACIS/HETG 0th order 3\farc{}--5\farc{} annular spectra. The upper panel shows all the calibration constants, and the lower panel shows the calibration constants consistent with the "floor" found at $0.424\pm 0.005$, marked by a red dashed line. }{\label{fig:cal_cte}}  
 \end{figure}

The upper panel of Figure~\ref{fig:cal_cte} shows the calibration constants calculated for all the BASS sources with HEG spectra. We obtained an evident consistency for 43/56 galaxies. For 13 sources, higher values of the calibration constant are found, and therefore we did not use the 3\farc{}--5\farc{} annular spectra for those cases. A possible explanation for these outliers is the presence of extended emission, either from the AGN or circumnuclear contamination. Some of the sources with a constant higher than one are already claimed to present extended emission, such as the Circinus Galaxy \citep{2014ApJ...791...81A}, NGC\,1068 \citep{2015ApJ...812..116B} and NGC\,4945 \citep{2017MNRAS.470.4039M}. A deeper analysis of the rest of the sources, beyond the scope of the present work, is needed to confirm the possibility of extended, but for now this is the most probable explanation.

The lower panel of Fig.~\ref{fig:cal_cte} shows the values of the calibration constants for the 43 consistent sources. If the aperture corrections of both the HETG and annular spectra are accurate, we would expect a constant consistent with unity. However, the average value for the constant is $0.424\pm 0.005$, which is highly constrained. Presumably the few percent of remaining variation in the lower panel of Fig.~\ref{fig:cal_cte} is due to differences in underlying spectral shape amongst the sources.

The CIAO aperture corrections are based on a theoretical model of the \textit{Chandra} point spread function (PSF), and hence the deviation from unity for the above constant implies a likely problem with the wings of the model. This inconsistency between the theoretical/model and real PSFs is discussed on the Chandra X-ray Center website, in $\$$4.2.3 of the Proposer's Observatory Guide\footnote{\url{https://asc.harvard.edu/proposer/POG/html/chap4.html}} and the ChaRT webpages,\footnote{see \url{https://cxc.cfa.harvard.edu/ciao/PSFs/chart2/caveats.html}}. In particular, uncertainties of 30--50\% are thought to exist at high energies, consistent with our findings.

We use the average value of the constant to rescale the fluxes obtained for all the annular spectra. We further compare these against the 1\farcs5 aperture fluxes obtained from the non-piled-up spectra, and confirm consistency within statistical errors. We revisit the issue of the PSF wings in the analysis of the \textit{Chandra} images, presented in $\S$\ref{sec:image}.

\section{Notes on individual sources} \label{ap:individual_sources}

\begin{itemize}
\item {\it 3C\,120}: this radio-loud sources has $\sigma_{\rm c}/\sigma_{\rm c, sim}{<}1$, suggesting that the X-ray emission of this object is not jet-dominated but coronal-dominated, as predicted by \citet{2021Zhu}, which also found that radio-loud objects coronal-dominated are $20\%$ less variable than radio-quiet objects. Additionally, \citet{2013Lohfink} argue that 3C\,120 may have a truncated disk, depleted by the ejection of a new superluminal knot.

\item {\it MRK\,273}: ultra-luminous infrared galaxy with dual obscured nuclei separated by $\approx$1\farc{}. Our extraction and spectral model does not attempt to distinguish between the NE and SW components, which have distinct spectral shapes and variability properties \citep{2018Iwasawa,2019Liu}; while the more variable obscured SW component dominates the <10 keV emission, the Compton-thick NE component dominates above $\sim$12 keV. This added complexity may lead to its moderately high $\sigma_{\rm c}/\sigma_{\rm c, sim}$ value.

\item {\it NGC 1365}: this source has a $\sigma_{\rm c}/\sigma_{\rm c, sim}{>}1$, not following the typical radio-quiet power spectrum. It is a well-known changing-look AGN, with $N_{\rm H}$ and covering factor variations between ${\sim}10^{22}$--$10^{24}$ cm$^{-2}$ and $\sim$0.2--1.0, respectively \citep[e.g.,][]{2015Rivers}; while our simple spectral model accounts for the variable $N_{\rm H}$, it does not incorporate covering factor changes, which appear to increase the continuum variability over expectations from the X-ray corona. 

\item {\it NGC 2992}: this is the only source in the sample with a reflector size estimate from the simulations and \kalfa{} FWHM consistently larger than the dust sublimation radius. Also, $\sigma_{\rm c}/\sigma_{\rm c, sim}\sim 50$, being much more variable than expected for a radio-quiet AGN. This intriguing AGN has optically changed from Seyfert 1.5 to Seyfert 2 \citep{2008Trippe}, and also has transitioned from a passive state to an active state in the X-ray band \citep{2000Gilli}. Long- and short-term variability with large amplitude ($\gtrsim 1$ order of magnitude) has already been reported for this source by many studies \citep[e.g.,][]{2007Murphy,2021Guolo,2021Ghosh}. 

\item {\it NGC\,7469}: this is the only source that does not reproduce the well-known radio-quiet power spectrum which seems to defy explanation, which by most accounts appears to be a fairly typical unobscured AGN, yet appears to have the lowest measure of $\sigma_{\rm c}/\sigma_{\rm c, sim}$ on record. Based on Fig.~\ref{fig:lc_cont}, NGC\,7469's 2--10\,keV continuum flux appears to vary by a factor of $\approx$1.4 on 1-year timescales, but remains within that window over the full extent of the light curve. The measurements from 2015-2016 are fully consistent with the {\it Swift} XRT light curve published by \citet{2020Behar}, who find a factor of $\approx$2 variation in 2--10\,keV count rate on 80-day timescales. The only "peculiarity" is that the AGN in NGC\,7469 appears to show two UV to X-ray delay timescales of 0.37 and $-$3.5 days \citep[e.g.,][]{2020Pahari}. It could be that there is unresolved emission contaminating the aperture, damping the expected variability, although its hard X-ray luminosity is sufficiently high that any contaminant would have to be very atypically bright (and constant) itself.

\end{itemize}

\section{Light curves} \label{ap:lc}

\begin{figure*}
%\centering
\hspace*{-0.6cm}\includegraphics[trim={0cm 18cm 6cm 0cm},clip,width=1.05\textwidth]{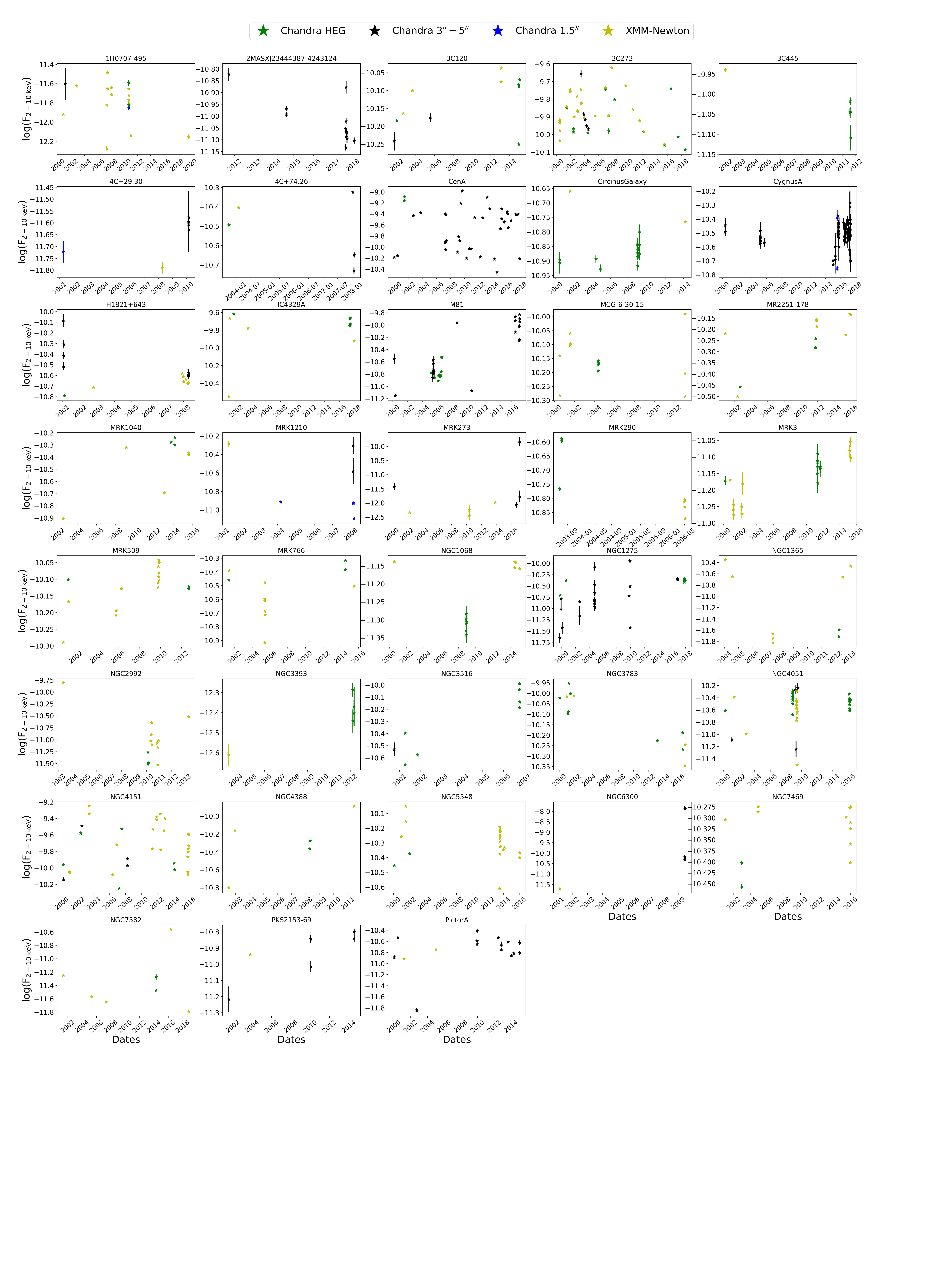}
\caption{Light curves for the 2--10 keV continuum flux ($\rm F_{2-10\: keV}$) for all sources in the sample. In each subplot, green stars denote \textit{Chandra} HEG spectra, black stars denote \textit{Chandra} ACIS 3\farc-5\farc{} annular spectra, blue stars denote 1.5\farc{} circular \textit{Chandra} ACIS spectra, and yellow stars denote \textit{XMM-Newton} pn spectra. The units of the flux are $\rm erg\:cm^{-2}\: s^{-1}$.}
\label{fig:lc_cont}	
% Agrandar ejes y labels
\end{figure*}

\begin{figure*}
%\centering
\hspace*{-0.6cm}\includegraphics[trim={0cm 18cm 6cm 0cm},clip,width=1.05\textwidth]{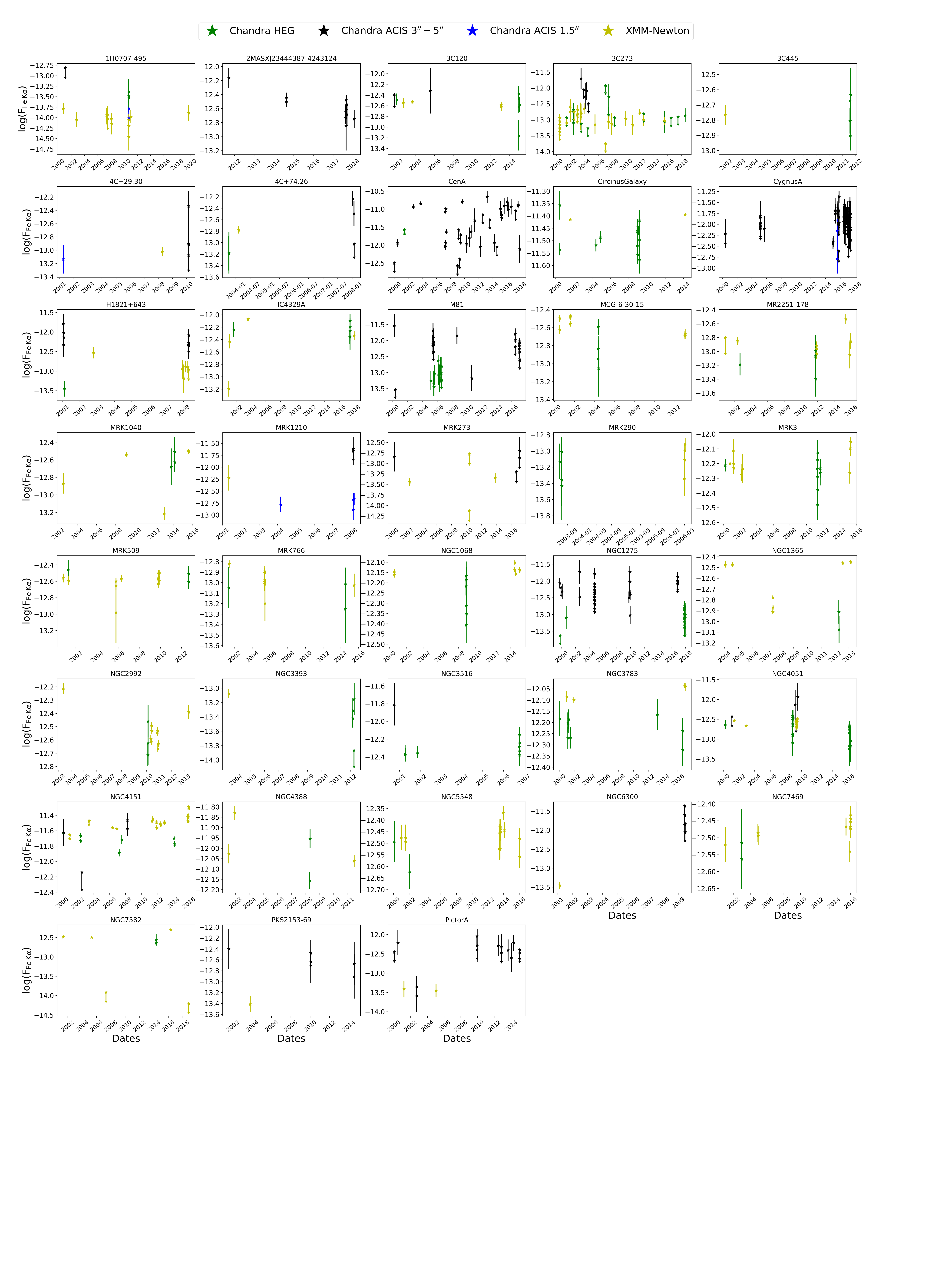}
\caption{Light curves for the Fe K$\alpha$ line flux ($\rm F_{\rm Fe\:K\alpha}$) for all sources in the sample.  In each subplot, green stars denote \textit{Chandra} HEG spectra, black stars denote \textit{Chandra} ACIS 3\farc-5\farc{} annular spectra, blue stars denote 1.5\farc{} circular \textit{Chandra} ACIS spectra, and yellow stars denote \textit{XMM-Newton} pn spectra.  The units of the flux are $\rm erg\:cm^{-2}\: s^{-1}$.}
\label{fig:lc_fe}	
% Agrandar ejes y labels
\end{figure*}

Figures~\ref{fig:lc_cont} and \ref{fig:lc_fe} show the continuum and \kalfa{} light curves, respectively, for all of the sources in our sample. Machine-readable files for each source will be provided upon publication.

\begin{figure*}
\centering
% \hbox{\hspace{-2.5em}}
\hspace*{-0.6cm}\includegraphics[trim={0cm 18cm 6cm 1cm},clip,width=1.0\textwidth]{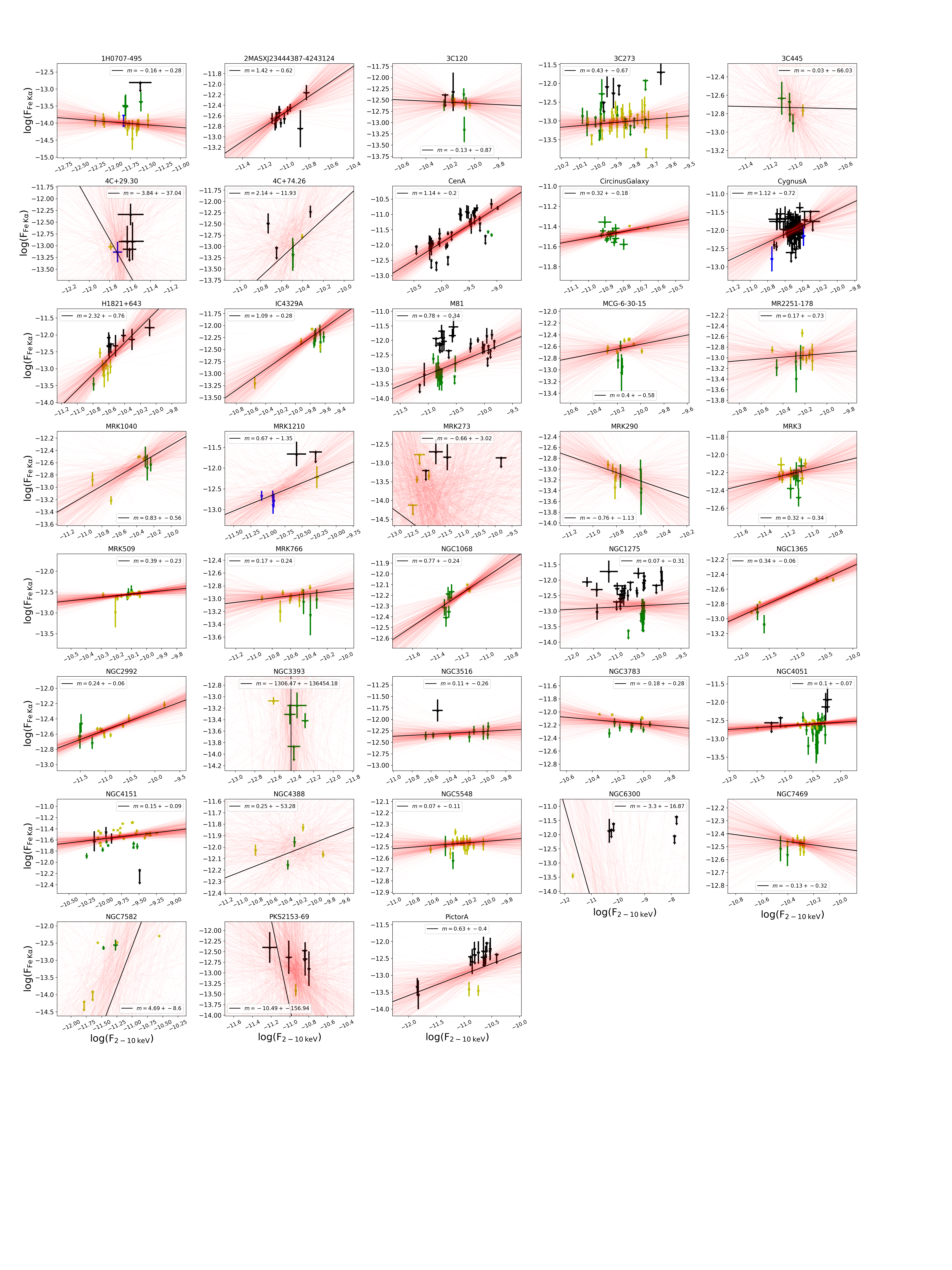}
\caption{Comparison between Fe K$\alpha$ line ($\rm F_{6.4\:keV}$) and 2--10 keV continuum ($\rm F_{2-10\: keV}$) fluxes, for all sources in our sample. In each subplot, green stars denote \textit{Chandra} HEG spectra, black stars denote \textit{Chandra} ACIS 3\farc-5\farc{} annular spectra, blue stars denote 1.5\farc{} circular \textit{Chandra} ACIS spectra, and yellow stars denote \textit{XMM-Newton} pn spectra. The thick red line shows the best-fit linear regression, while thin red lines represent fits from Monte Carlo resampling to demonstrate uncertainties; the best-fit slope and error are indicated in the legend of each plot. Fits to sources with fewer than five measurements in both fluxes are often poorly constrained. The values for the slopes are reported in the label of each plot. The flux units are $\rm erg\:cm^{-2}\: s^{-1}$.}
\label{fig:fluxes_vs_q}	

\end{figure*}

\section{Variability versus AGN and host galaxy properties} \label{ap:var_prop}

\begin{figure*} 
 \centering
 $\begin{subfigure}[b]{0.45\textwidth}  
 \includegraphics[trim=40 20 50 50,clip,width=\textwidth]{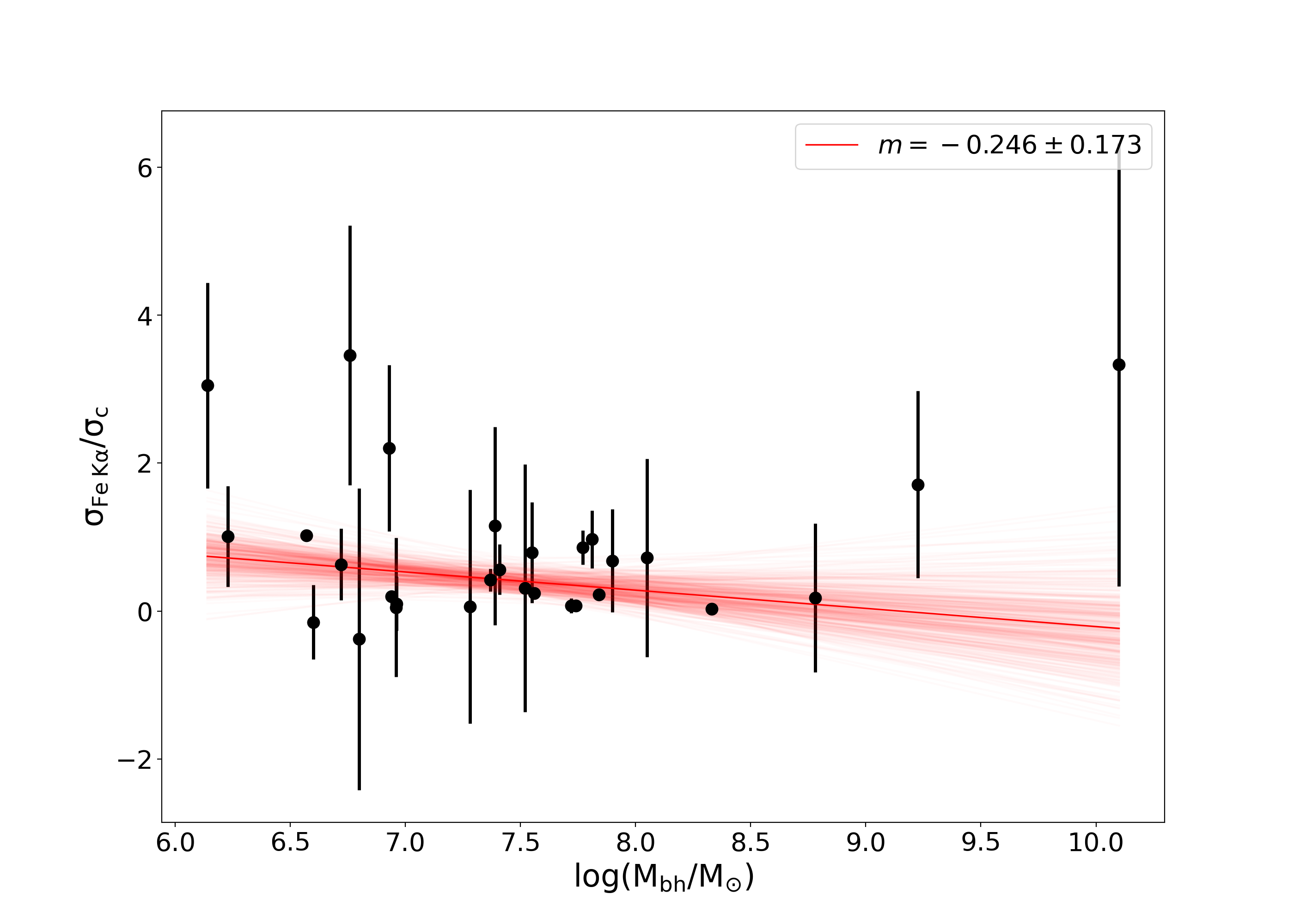}
 \caption{}
 \label{subfig:ratio-mbh}
 \end{subfigure}$ 
 $\begin{subfigure}[b]{0.45\textwidth}  
 \includegraphics[trim=40 20 50 50,clip,width=\textwidth]{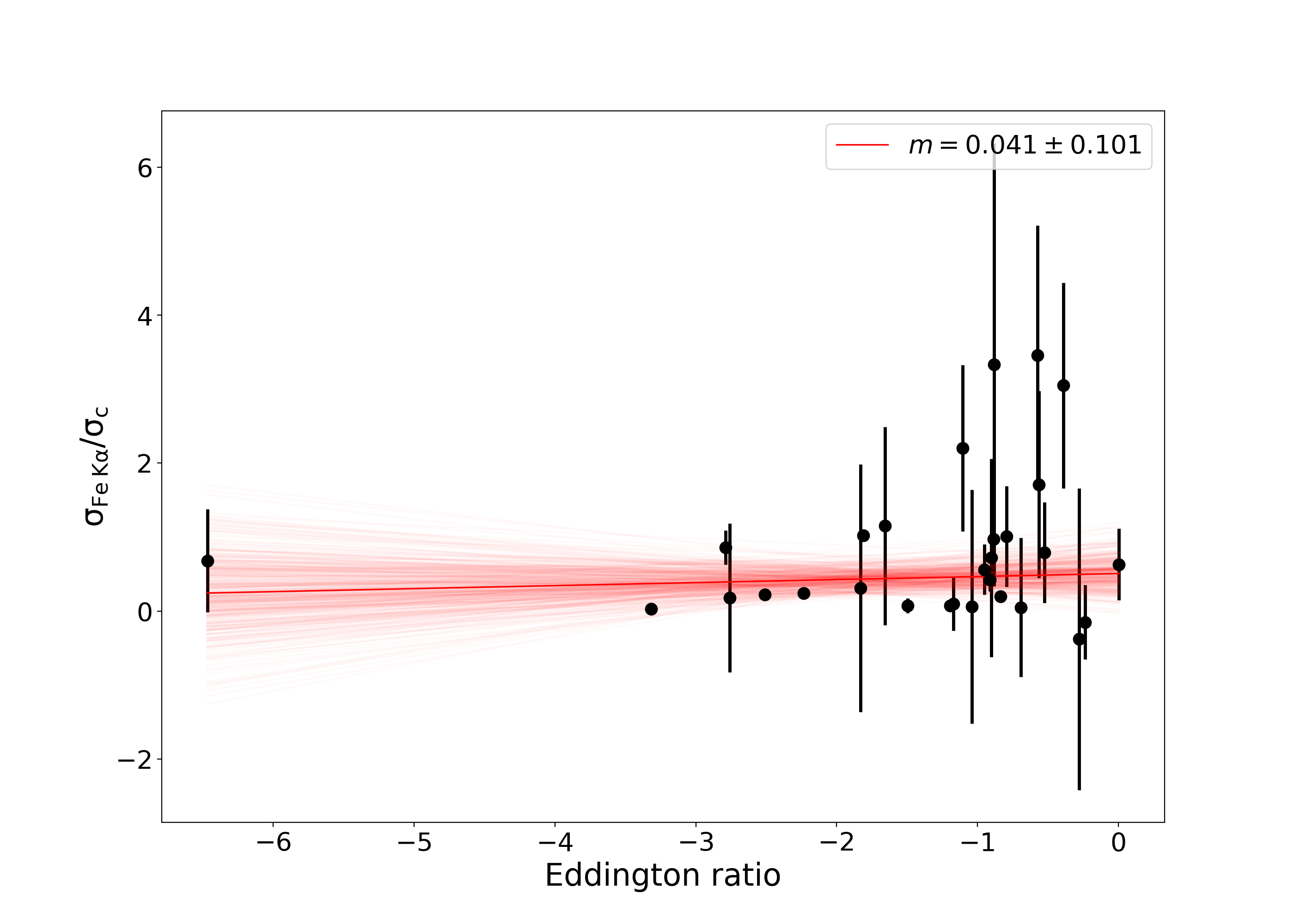}
 \caption{}
 \label{subfig:ratio-er}
 \end{subfigure}$ 
$\begin{subfigure}[b]{0.45\textwidth}  
 \includegraphics[trim=40 20 50 50,clip,width=\textwidth]{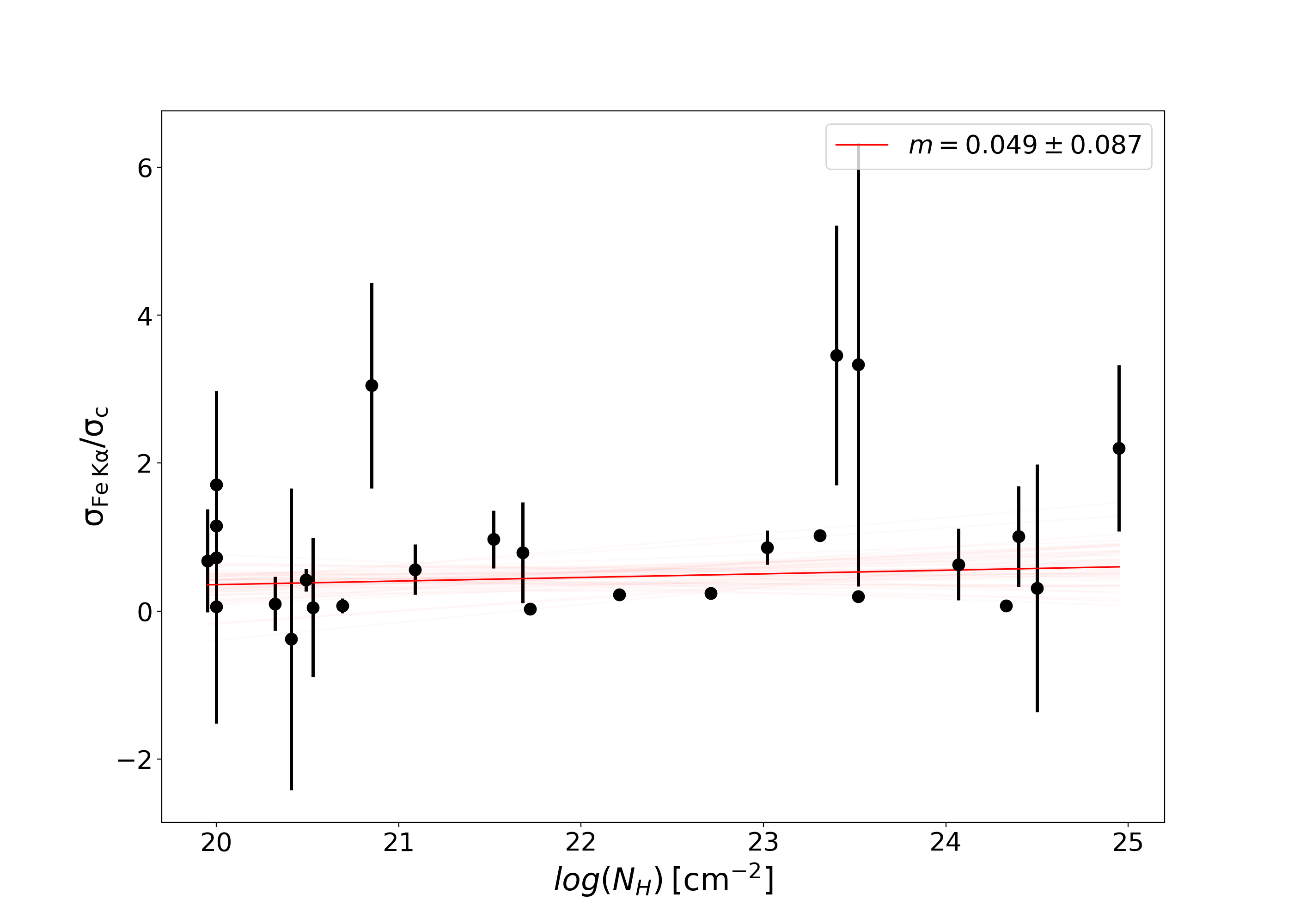}
 \caption{}
 \label{subfig:ratio-nh}
 \end{subfigure}$ 
 $\begin{subfigure}[b]{0.45\textwidth}  
 \includegraphics[trim=40 20 50 50,clip,width=\textwidth]{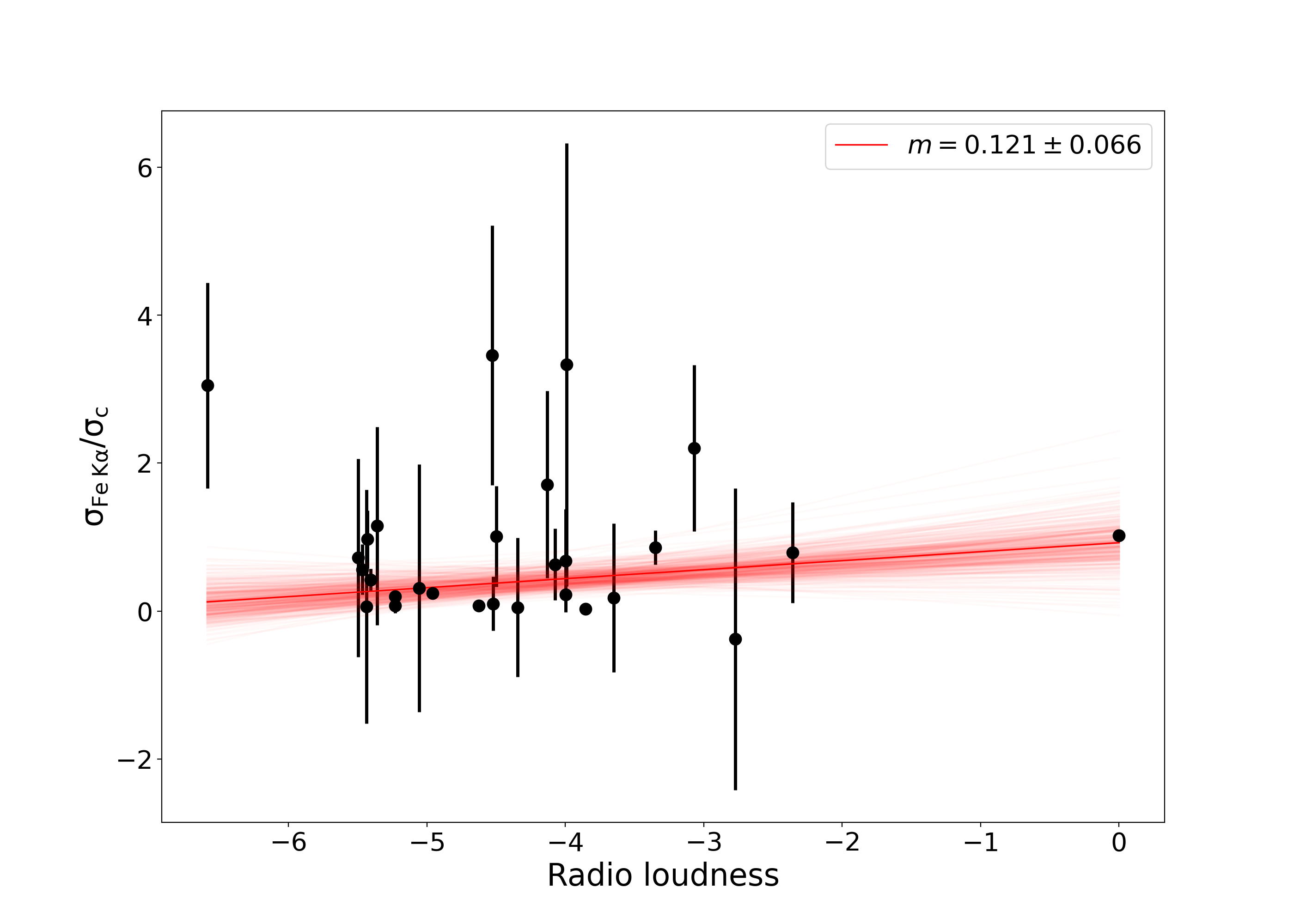}
 \caption{}
 \label{subfig:ratio-RL}
 \end{subfigure}$ 

 $\begin{subfigure}[b]{0.45\textwidth}  
 \includegraphics[trim=40 20 50 50,clip,width=\textwidth]{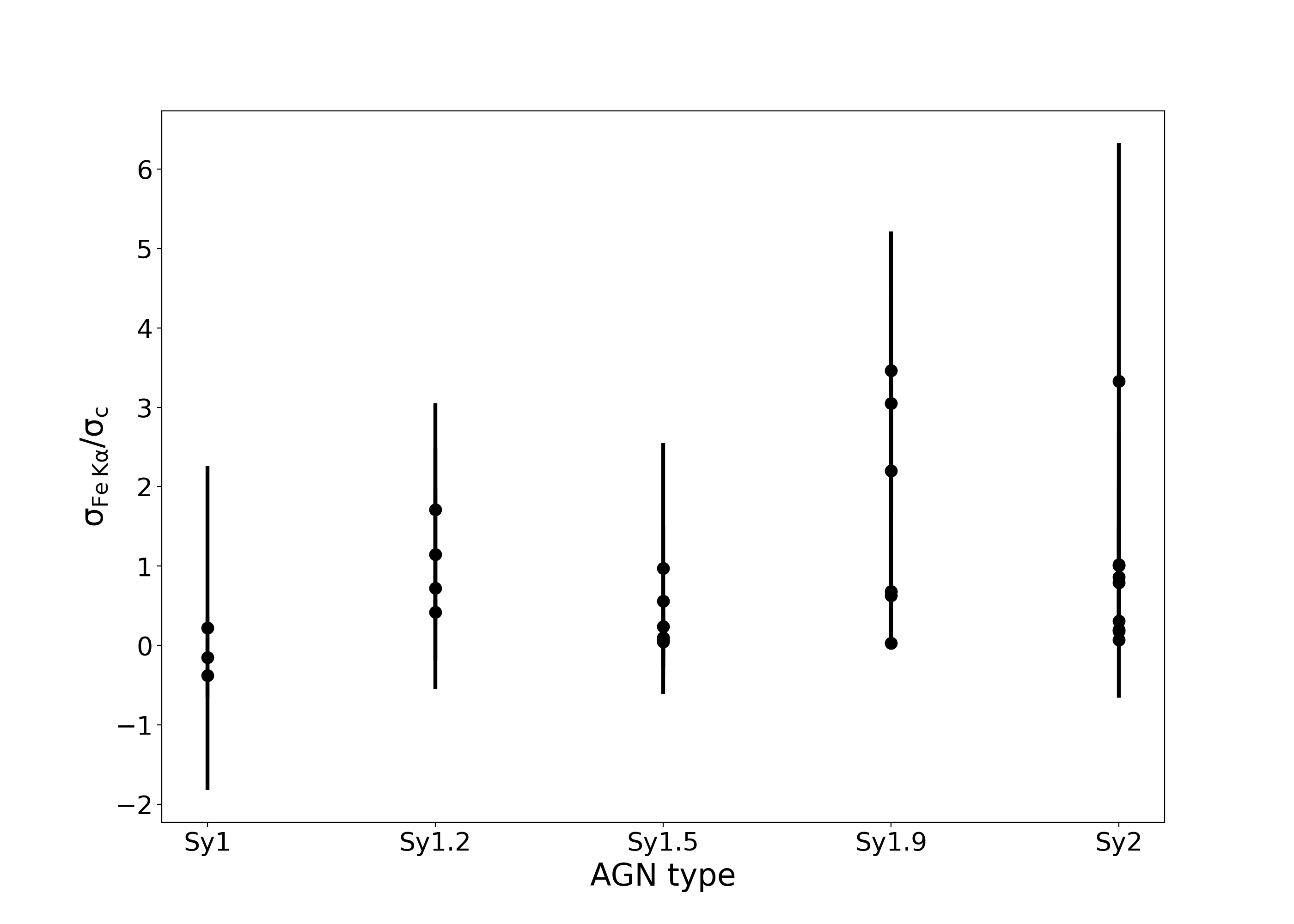}
 \caption{}
 \label{subfig:ratio-type}
 \end{subfigure}$ 
 %\captionsetup{width=1.5\linewidth}  
 \caption{Comparison of the $\sigma_{\rm Fe\:K\alpha}/\sigma_{c}$ slopes with respect to (a) SMBH $M_{\rm BH}$, (b) Eddington ratio, (c) line-of-sight column density, (d) radio-loudness, and (e) AGN type.  For the first five plots, the thick red line denotes the best-fit linear regression, while thin red lines denote fits from Monte Carlo resampling to estimate uncertainties; the best-fit slope and error are indicated in the legend of each plot.}\label{fig:ratio_rel}
 \end{figure*}

\begin{figure*} 
 \centering
 $\begin{subfigure}[b]{0.45\textwidth}  
 \includegraphics[trim=40 20 50 50,clip,width=\textwidth]{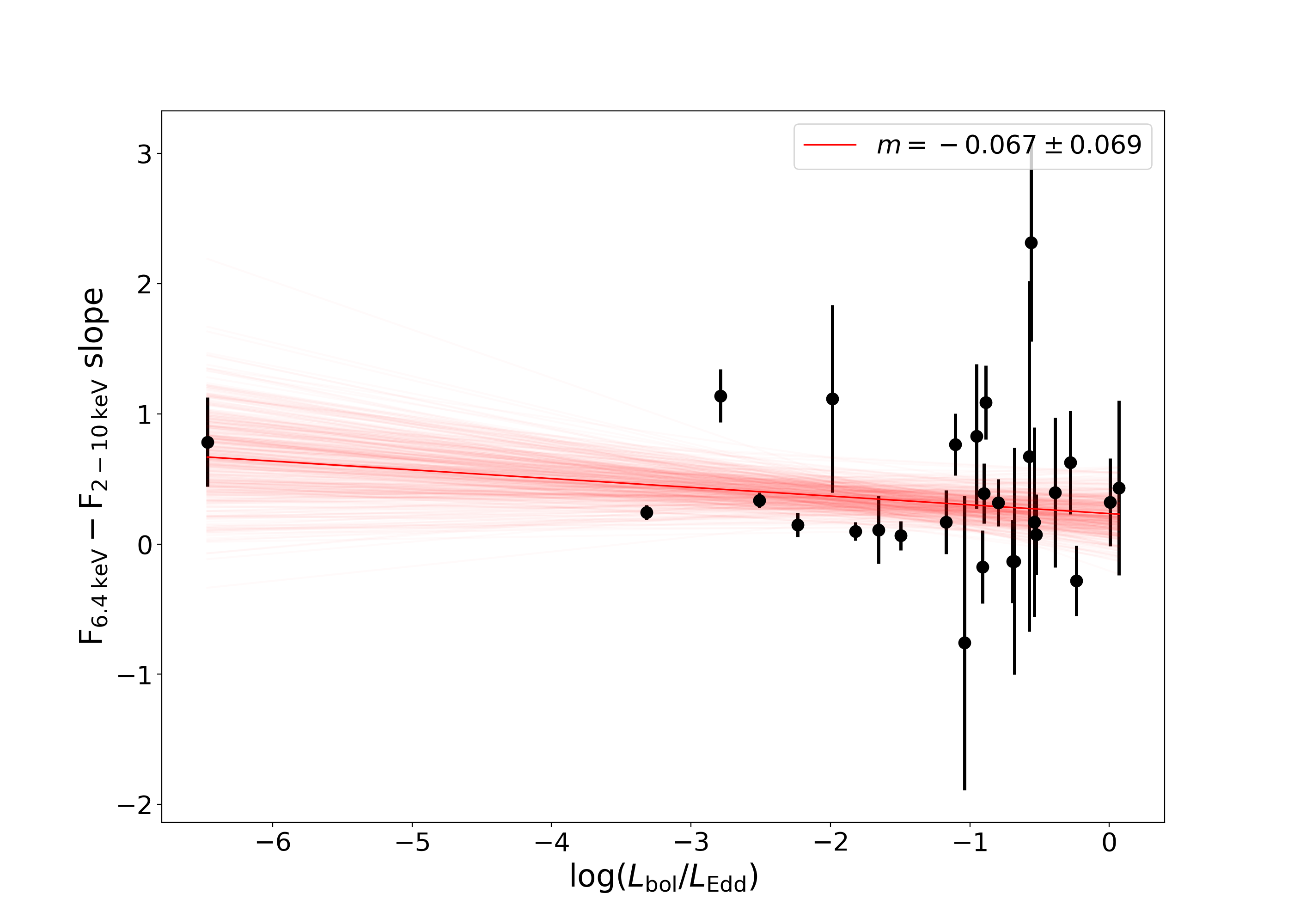}
 \caption{}
 \label{subfig:rel-er}
 \end{subfigure}$ 
$\begin{subfigure}[b]{0.45\textwidth}  
 \includegraphics[trim=40 20 50 50,clip,width=\textwidth]{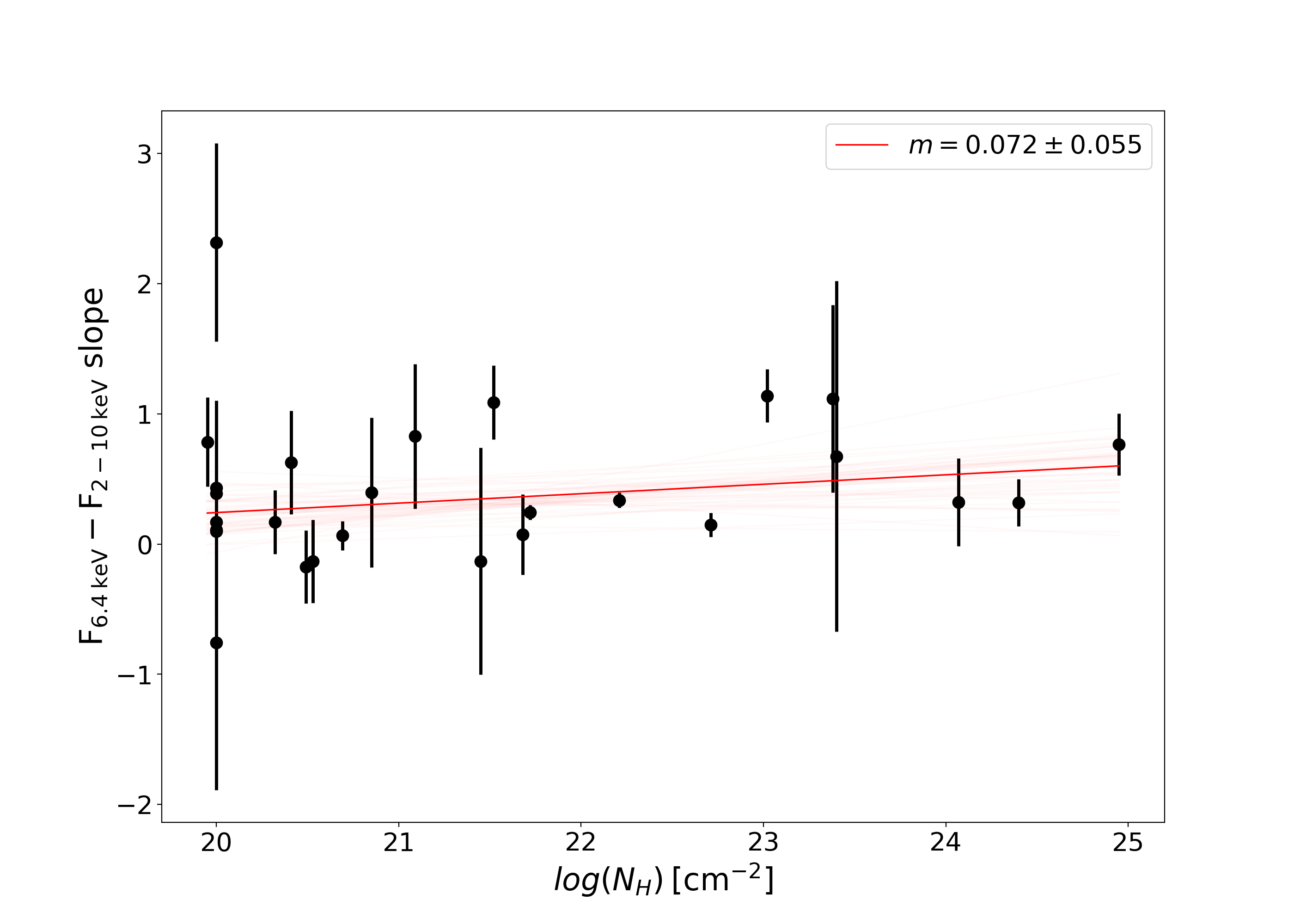}
 \caption{}
 \label{subfig:rel-nh}
 \end{subfigure}$ 
%  \centering
 $\begin{subfigure}[b]{0.45\textwidth}  
 \includegraphics[trim=40 20 50 50,clip,width=\textwidth]{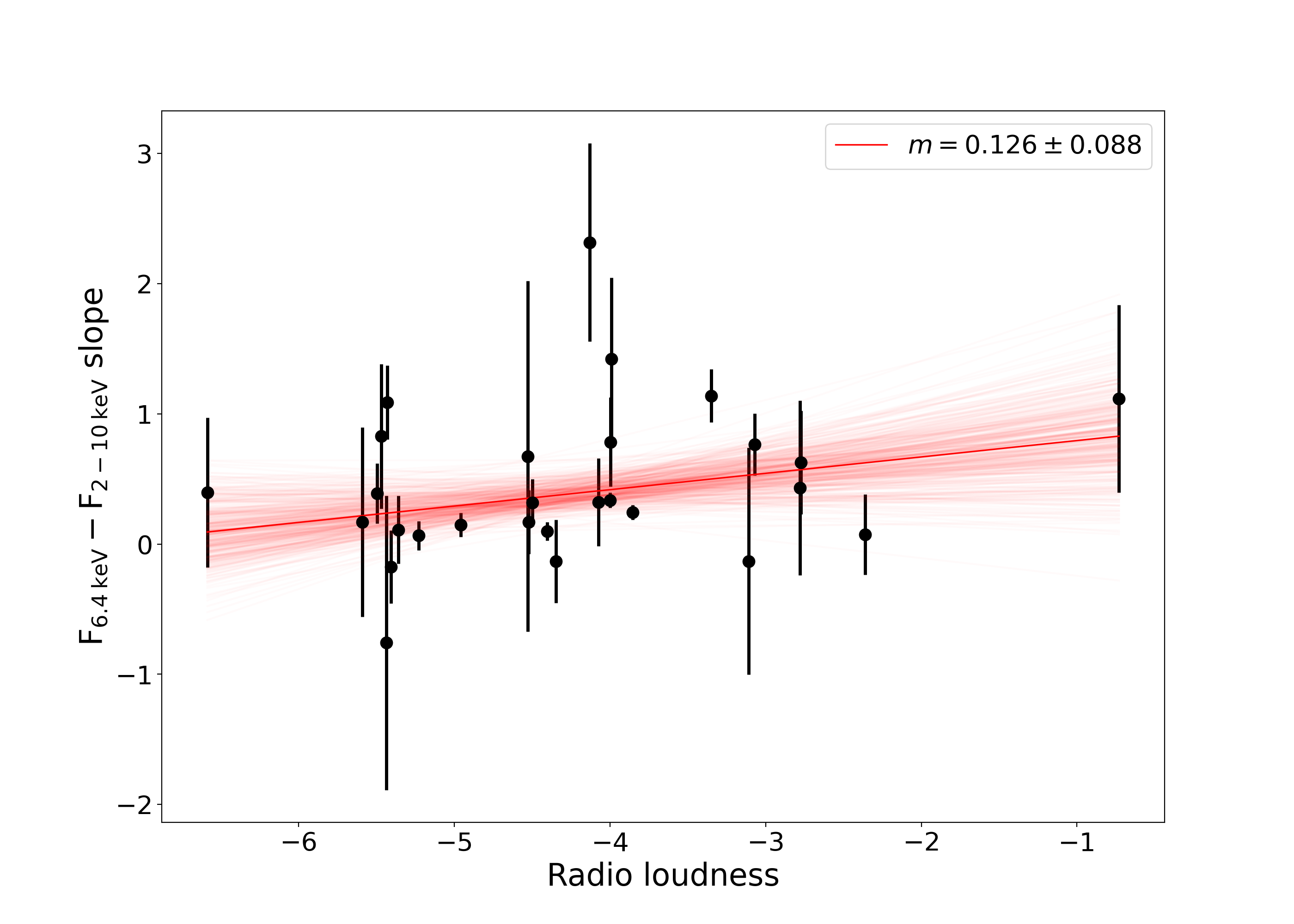}
 \caption{}
 \label{subfig:rel-RL}
 \end{subfigure}$ 
$\begin{subfigure}[b]{0.45\textwidth}  
 \includegraphics[trim=40 20 50 50,clip,width=\textwidth]{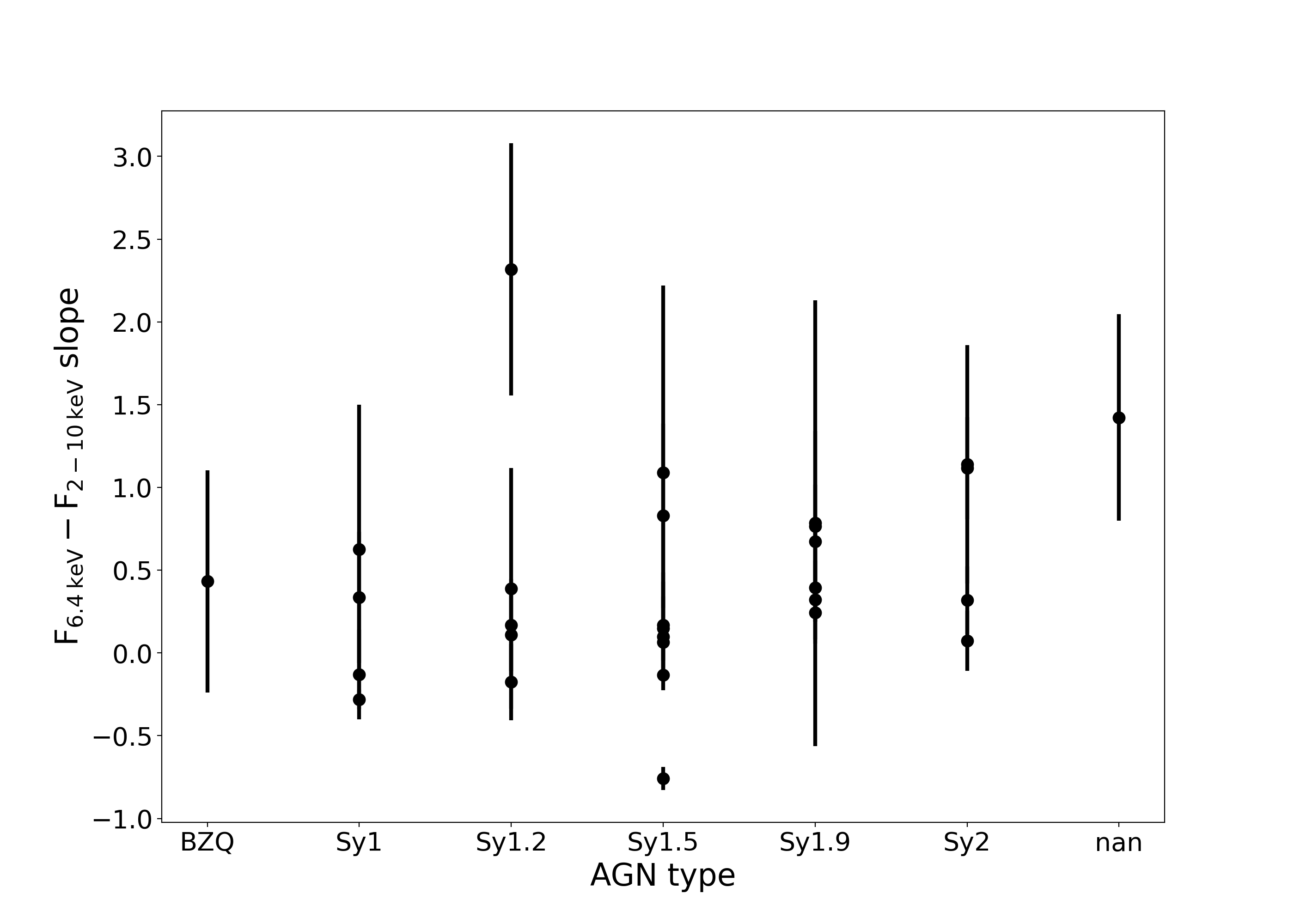}
 \caption{}
 \label{subfig:rel-type}
 \end{subfigure}$ 
 %\captionsetup{width=1.5\linewidth}  
 \caption{Comparison of the $\rm F_{\rm Fe\:K\alpha}$--$\rm F_{2-10\:keV}$ slopes with respect to (a) Eddington ratio, (b) line-of-sight column density, (c) radio-loudness, and (d) AGN type.  For the first three plots, the thick red line denotes the best-fit linear regression, while thin red lines denote fits from Monte Carlo resampling to estimate uncertainties; the best-fit slope and error are indicated in the legend of each plot.}\label{fig:slope_rel}
 \end{figure*}

Figures~\ref{fig:ratio_rel} and \ref{fig:slope_rel} compare the properties of the continuum and Fe K$\alpha$ variability to various AGN and host galaxy properties.

\section{Radial profiles} \label{ap:rprof}

Figure~\ref{fig:rprof_fe} shows a comparison of the continuum-subtracted \kalfa{} radial profiles for each source in our sample with the continuum-subtracted \kalfa{} radial profile of HERX1, which serves as our nominal PSF. The HERX1 radial profile is renormalized to match each source light curve at 2\farcs5, as pixels inside this radius are potentially affected by pileup both in the sample sources and HERX1; we renormalized at 4\farc{} in the case of Cen\,A, as the pileup is worse for that source. In a few cases, the sample AGN are embedded in strong, diffuse cluster emission, and thus we crop the radial profiles at the point where the cluster background starts to dominate.

\begin{figure*}
%\centering
\hspace*{-1cm}\includegraphics[width=1.05\textwidth,trim={2cm 6cm 6cm 8cm},clip]{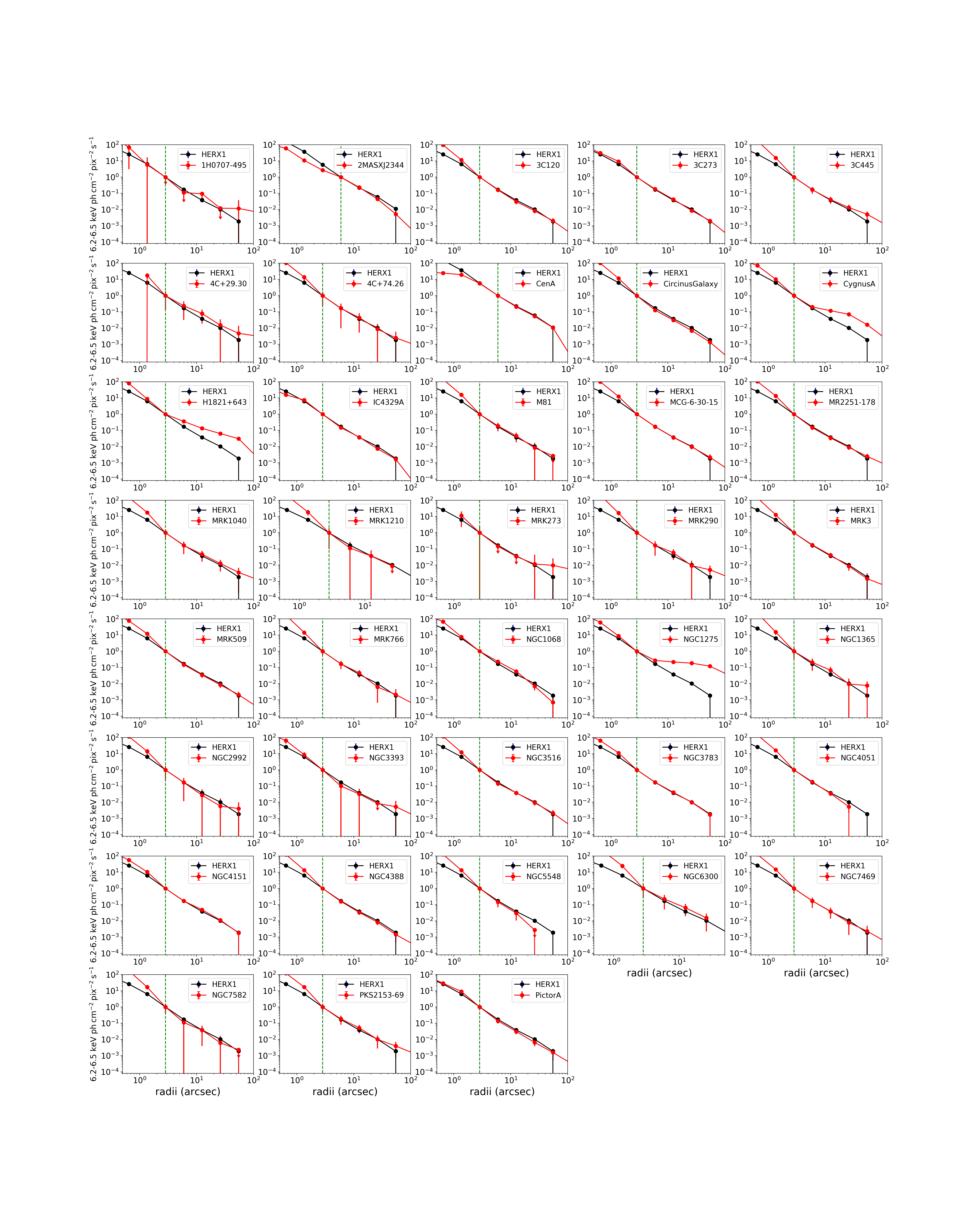}
\caption{Comparison between the 6.2--6.5 keV radial profile of HERX1 (black) and all the sources in the sample (red). The green dotted line on each plot represents the renormalization radius (i.e., pileup-free radius), which is 2\farcs5 for all the sources except Cen\,A and 2MASXJ23444, where we adopt 4\farc{}. The errors plotted are 99.987\% confidence.}\label{fig:rprof_62-65}	
% Agrandar ejes y labels
\end{figure*}

\begin{figure*}
%\centering
\hspace*{-1cm}\includegraphics[width=1.0\textwidth,trim={2cm 6cm 6cm 8cm},clip]{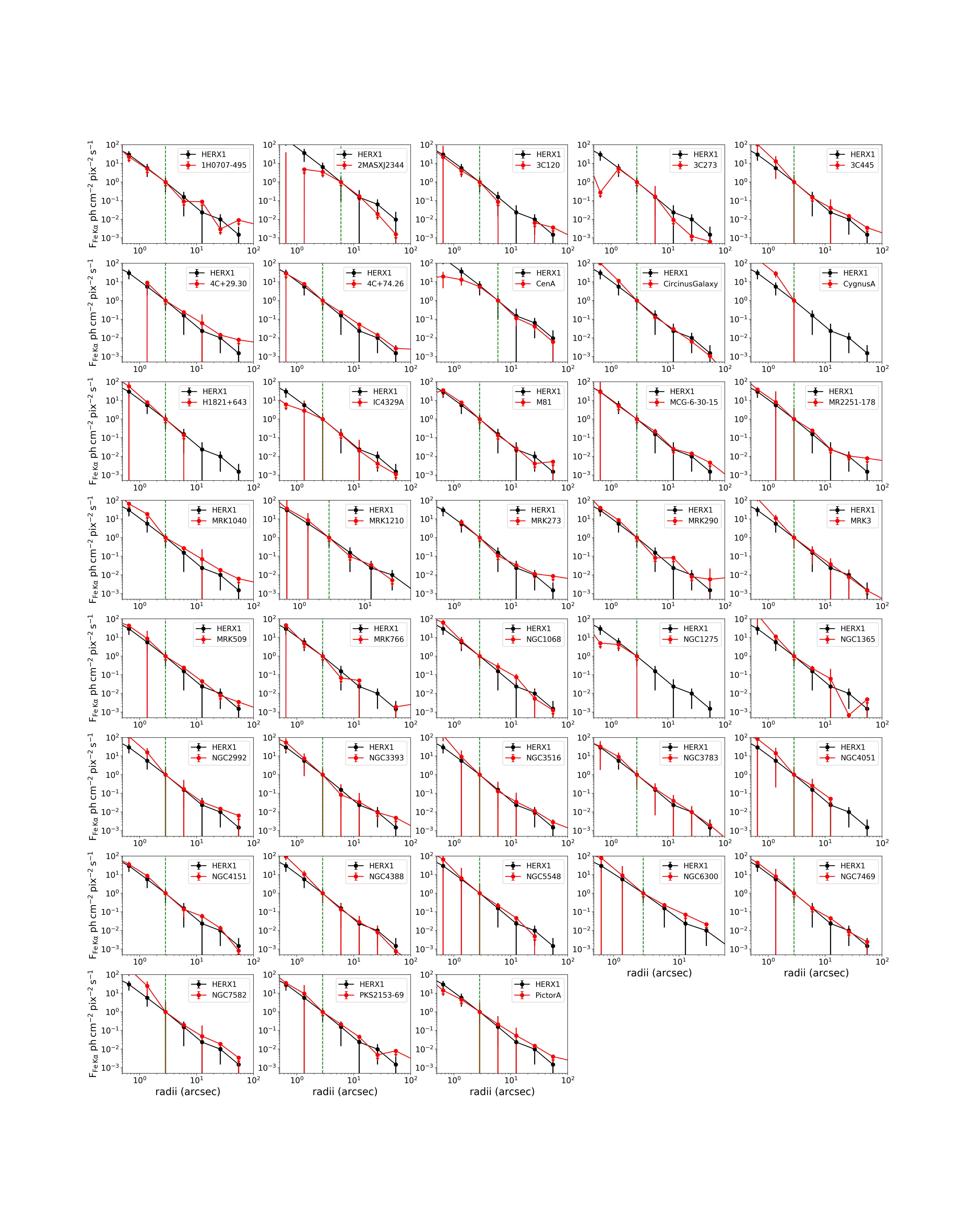}
\caption{Comparison between the Fe K$\alpha$ radial profile of HERX1 (black) and all the sources in the sample
(red). The green dotted line on each plot represents the renormalization radius (i.e., pileup-free radius), which is 2\farcs5 for all the sources except Cen\,A and 2MASXJ23444, where we adopt 4\farc{}. The errors plotted are 99.987\% confidence.}\label{fig:rprof_fe}	
% Agrandar ejes y labels
\end{figure*}

Similarly, Fig.~\ref{fig:rprof_62-65} shows a comparison of the 6.2--6.5\,keV\, band (\kalfa{} line and continuum) for each source in our sample and the 6.2--6.5\,keV\, profile of HERX1, renormalized at 2\farcs5.

\begin{figure*}
\centering
\hspace*{-0.2cm}\includegraphics[width=1.05\textwidth,trim={4cm 6cm 1cm 6cm},clip]{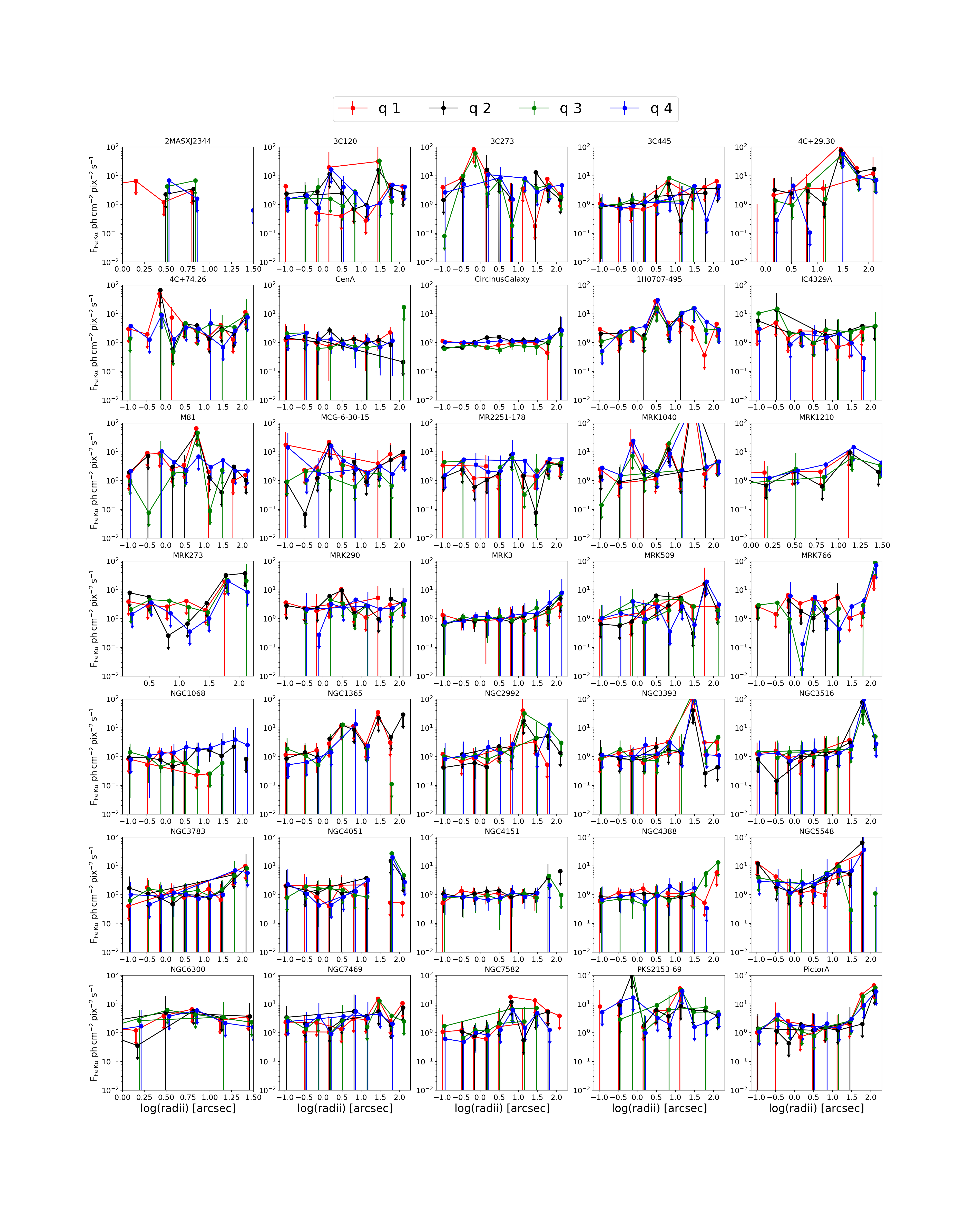}
\caption{Comparison of continuum-subtracted Fe K$\alpha$ radial profiles per quadrant (q1, q2, q3, and q4, as defined in Appendix Fig.~\ref{fig:Circinus1068}) for all the sources in the sample. The profiles have been divided by 25\% of the renormlized HERX1 radial profile used in Fig.~\ref{fig:rprof_fe}, such that azimuthally symmetric profiles consistent with the PSF should have values consistent with 1 at all radii. The errors plotted are 99.987\% confidence. The radial bins of the different quadrants have been shifted slightly so that the error bars are easier to compare.
}\label{fig:all_quarters_line}	
% Agrandar ejes y labels
\end{figure*}

Finally, Fig.~\ref{fig:all_quarters_line} compares the continuum-subtracted \kalfa{} radial profiles split into four quadrants for all of the sources in our sample. The profiles have been divided by 25\% of the renormlized HERX1 radial profile used in Fig.~\ref{fig:rprof_fe}, such that azimuthally symmetric profiles consistent with the PSF should have values consistent with 1 at all radii.

\begin{figure*}
\centering
\includegraphics[trim=0 0 0 30,clip,width=0.7\textwidth]{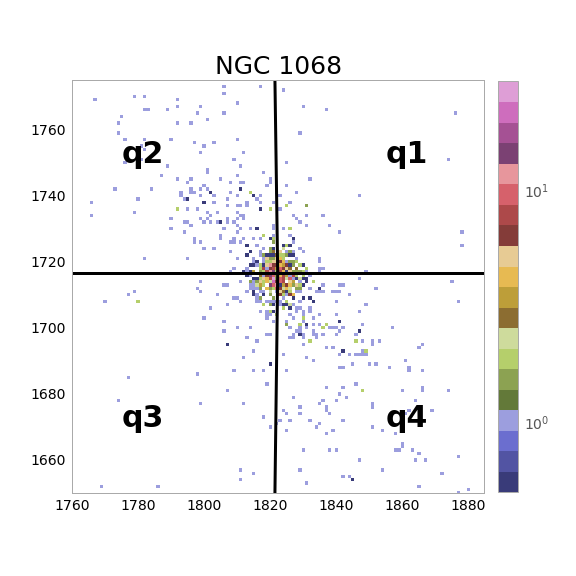}
\includegraphics[trim=0 60 0 30,clip,width=0.7\textwidth]{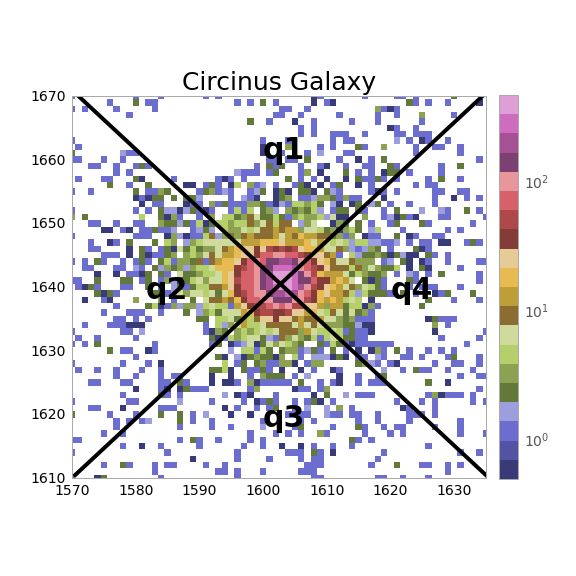}
\caption{\kalfa{} images of the central 18\farc{}$\times$ 18\farc{} of NGC\,1068 (top) and the central 8\farc{}$\times$ 8\farc{} of Circinus Galaxy  (bottom). The quadrants are denoted in the image. }\label{fig:Circinus1068}	
% Agrandar ejes y labels
\end{figure*}

\section{Contribution from multiple reflectors}\label{ap:mult_ref}

The size of the reflector, which is estimated through the relative variance of the \kalfa\ light curve compared to the continuum light curve, can be viewed as a weighted average of the distances of different reflecting structures. This method uses the ratio between the flux-normalized variances of the intrinsic (continuum) and reflected (\kalfa) light curves to estimate the size of the reflector. Therefore when comparing the contribution of different reflectors, the key is to understand how they all affect the measured variance of the reflected light curve.  

Assuming only two different reflecting regions for simplicity of the argument, we can construct the \kalfa\ light curve as the sum of two reflected light curves $R_1$ and $R_2$. The flux-normalized variance of this light curve will be 
\begin{eqnarray}
\sigma_{\rm Fe}^2=\frac{<(R_1+R_2)^2>-<R_1+R_2>^2}{<R_1+R_2>^2}\\ =\frac{<R_1^2>-<R_1>^2+<R_2^2>-<R_2>^2}{<R_1+R_2>^2}\\+\frac{2}{<R_1+R_2>^2}(<R_1R_2>-<R_1><R_2>)\\
=\frac{<R_1>^2}{<R_1+R_2>^2}\sigma^2_{\rm Fe,1}+\frac{<R_2>^2}{<R_1+R_2>^2}\sigma^2_{\rm Fe,2} \\ +\frac{2}{<R_1+R_2>^2}(<R_1R_2>-<R_1><R_2>),
\end{eqnarray}
i.e., the weighted average of the flux-normalized variances of each reflected light curve, where the weighting factor is the square of the fraction of \kalfa\ flux contributed by each reflector, plus a mixed term. Separating the reflected light curves into their mean fluxes $\bar R_{1,2}$ and their deviations around the mean $\delta_{1,2}$, the part within parentheses of the mixed term (G.5) can be expressed as 
\begin{eqnarray}
<R_1R_2>-<R_1><R_2>\\
= <\bar R_1(1+\delta_1) \bar R_2(1+\delta_2)>-\bar R_1 \bar R_2\\
= \bar R_1\bar R_2<(1+\delta_1)(1+\delta_2)>-\bar R_1 \bar R_2\\
= \bar R_1\bar R_2<(1+\delta_1+\delta_2+\delta_1\delta_2)>-\bar R_1 \bar R_2\\
= \bar R_1\bar R_2(1+<\delta_1\delta_2>-1)\\
= \bar R_1\bar R_2<\delta_1\delta_2>.
\end{eqnarray}
If the reflectors are sufficiently different, the deviations around the mean fluxes of their reflected light curves will be shifted in time by different delay times and therefore will tend to be uncorrelated. In this case, the mixed term will tend to vanish and the variance of the \kalfa\ light curve tends toward the weighted average discussed above.

\section{Poisson noise estimate using simulations}\label{ap:noise}
As described in the main text we used flux-randomization to estimate the contribution of observational noise to the variance. The variance of the flux-randomized light curve can be viewed as 
\begin{equation}
    \sigma^2_{\rm sim} =\ <lc^2>-<lc>^2\ =\ <(s+e+n)^2>-<s+e+n>^2,
\end{equation} where $lc$ is the flux-randomized light curve, $s$ is the intrinsic variability signal, $e$ are the deviations of each flux measurement due to observation error and $n$ are the flux deviates from the flux randomization. We have avoided the explicit time dependence of all these quantities for simplicity. Assuming that the intrinsic signal, the sign of the observational error and of the additional noise are uncorrelated and that the last two are distributed around zero, the cross terms and the expectation value of $n$ and $e$ can be canceled, which  results in 
\begin{eqnarray}
\sigma^2_{\rm sim}=<s^2>+<e^2>+<n^2> - <s>^2\\
 = <s^2>-<s>^2+<e^2>+<n^2>,
\end{eqnarray}
i.e., the intrinsic variance, $\sigma^2_{\rm intrinsic}$ plus the variance of the noise, $<e^2>=\sigma^2_{\rm noise}$, and the variance of the additional noise, $<n^2>=\sigma^2_{\rm noise, estimate}$. Therefore, taking the difference between the variance of the flux-randomized light curve and the original light curve results in an estimate of the variance of the noise, i.e., ($\sigma^2_{\rm sim})-(\sigma^2_{\rm original})=(\sigma^2_{\rm intrinsic}+\sigma^2_{\rm noise}+\sigma^2_{\rm noise,estimate}) - (\sigma^2_{\rm intrinsic}+\sigma^2_{\rm noise}) = \sigma^2_{\rm noise,estimate}$. Repeating the flux-randomization many times then produces the distribution of possible contributions of observational noise to the variance, for the particular flux and error values of a given light curve.

\end{document}